\documentclass[11pt]{article}
\pdfoutput=1

\usepackage{jheppub}
\usepackage[utf8]{inputenc}
\usepackage{booktabs}
\usepackage{graphicx}
\usepackage{bbold}
\usepackage{subcaption}

\input{widebar}
\usepackage{xkeyval}

\usepackage{tikz,fp}
\usetikzlibrary{external,fixedpointarithmetic,decorations.pathmorphing,positioning,calc,fadings,decorations.markings,decorations.pathreplacing,patterns,arrows}

\tikzset{
  label distance=-3pt,
  >=stealth,
  inner sep=1.5pt,
  boson/.style={
    decoration={snake, segment length=2mm, amplitude=0.5mm},
    decorate,
  },
  quark/.style={
    postaction={decorate},
    decoration={markings,mark=at position .5 with {\draw[->] (-0.01,0) -- (0.07,0);}}
  },
  aquark/.style={
    postaction={decorate},
    decoration={markings,mark=at position .5 with {\draw[->] (0.01,0) -- (-0.07,0);}}
  },
  gluonOld/.style={
    line width=0.7pt,
    decorate,
    decoration={coil,aspect=-0.5,amplitude=-2pt,segment length=2.2pt}
  },
  gluon/.style={
    line width=0.5pt,
    decorate,
    decoration={coil,aspect=-0.75,amplitude=-1.5pt,segment length=2pt}
  },
  agluon/.style={
    line width=0.5pt,
    decorate,
    decoration={coil,aspect=0.75,amplitude=1.5pt,segment length=2pt}
  },
  scalar/.style={
    dashed
  },
  dscalar/.style={
    dashed,
    postaction={decorate},
    decoration={markings,mark=at position 0.65 with {\draw[->] (-0.01,0) -- (0.07,0);}}
  },
  adscalar/.style={
    dashed,
    postaction={decorate},
    decoration={markings,mark=at position 0.65 with {\draw[->] (-0.01,0) -- (0.07,0);}}
  },
  line/.style={
  },
  doubleline/.style={
    double
  },
  middleArrow/.style={
    draw=white,
    decoration={markings,mark=at position 0.65 with{\arrow[scale=1,black]{>}}},
    decorate
  },
  dot/.style={
  	postaction={decorate},
    decoration={markings,mark=between positions 0.5 and 0.5 step 1 with {\draw[fill] (0,0) circle [radius=1.5pt];}}  
  },
  dots/.style={
  	postaction={decorate},
    decoration={markings,mark=between positions 0.3 and 0.6 step 0.3 with {\draw[fill] (0,0) circle [radius=1.5pt];}}  
  },
  dotss/.style={
  	postaction={decorate},
    decoration={markings,mark=between positions 0.25 and 0.75 step 0.25 with {\draw[fill] (0,0) circle [radius=1.5pt];}}  
  },
  blob/.style={
  	pattern=north west lines,
    draw=black
  },
  blobBlack/.style={
    draw=black,
    fill=black
  },
  blobWhite/.style={
    draw=black,
    fill=white
  }
}

\makeatletter
\def\gSetStdExtTypes#1{\presetkeys{gTypes}{eA=#1,eB=#1,eC=#1,eD=#1,eE=#1}{}}
\def\gSetStdIntTypes#1{\presetkeys{gTypes}{iA=#1,iB=#1,iC=#1,iD=#1,iE=#1,iF=#1,iG=#1,iH=#1,iI=#1,iJ=#1}{}}
\def\gSetStdTypes#1{\gSetStdExtTypes{#1}\gSetStdIntTypes{#1}}

\define@cmdkeys{general}{scale}
\define@cmdkeys{general}{rotate}
\define@cmdkeys{general}{font}
\define@cmdkeys{general}{yshift}
\presetkeys{general}{scale=1, rotate=0, font=\scriptsize, yshift=0pt}{}

\define@cmdkeys{gTypes}{eA,eB,eC,eD,eE,iA,iB,iC,iD,iE,iF,iG,iH,iI,iJ}
\gSetStdTypes{line}
\define@key{pre}{all}{\gSetStdTypes{#1}}
\define@key{pre}{allE}{\gSetStdExtTypes{#1}}
\define@key{pre}{allI}{\gSetStdIntTypes{#1}}

\define@cmdkeys{gLabels}{eLA,eLB,eLC,eLD,eLE,iLA,iLB,iLC,iLD,iLE,iLF,iLG,iLH,iLI,iLJ}
\presetkeys{gLabels}{eLA={},eLB={},eLC={},eLD={},eLE={},iLA={},iLB={},iLC={},iLD={},iLE={},iLF={},iLG={},iLH={},iLI={},iLJ={}}{}

\define@boolkeys{dots}{lDots,rDots}
\presetkeys{dots}{lDots=false, rDots=false}{}

\define@cmdkeys{blobs}{blobA,blobB,blobC}
\presetkeys{blobs}{blobA=blob,blobB=blob,blobC=blob}{}

\def\gTreeTri{\@ifnextchar[\@gTreeTri{\@gTreeTri[]}}
\def\@gTreeTri[#1]#2{%
  \setkeys*{pre}{#1}%
  \setrmkeys{general,gTypes,gLabels}%
  \begin{tikzpicture}
    [line width=1pt,
    baseline={([yshift=-0.5ex+\cmdKV@general@yshift]current bounding box.center)},
    scale=\cmdKV@general@scale,
    rotate=\cmdKV@general@rotate,
    font=\cmdKV@general@font]
    \draw[\cmdKV@gTypes@eA] (0.7,0.3) -- (0.4,0.3) node[label=(-180+\cmdKV@general@rotate):\cmdKV@gLabels@eLA] {};
    \draw[\cmdKV@gTypes@eB] (0.7,0.3) -- (0.9,0.6) node[label=(\cmdKV@general@rotate):\cmdKV@gLabels@eLB] {};
    \draw[\cmdKV@gTypes@eC] (0.7,0.3) -- (0.9,0) node[label=(\cmdKV@general@rotate):\cmdKV@gLabels@eLC] {};
    #2
  \end{tikzpicture}
  \gSetStdTypes{line}
}

\def\gTreeS{\@ifnextchar[\@gTreeS{\@gTreeS[]}}
\def\@gTreeS[#1]#2{%
  \setkeys*{pre}{#1}%
  \setrmkeys{general,gTypes,gLabels}%
  \begin{tikzpicture}
    [line width=1pt,
    baseline={([yshift=-0.5ex+\cmdKV@general@yshift]current bounding box.center)},
    scale=\cmdKV@general@scale,
    rotate=\cmdKV@general@rotate,
    font=\cmdKV@general@font]
    \draw[\cmdKV@gTypes@eA] (0.2,0.3) -- (0,0) node[label=(-180+\cmdKV@general@rotate):\cmdKV@gLabels@eLA] {};
    \draw[\cmdKV@gTypes@eB] (0.2,0.3) -- (0,0.6) node[label=(180+\cmdKV@general@rotate):\cmdKV@gLabels@eLB] {};
    \draw[\cmdKV@gTypes@eC] (0.7,0.3) -- (0.9,0.6) node[label=(\cmdKV@general@rotate):\cmdKV@gLabels@eLC] {};
    \draw[\cmdKV@gTypes@eD] (0.7,0.3) -- (0.9,0) node[label=(\cmdKV@general@rotate):\cmdKV@gLabels@eLD] {};    
    \draw[label distance=0,\cmdKV@gTypes@iA] (0.2,0.3) -- node[label=(90+\cmdKV@general@rotate):\cmdKV@gLabels@iLA] {} (0.7,0.3);
    #2
  \end{tikzpicture}
  \gSetStdTypes{line}
}

\def\gTreeT{\@ifnextchar[\@gTreeT{\@gTreeT[]}}
\def\@gTreeT[#1]#2{%
  \setkeys*{pre}{#1}%
  \setrmkeys{general,gTypes,gLabels}%
  \begin{tikzpicture}
    [line width=1pt,
    baseline={([yshift=-0.5ex+\cmdKV@general@yshift]current bounding box.center)},
    scale=\cmdKV@general@scale,
    rotate=\cmdKV@general@rotate,
    font=\cmdKV@general@font]
    \draw[\cmdKV@gTypes@eA] (0.45,0.1) -- (0,0) node[label=(-180+\cmdKV@general@rotate):\cmdKV@gLabels@eLA] {};
    \draw[\cmdKV@gTypes@eB] (0.45,0.5) -- (0,0.6) node[label=(180+\cmdKV@general@rotate):\cmdKV@gLabels@eLB] {};
    \draw[\cmdKV@gTypes@eC] (0.45,0.5) -- (0.9,0.6) node[label=(\cmdKV@general@rotate):\cmdKV@gLabels@eLC] {};
    \draw[\cmdKV@gTypes@eD] (0.45,0.1) -- (0.9,0) node[label=(\cmdKV@general@rotate):\cmdKV@gLabels@eLD] {};
    \draw[label distance=0,\cmdKV@gTypes@iA] (0.45,0.5) -- node[label=(\cmdKV@general@rotate):\cmdKV@gLabels@iLA] {} (0.45,0.1);
    #2
  \end{tikzpicture}
  \gSetStdTypes{line}
}

\def\gTreeU{\@ifnextchar[\@gTreeU{\@gTreeU[]}}
\def\@gTreeU[#1]#2{%
  \setkeys*{pre}{#1}%
  \setrmkeys{general,gTypes,gLabels}%
  \begin{tikzpicture}
    [line width=1pt,
    baseline={([yshift=-0.5ex+\cmdKV@general@yshift]current bounding box.center)},
    scale=\cmdKV@general@scale,
    rotate=\cmdKV@general@rotate,
    font=\cmdKV@general@font]
    \draw[\cmdKV@gTypes@eA] (0.45,0.5) -- (0,0) node[label=(-180+\cmdKV@general@rotate):\cmdKV@gLabels@eLA] {};
    \foreach \x in {5,...,14} \fill[opacity=0.12,white] (0.27,0.3) circle (\x/100);
    \draw[\cmdKV@gTypes@eB] (0.45,0.1) -- (0,0.6) node[label=(180+\cmdKV@general@rotate):\cmdKV@gLabels@eLB] {};
    \draw[\cmdKV@gTypes@eC] (0.45,0.5) -- (0.9,0.6) node[label=(\cmdKV@general@rotate):\cmdKV@gLabels@eLC] {};
    \draw[\cmdKV@gTypes@eD] (0.45,0.1) -- (0.9,0) node[label=(\cmdKV@general@rotate):\cmdKV@gLabels@eLD] {};
    \draw[label distance=0,\cmdKV@gTypes@iA] (0.45,0.5) -- node[label=(\cmdKV@general@rotate):\cmdKV@gLabels@iLA] {} (0.45,0.1);
    #2
  \end{tikzpicture}
  \gSetStdTypes{line}
}

\def\gTreeFourPt{\@ifnextchar[\@gTreeFourPt{\@gTreeFourPt[]}}
\def\@gTreeFourPt[#1]#2{%
  \setkeys*{pre}{#1}%
  \setrmkeys{general,gTypes,gLabels}%
  \begin{tikzpicture}
    [line width=1pt,
    baseline={([yshift=-0.5ex+\cmdKV@general@yshift]current bounding box.center)},
    scale=\cmdKV@general@scale,
    rotate=\cmdKV@general@rotate,
    font=\cmdKV@general@font]
    \draw[\cmdKV@gTypes@eA] (0,0) -- (-0.4,-0.4) node[label=(-180+\cmdKV@general@rotate):\cmdKV@gLabels@eLA] {};
    \draw[\cmdKV@gTypes@eB] (0,0) -- (-0.4,0.4) node[label=(180+\cmdKV@general@rotate):\cmdKV@gLabels@eLB] {};
    \draw[\cmdKV@gTypes@eC] (0,0) -- (0.4,0.4) node[label=(\cmdKV@general@rotate):\cmdKV@gLabels@eLC] {};
    \draw[\cmdKV@gTypes@eD] (0,0) -- (0.4,-0.4) node[label=(\cmdKV@general@rotate):\cmdKV@gLabels@eLD] {};
    #2
  \end{tikzpicture}
  \gSetStdTypes{line}
}

\def\gTadA{\@ifnextchar[\@gTadA{\@gTadA[]}}
\def\@gTadA[#1]#2{%
  \setkeys*{pre}{#1}%
  \setrmkeys{general,gTypes,gLabels}%
  \begin{tikzpicture}
    [line width=1pt,
    baseline={([yshift=-0.5ex+\cmdKV@general@yshift]current bounding box.center)},
    scale=\cmdKV@general@scale,
    rotate=\cmdKV@general@rotate,
    font=\cmdKV@general@font]
    \draw[\cmdKV@gTypes@eA] (-0.7,0.3) -- (-0.9,0) node[label=(180+\cmdKV@general@rotate):\cmdKV@gLabels@eLA] {};
    \draw[\cmdKV@gTypes@eB] (-0.7,0.3) -- (-0.9,0.6) node[label=(180+\cmdKV@general@rotate):\cmdKV@gLabels@eLB] {};
    \draw[\cmdKV@gTypes@eC] (0.15,0.45) -- (0.5,0.6) node[label=(\cmdKV@general@rotate):\cmdKV@gLabels@eLC] {};
    \draw[\cmdKV@gTypes@eD] (-0.3,0.3) -- (-0.2,0) node[label=(\cmdKV@general@rotate):\cmdKV@gLabels@eLD] {};
    \draw[label distance=0,\cmdKV@gTypes@iA] (-0.7,0.3) -- node[label=(90+\cmdKV@general@rotate):\cmdKV@gLabels@iLA] {} (-0.3,0.3);
    \draw[label distance=0,\cmdKV@gTypes@iB] (-0.3,0.3) -- node[label=(-80+\cmdKV@general@rotate):\cmdKV@gLabels@iLB] {} (0.15,0.45);
    \draw[label distance=0,\cmdKV@gTypes@iC] (0.15,0.45) -- node[label=(45+\cmdKV@general@rotate):\cmdKV@gLabels@iLC] {} (-0.18,0.69);
    \draw[label distance=-0.5,\cmdKV@gTypes@iD] (-0.18,0.69) .. controls ($(-0.18,0.69) + (-0.12,-0.12)$) and ($(-0.35,0.87) + (-0.12,-0.12)$) .. (-0.35,0.87) node[label=(180+\cmdKV@general@rotate):\cmdKV@gLabels@iLD] {} .. controls ($(-0.35,0.87) + (0.12,0.12)$) and ($(-0.18,0.69) + (0.12,0.12)$) .. (-0.18,0.69);
    #2
  \end{tikzpicture}
  \gSetStdTypes{line}
}

\def\gTadB{\@ifnextchar[\@gTadB{\@gTadB[]}}
\def\@gTadB[#1]#2{%
  \setkeys*{pre}{#1}%
  \setrmkeys{general,gTypes,gLabels}%
  \begin{tikzpicture}
    [line width=1pt,
    baseline={([yshift=-0.5ex+\cmdKV@general@yshift]current bounding box.center)},
    scale=\cmdKV@general@scale,
    rotate=\cmdKV@general@rotate,
    font=\cmdKV@general@font]
    \draw[\cmdKV@gTypes@eA] (0,0.3) -- (-0.2,0) node[label=(-180+\cmdKV@general@rotate):\cmdKV@gLabels@eLA] {};
    \draw[\cmdKV@gTypes@eB] (0,0.3) -- (-0.2,0.6) node[label=(180+\cmdKV@general@rotate):\cmdKV@gLabels@eLB] {};
    \draw[\cmdKV@gTypes@eC] (0.7,0.3) -- (0.9,0.6) node[label=(\cmdKV@general@rotate):\cmdKV@gLabels@eLC] {};
    \draw[\cmdKV@gTypes@eD] (0.7,0.3) -- (0.9,0) node[label=(\cmdKV@general@rotate):\cmdKV@gLabels@eLD] {};
    \draw[label distance=0,\cmdKV@gTypes@iA] (0,0.3) -- node[label=(-90+\cmdKV@general@rotate):\cmdKV@gLabels@iLA] {} (0.35,0.3);
    \draw[label distance=0,\cmdKV@gTypes@iB] (0.35,0.3) -- node[label=(\cmdKV@general@rotate):\cmdKV@gLabels@iLB] {} (0.35,0.6);
    \draw[label distance=-0.5,\cmdKV@gTypes@iC] (0.35,0.6) .. controls (0.18,0.6) and (0.18,0.85) .. (0.35,0.85) .. controls (0.52,0.85) and (0.52,0.6) .. node[label=(\cmdKV@general@rotate):\cmdKV@gLabels@iLC] {} (0.35,0.6);
    \draw[label distance=0,\cmdKV@gTypes@iD] (0.35,0.3) -- node[label=(-90+\cmdKV@general@rotate):\cmdKV@gLabels@iLD] {} (0.7,0.3);
    #2
  \end{tikzpicture}
  \gSetStdTypes{line}
}
\def\gBubA{\@ifnextchar[\@gBubA{\@gBubA[]}}
\def\@gBubA[#1]#2{%
  \setkeys*{pre}{#1}%
  \setrmkeys{general,gTypes,gLabels}%
  \begin{tikzpicture}
    [line width=1pt,
    baseline={([yshift=-0.5ex+\cmdKV@general@yshift]current bounding box.center)},
    scale=\cmdKV@general@scale,
    rotate=\cmdKV@general@rotate,
    font=\cmdKV@general@font]
    \draw[\cmdKV@gTypes@eA] (0.5,0.7) -- (0.2,0.7) node[label=(180+\cmdKV@general@rotate):\cmdKV@gLabels@eLA] {};
    \draw[\cmdKV@gTypes@eB] (0.9,0.7) -- (1.2,0.7) node[label=(\cmdKV@general@rotate):\cmdKV@gLabels@eLB] {};
    \draw[label distance=0,\cmdKV@gTypes@iA] (0.5,0.7) .. controls (0.5,0.95) and (0.9,0.95) .. node[label=(90+\cmdKV@general@rotate):\cmdKV@gLabels@iLA] {} (0.9,0.7);
    \draw[label distance=0,\cmdKV@gTypes@iB] (0.9,0.7) .. controls (0.9,0.45) and (0.5,0.45) .. node[label=(-90+\cmdKV@general@rotate):\cmdKV@gLabels@iLB] {} (0.5,0.7);
    #2
  \end{tikzpicture}
  \gSetStdTypes{line}
}
\def\gBubB{\@ifnextchar[\@gBubB{\@gBubB[]}}
\def\@gBubB[#1]#2{%
  \setkeys*{pre}{#1}%
  \setrmkeys{general,gTypes,gLabels}%
  \begin{tikzpicture}
    [line width=1pt,
    baseline={([yshift=-0.5ex+\cmdKV@general@yshift]current bounding box.center)},
    scale=\cmdKV@general@scale,
    rotate=\cmdKV@general@rotate,
    font=\cmdKV@general@font]
    \draw[\cmdKV@gTypes@eA] (0.2,0.7) -- (0,0.4) node[label=(-180+\cmdKV@general@rotate):\cmdKV@gLabels@eLA] {};
    \draw[\cmdKV@gTypes@eB] (0.2,0.7) -- (0,1) node[label=(180+\cmdKV@general@rotate):\cmdKV@gLabels@eLB] {};
    \draw[\cmdKV@gTypes@eC] (1.2,0.7) -- (1.4,1) node[label=(\cmdKV@general@rotate):\cmdKV@gLabels@eLC] {};
    \draw[\cmdKV@gTypes@eD] (1.2,0.7) -- (1.4,0.4) node[label=(\cmdKV@general@rotate):\cmdKV@gLabels@eLD] {};
    \draw[label distance=0,\cmdKV@gTypes@iA] (0.2,0.7) -- node[label=(90+\cmdKV@general@rotate):\cmdKV@gLabels@iLA] {} (0.5,0.7);
    \draw[label distance=0,\cmdKV@gTypes@iB] (0.5,0.7) .. controls (0.5,0.95) and (0.9,0.95) .. node[label=(90+\cmdKV@general@rotate):\cmdKV@gLabels@iLB] {} (0.9,0.7);
    \draw[label distance=0,\cmdKV@gTypes@iC] (0.9,0.7) .. controls (0.9,0.45) and (0.5,0.45) .. node[label=(-90+\cmdKV@general@rotate):\cmdKV@gLabels@iLC] {} (0.5,0.7);
    \draw[label distance=0,\cmdKV@gTypes@iD] (0.9,0.7) -- node[label=(90+\cmdKV@general@rotate):\cmdKV@gLabels@iLD] {} (1.2,0.7);
    #2
  \end{tikzpicture}
  \gSetStdTypes{line}
}

\def\gBubC{\@ifnextchar[\@gBubC{\@gBubC[]}}
\def\@gBubC[#1]#2{%
  \setkeys*{pre}{#1}%
  \setrmkeys{general,gTypes,gLabels}%
  \begin{tikzpicture}
    [line width=1pt,
    baseline={([yshift=-0.5ex+\cmdKV@general@yshift]current bounding box.center)},
    scale=\cmdKV@general@scale,
    rotate=\cmdKV@general@rotate,
    font=\cmdKV@general@font]
    \draw[\cmdKV@gTypes@eA] (-0.7,0.3) -- (-0.9,0) node[label=(180+\cmdKV@general@rotate):\cmdKV@gLabels@eLA] {};
    \draw[\cmdKV@gTypes@eB] (-0.7,0.3) -- (-0.9,0.6) node[label=(180+\cmdKV@general@rotate):\cmdKV@gLabels@eLB] {};
    \draw[\cmdKV@gTypes@eC] (0.3,0.5) -- (0.5,0.6) node[label=(\cmdKV@general@rotate):\cmdKV@gLabels@eLC] {};
    \draw[\cmdKV@gTypes@eD] (-0.3,0.3) -- (-0.2,0) node[label=(\cmdKV@general@rotate):\cmdKV@gLabels@eLD] {};
    \draw[label distance=0,\cmdKV@gTypes@iA] (-0.7,0.3) -- node[label=(90+\cmdKV@general@rotate):\cmdKV@gLabels@iLA] {} (-0.3,0.3);
    \draw[label distance=0,\cmdKV@gTypes@iB] (-0.3,0.3) -- node[label=(90+\cmdKV@general@rotate):\cmdKV@gLabels@iLB] {} (-0,0.4);
    \draw[label distance=0,\cmdKV@gTypes@iC] (0,0.4) .. controls (-0.03,0.6) and (0.22,0.65) .. node[label=(90+\cmdKV@general@rotate):\cmdKV@gLabels@iLC] {} (0.3,0.5);
    \draw[label distance=0,\cmdKV@gTypes@iD] (0.3,0.5) .. controls (0.38,0.3) and (0.08,0.21) .. node[label=(-90+\cmdKV@general@rotate):\cmdKV@gLabels@iLD] {} (0,0.4);
    #2
  \end{tikzpicture}
  \gSetStdTypes{line}
}
\def\gTriA{\@ifnextchar[\@gTriA{\@gTriA[]}}
\def\@gTriA[#1]#2{%
  \setkeys*{pre}{#1}%
  \setrmkeys{general,gTypes,gLabels}%
  \begin{tikzpicture}
    [line width=1pt,
    baseline={([yshift=-0.5ex+\cmdKV@general@yshift]current bounding box.center)},
    scale=\cmdKV@general@scale,
    rotate=\cmdKV@general@rotate,
    font=\cmdKV@general@font]
    \draw[\cmdKV@gTypes@eC] (0:0.3) -- (0:0.55) node[label=(\cmdKV@general@rotate):\cmdKV@gLabels@eLC] {};
    \draw[\cmdKV@gTypes@eA] (-120:0.3) -- (-120:0.55) node[label=(-180+\cmdKV@general@rotate):\cmdKV@gLabels@eLA] {};
    \draw[\cmdKV@gTypes@eB] (120:0.3) -- (120:0.55) node[label=(180+\cmdKV@general@rotate):\cmdKV@gLabels@eLB] {};    
    \draw[label distance=0,\cmdKV@gTypes@iA] (-120:0.3) -- node[label=(180+\cmdKV@general@rotate):\cmdKV@gLabels@iLA] {} (120:0.3);
    \draw[label distance=0,\cmdKV@gTypes@iB] (120:0.3) -- node[label=(60+\cmdKV@general@rotate):\cmdKV@gLabels@iLB] {} (0:0.3);
    \draw[label distance=0,\cmdKV@gTypes@iC] (0:0.3) -- node[label=(-60+\cmdKV@general@rotate):\cmdKV@gLabels@iLC] {} (-120:0.3);
    #2
  \end{tikzpicture}
  \gSetStdTypes{line}
}
\def\gTriB{\@ifnextchar[\@gTriB{\@gTriB[]}}
\def\@gTriB[#1]#2{%
  \setkeys*{pre}{#1}%
  \setrmkeys{general,gTypes,gLabels}%
  \begin{tikzpicture}
    [line width=1pt,
    baseline={([yshift=-0.5ex+\cmdKV@general@yshift]current bounding box.center)},
    scale=\cmdKV@general@scale,
    rotate=\cmdKV@general@rotate,
    font=\cmdKV@general@font]
    \draw[\cmdKV@gTypes@eA] (0.3,0.45) -- (0,0.3) node[label=(-180+\cmdKV@general@rotate):\cmdKV@gLabels@eLA] {};
    \draw[\cmdKV@gTypes@eB] (0.3,0.95) -- (0,1.1) node[label=(180+\cmdKV@general@rotate):\cmdKV@gLabels@eLB] {};
    \draw[\cmdKV@gTypes@eC] (1,0.7) -- (1.3,1.1) node[label=(\cmdKV@general@rotate):\cmdKV@gLabels@eLC] {};
    \draw[\cmdKV@gTypes@eD] (1,0.7) -- (1.3,0.3) node[label=(\cmdKV@general@rotate):\cmdKV@gLabels@eLD] {};
    \draw[label distance=0,\cmdKV@gTypes@iA] (0.3,0.45) -- node[label=(180+\cmdKV@general@rotate):\cmdKV@gLabels@iLA] {} (0.3,0.95);
    \draw[label distance=0,\cmdKV@gTypes@iB] (0.3,0.95) -- node[label=(70+\cmdKV@general@rotate):\cmdKV@gLabels@iLB] {} (0.7,0.7);
    \draw[label distance=0,\cmdKV@gTypes@iC] (0.7,0.7) -- node[label=(-70+\cmdKV@general@rotate):\cmdKV@gLabels@iLC] {} (0.3,0.45);
    \draw[label distance=0,\cmdKV@gTypes@iD] (1,0.7) -- node[label=(90+\cmdKV@general@rotate):\cmdKV@gLabels@iLD] {} (0.7,0.7);
    #2
  \end{tikzpicture}
  \gSetStdTypes{line}
}
\def\gBox{\@ifnextchar[\@gBox{\@gBox[]}}
\def\@gBox[#1]#2{%
  \setkeys*{pre}{#1}%
  \setrmkeys{general,gTypes,gLabels}%
  \begin{tikzpicture}
    [line width=1pt,
    baseline={([yshift=-0.5ex+\cmdKV@general@yshift]current bounding box.center)},
    scale=\cmdKV@general@scale,
    rotate=\cmdKV@general@rotate,
    font=\cmdKV@general@font]
    \draw[\cmdKV@gTypes@eA] (0.25,0.25) -- (0,0) node[label=(-180+\cmdKV@general@rotate):\cmdKV@gLabels@eLA] {};
    \draw[\cmdKV@gTypes@eB] (0.25,0.75) -- (0,1) node[label=(180+\cmdKV@general@rotate):\cmdKV@gLabels@eLB] {};
    \draw[\cmdKV@gTypes@eC] (0.75,0.75) -- (1,1) node[label=(\cmdKV@general@rotate):\cmdKV@gLabels@eLC] {};
    \draw[\cmdKV@gTypes@eD] (0.75,0.25) -- (1,0) node[label=(\cmdKV@general@rotate):\cmdKV@gLabels@eLD] {};
    \draw[label distance=0,\cmdKV@gTypes@iA] (0.25,0.25) -- node[label=(180+\cmdKV@general@rotate):\cmdKV@gLabels@iLA] {} (0.25,0.75);
    \draw[label distance=0,\cmdKV@gTypes@iB] (0.25,0.75) -- node[label=(90+\cmdKV@general@rotate):\cmdKV@gLabels@iLB] {} (0.75,0.75);
    \draw[label distance=0,\cmdKV@gTypes@iC] (0.75,0.75) -- node[label=(\cmdKV@general@rotate):\cmdKV@gLabels@iLC] {} (0.75,0.25);
    \draw[label distance=0,\cmdKV@gTypes@iD] (0.75,0.25) -- node[label=(-90+\cmdKV@general@rotate):\cmdKV@gLabels@iLD] {} (0.25,0.25);
    #2
  \end{tikzpicture}
  \gSetStdTypes{line}
}
\def\gPenta{\@ifnextchar[\@gPenta{\@gPenta[]}}
\def\@gPenta[#1]#2{%
  \setkeys*{pre}{#1}%
  \setrmkeys{general,gTypes,gLabels}%
  \begin{tikzpicture}
    [line width=1pt,
    baseline={([yshift=-0.5ex+\cmdKV@general@yshift]current bounding box.center)},
    scale=\cmdKV@general@scale,
    rotate=\cmdKV@general@rotate,
    font=\cmdKV@general@font]
    \draw[\cmdKV@gTypes@eA] (-144:0.35) -- (-144:0.75) node[label=(-144+\cmdKV@general@rotate):\cmdKV@gLabels@eLA] {};
    \draw[\cmdKV@gTypes@eB] (144:0.35) -- (144:0.75) node[label=(144+\cmdKV@general@rotate):\cmdKV@gLabels@eLB] {};
    \draw[\cmdKV@gTypes@eC] (72:0.35) -- (72:0.75) node[label=(72+\cmdKV@general@rotate):\cmdKV@gLabels@eLC] {};
    \draw[\cmdKV@gTypes@eD] (0:0.35) -- (0:0.75) node[label=(\cmdKV@general@rotate):\cmdKV@gLabels@eLD] {};
    \draw[\cmdKV@gTypes@eE] (-72:0.35) -- (-72:0.75) node[label=(-72+\cmdKV@general@rotate):\cmdKV@gLabels@eLE] {};
    \draw[label distance=0,\cmdKV@gTypes@iA] (-144:0.35) -- node[label=(180+\cmdKV@general@rotate):\cmdKV@gLabels@iLA] {} (144:0.35);
    \draw[label distance=0,\cmdKV@gTypes@iB] (144:0.35) -- node[label=(108+\cmdKV@general@rotate):\cmdKV@gLabels@iLB] {} (72:0.35);
    \draw[label distance=0,\cmdKV@gTypes@iC] (72:0.35) -- node[label=(36+\cmdKV@general@rotate):\cmdKV@gLabels@iLC] {} (0:0.35);
    \draw[label distance=0,\cmdKV@gTypes@iD] (0:0.35) -- node[label=(-36+\cmdKV@general@rotate):\cmdKV@gLabels@iLD] {} (-72:0.35);
    \draw[label distance=0,\cmdKV@gTypes@iE] (-72:0.35) -- node[label=(-108+\cmdKV@general@rotate):\cmdKV@gLabels@iLE] {} (-144:0.35);
    #2
  \end{tikzpicture}
  \gSetStdTypes{line}
}

\def\gBoxBox{\@ifnextchar[\@gBoxBox{\@gBoxBox[]}}
\def\@gBoxBox[#1]#2{%
  \setkeys*{pre}{#1}%
  \setrmkeys{general,gTypes,gLabels}%
  \begin{tikzpicture}
    [line width=1pt,
    baseline={([yshift=-0.5ex+\cmdKV@general@yshift]current bounding box.center)},
    scale=\cmdKV@general@scale,
    rotate=\cmdKV@general@rotate,
    font=\cmdKV@general@font]
    \draw[\cmdKV@gTypes@eA] (0.3,0.3) -- (0,0) node[label=(-180+\cmdKV@general@rotate):\cmdKV@gLabels@eLA] {};
    \draw[\cmdKV@gTypes@eB] (0.3,0.9) -- (0,1.2) node[label=(180+\cmdKV@general@rotate):\cmdKV@gLabels@eLB] {};
    \draw[\cmdKV@gTypes@eC] (1.5,0.9) -- (1.8,1.2) node[label=(\cmdKV@general@rotate):\cmdKV@gLabels@eLC] {};
    \draw[\cmdKV@gTypes@eD] (1.5,0.3) -- (1.8,0) node[label=(\cmdKV@general@rotate):\cmdKV@gLabels@eLD] {};
    \draw[label distance=0,\cmdKV@gTypes@iA] (0.3,0.3) -- node[label=(180+\cmdKV@general@rotate):\cmdKV@gLabels@iLA] {} (0.3,0.9);
    \draw[label distance=0,\cmdKV@gTypes@iB] (0.3,0.9) -- node[label=(90+\cmdKV@general@rotate):\cmdKV@gLabels@iLB] {} (0.9,0.9);
    \draw[label distance=0,\cmdKV@gTypes@iC] (0.9,0.9) -- node[label=(90+\cmdKV@general@rotate):\cmdKV@gLabels@iLC] {} (1.5,0.9);
    \draw[label distance=0,\cmdKV@gTypes@iD] (1.5,0.9) -- node[label=(\cmdKV@general@rotate):\cmdKV@gLabels@iLD] {} (1.5,0.3);
    \draw[label distance=0,\cmdKV@gTypes@iE] (1.5,0.3) -- node[label=(-90+\cmdKV@general@rotate):\cmdKV@gLabels@iLE] {} (0.9,0.3);
    \draw[label distance=0,\cmdKV@gTypes@iF] (0.9,0.3) -- node[label=(-90+\cmdKV@general@rotate):\cmdKV@gLabels@iLF] {} (0.3,0.3);
    \draw[label distance=0,\cmdKV@gTypes@iG] (0.9,0.9) -- node[label=(\cmdKV@general@rotate):\cmdKV@gLabels@iLG] {} (0.9,0.3);
    #2
  \end{tikzpicture}
  \gSetStdTypes{line}
}
\def\gBoxBoxNP{\@ifnextchar[\@gBoxBoxNP{\@gBoxBoxNP[]}}
\def\@gBoxBoxNP[#1]#2{%
  \setkeys*{pre}{#1}%
  \setrmkeys{general,gTypes,gLabels}%
  \begin{tikzpicture}
    [line width=1pt,
    baseline={([yshift=-0.5ex+\cmdKV@general@yshift]current bounding box.center)},
    scale=\cmdKV@general@scale,
    rotate=\cmdKV@general@rotate,
    font=\cmdKV@general@font]
    \draw[\cmdKV@gTypes@eA] (-1.9,0.3) -- (-2.2,0) node[label=(180+\cmdKV@general@rotate):\cmdKV@gLabels@eLA] {};
    \draw[\cmdKV@gTypes@eB] (-1.9,1.1) -- (-2.2,1.4) node[label=(180+\cmdKV@general@rotate):\cmdKV@gLabels@eLB] {};
    \draw[\cmdKV@gTypes@eC] (-0.3,0.7) -- (0,0.7) node[label=(\cmdKV@general@rotate):\cmdKV@gLabels@eLC] {};
    \draw[\cmdKV@gTypes@eD] (-1.1,0.7) -- (-1.4,0.7) node[label=(180+\cmdKV@general@rotate):\cmdKV@gLabels@eLD] {};
    \draw[label distance=0,\cmdKV@gTypes@iA] (-1.9,0.3) -- node[label=(180+\cmdKV@general@rotate):\cmdKV@gLabels@iLA] {} (-1.9,1.1);
    \draw[label distance=0,\cmdKV@gTypes@iB] (-1.9,1.1) -- node[label=(90+\cmdKV@general@rotate):\cmdKV@gLabels@iLB] {} (-0.7,1.1);
    \draw[label distance=0,\cmdKV@gTypes@iC] (-0.7,1.1) -- node[label=(45+\cmdKV@general@rotate):\cmdKV@gLabels@iLC] {} (-0.3,0.7);
    \draw[label distance=0,\cmdKV@gTypes@iD] (-0.3,0.7) -- node[label=(-45+\cmdKV@general@rotate):\cmdKV@gLabels@iLD] {} (-0.7,0.3);
    \draw[label distance=0,\cmdKV@gTypes@iE] (-0.7,0.3) -- node[label=(-90+\cmdKV@general@rotate):\cmdKV@gLabels@iLE] {} (-1.9,0.3);
    \draw[label distance=0,\cmdKV@gTypes@iF] (-0.7,1.1) -- node[label=(180+\cmdKV@general@rotate):\cmdKV@gLabels@iLF] {} (-1.1,0.7);
    \draw[label distance=0,\cmdKV@gTypes@iG] (-1.1,0.7) -- node[label=(-180+\cmdKV@general@rotate):\cmdKV@gLabels@iLG] {} (-0.7,0.3);
    #2
  \end{tikzpicture}
  \gSetStdTypes{line}
}
\def\gTriPenta{\@ifnextchar[\@gTriPenta{\@gTriPenta[]}}
\def\@gTriPenta[#1]#2{%
  \setkeys*{pre}{#1}%
  \setrmkeys{general,gTypes,gLabels}%
  \begin{tikzpicture}
    [line width=1pt,
    baseline={([yshift=-0.5ex+\cmdKV@general@yshift]current bounding box.center)},
    scale=\cmdKV@general@scale,
    rotate=\cmdKV@general@rotate,
    font=\cmdKV@general@font]
    \draw[\cmdKV@gTypes@eA] (-108:0.5) -- (-135:1) node[label=(145+\cmdKV@general@rotate):\cmdKV@gLabels@eLA] {};
    \draw[\cmdKV@gTypes@eB] (180:0.5) -- (180:1) node[label=(180+\cmdKV@general@rotate):\cmdKV@gLabels@eLB] {};
    \draw[\cmdKV@gTypes@eC] (108:0.5) -- (135:1) node[label=(-145+\cmdKV@general@rotate):\cmdKV@gLabels@eLC] {};
    \draw[\cmdKV@gTypes@eD] (0:1.2) -- (0:1.5) node[label=(\cmdKV@general@rotate):\cmdKV@gLabels@eLD] {};
    \draw[label distance=0,\cmdKV@gTypes@iA] (-108:0.5) -- node[label=(-135+\cmdKV@general@rotate):\cmdKV@gLabels@iLA] {} (-180:0.5);
    \draw[label distance=0,\cmdKV@gTypes@iB] (180:0.5) -- node[label=(135+\cmdKV@general@rotate):\cmdKV@gLabels@iLB] {} (108:0.5);
    \draw[label distance=0,\cmdKV@gTypes@iC] (108:0.5) -- node[label=(72+\cmdKV@general@rotate):\cmdKV@gLabels@iLC] {} (36:0.5);
    \draw[label distance=0,\cmdKV@gTypes@iD] (36:0.5) -- node[label=(72+\cmdKV@general@rotate):\cmdKV@gLabels@iLD] {} (0:1.2);
    \draw[label distance=2,\cmdKV@gTypes@iE] (0:1.2) -- node[label=(-90+\cmdKV@general@rotate):\cmdKV@gLabels@iLE] {} (-36:0.5);
    \draw[label distance=2,\cmdKV@gTypes@iF] (-36:0.5) -- node[label=(-90+\cmdKV@general@rotate):\cmdKV@gLabels@iLF] {} (-108:0.5);
    \draw[label distance=0,\cmdKV@gTypes@iG] (36:0.5) -- node[label=(180+\cmdKV@general@rotate):\cmdKV@gLabels@iLG] {} (-36:0.5);
    #2
  \end{tikzpicture}
  \gSetStdTypes{line}
}
\def\gTriTri{\@ifnextchar[\@gTriTri{\@gTriTri[]}}
\def\@gTriTri[#1]#2{%
  \setkeys*{pre}{#1}%
  \setrmkeys{general,gTypes,gLabels}%
  \begin{tikzpicture}
    [line width=1pt,
    baseline={([yshift=-0.5ex+\cmdKV@general@yshift]current bounding box.center)},
    scale=\cmdKV@general@scale,
    rotate=\cmdKV@general@rotate,
    font=\cmdKV@general@font]
    \draw[\cmdKV@gTypes@eA] (0.3,0.3) -- (0,0) node[label=(-180+\cmdKV@general@rotate):\cmdKV@gLabels@eLA] {};
    \draw[\cmdKV@gTypes@eB] (0.3,1.1) -- (0,1.4) node[label=(180+\cmdKV@general@rotate):\cmdKV@gLabels@eLB] {};
    \draw[\cmdKV@gTypes@eC] (1.9,1.1) -- (2.2,1.4) node[label=(\cmdKV@general@rotate):\cmdKV@gLabels@eLC] {};
    \draw[\cmdKV@gTypes@eD] (1.9,0.3) -- (2.2,0) node[label=(\cmdKV@general@rotate):\cmdKV@gLabels@eLD] {};
    \draw[label distance=0,\cmdKV@gTypes@iA] (0.3,0.3) -- node[label=(180+\cmdKV@general@rotate):\cmdKV@gLabels@iLA] {} (0.3,1.1);
    \draw[label distance=0,\cmdKV@gTypes@iB] (0.3,1.1) -- node[label=(45+\cmdKV@general@rotate):\cmdKV@gLabels@iLB] {} (0.9,0.7);
    \draw[label distance=0,\cmdKV@gTypes@iC] (0.9,0.7) -- node[label=(-45+\cmdKV@general@rotate):\cmdKV@gLabels@iLC] {} (0.3,0.3);
    \draw[label distance=0,\cmdKV@gTypes@iD] (0.9,0.7) -- node[label=(90+\cmdKV@general@rotate):\cmdKV@gLabels@iLD] {} (1.3,0.7);
    \draw[label distance=0,\cmdKV@gTypes@iE] (1.3,0.7) -- node[label=(135+\cmdKV@general@rotate):\cmdKV@gLabels@iLE] {} (1.9,1.1);
    \draw[label distance=0,\cmdKV@gTypes@iF] (1.9,1.1) -- node[label=(\cmdKV@general@rotate):\cmdKV@gLabels@iLF] {} (1.9,0.3);
    \draw[label distance=0,\cmdKV@gTypes@iG] (1.9,0.3) -- node[label=(-135+\cmdKV@general@rotate):\cmdKV@gLabels@iLG] {} (1.3,0.7);
    #2
  \end{tikzpicture}
  \gSetStdTypes{line}
}

\def\gBoxBubA{\@ifnextchar[\@gBoxBubA{\@gBoxBubA[]}}
\def\@gBoxBubA[#1]#2{%
  \setkeys*{pre}{#1}%
  \setrmkeys{general,gTypes,gLabels}%
  \begin{tikzpicture}
    [line width=1pt,
    baseline={([yshift=-0.5ex+\cmdKV@general@yshift]current bounding box.center)},
    scale=\cmdKV@general@scale,
    rotate=\cmdKV@general@rotate,
    font=\cmdKV@general@font]
    \draw[\cmdKV@gTypes@eA] (0.1,0.1) -- (-0.1,-0.1) node[label=(-180+\cmdKV@general@rotate):\cmdKV@gLabels@eLA] {};
    \draw[\cmdKV@gTypes@eB] (0.1,0.9) -- (-0.1,1.1) node[label=(180+\cmdKV@general@rotate):\cmdKV@gLabels@eLB] {};
    \draw[\cmdKV@gTypes@eC] (0.9,0.9) -- (1.1,1.1) node[label=(\cmdKV@general@rotate):\cmdKV@gLabels@eLC] {};
    \draw[\cmdKV@gTypes@eD] (0.9,0.1) -- (1.1,-0.1) node[label=(\cmdKV@general@rotate):\cmdKV@gLabels@eLD] {};
    \draw[label distance=0,\cmdKV@gTypes@iA] (0.1,0.1) -- node[label=(180+\cmdKV@general@rotate):\cmdKV@gLabels@iLA] {} (0.1,0.9);
    \draw[label distance=0,\cmdKV@gTypes@iB] (0.1,0.9) -- node[label=(90+\cmdKV@general@rotate):\cmdKV@gLabels@iLB] {} (0.32,0.9);
    \draw[label distance=0,\cmdKV@gTypes@iC] (0.32,0.9) .. controls (0.32,1.15) and (0.68,1.15) .. node[label=(90+\cmdKV@general@rotate):\cmdKV@gLabels@iLC] {} (0.68,0.9);
    \draw[label distance=0,\cmdKV@gTypes@iD] (0.68,0.9) .. controls (0.68,0.65) and (0.32,0.65) .. node[label=(-90+\cmdKV@general@rotate):\cmdKV@gLabels@iLD] {} (0.32,0.9);
    \draw[label distance=0,\cmdKV@gTypes@iE] (0.68,0.9) -- node[label=(90+\cmdKV@general@rotate):\cmdKV@gLabels@iLE] {} (0.9,0.9);
    \draw[label distance=0,\cmdKV@gTypes@iF] (0.9,0.9) -- node[label=(\cmdKV@general@rotate):\cmdKV@gLabels@iLF] {} (0.9,0.1);
    \draw[label distance=0,\cmdKV@gTypes@iG] (0.9,0.1) -- node[label=(-90+\cmdKV@general@rotate):\cmdKV@gLabels@iLG] {} (0.1,0.1);
    #2
  \end{tikzpicture}
  \gSetStdTypes{line}
}

\def\gBoxBubB{\@ifnextchar[\@gBoxBubB{\@gBoxBubB[]}}
\def\@gBoxBubB[#1]#2{%
  \setkeys*{pre}{#1}%
  \setrmkeys{general,gTypes,gLabels}%
  \begin{tikzpicture}
    [line width=1pt,
    baseline={([yshift=-0.5ex+\cmdKV@general@yshift]current bounding box.center)},
    scale=\cmdKV@general@scale,
    rotate=\cmdKV@general@rotate,
    font=\cmdKV@general@font]
    \draw[label distance=-1,\cmdKV@gTypes@eA] (0,-0.4) -- (0,-0.7) node[label=(180+\cmdKV@general@rotate):\cmdKV@gLabels@eLA] {};
    \draw[\cmdKV@gTypes@eB] (-0.4,0) -- (-0.7,0) node[label=(180+\cmdKV@general@rotate):\cmdKV@gLabels@eLB] {};
    \draw[label distance=-1,\cmdKV@gTypes@eC] (0,0.4) -- (0,0.7) node[label=(180+\cmdKV@general@rotate):\cmdKV@gLabels@eLC] {};
    \draw[\cmdKV@gTypes@eD] (1.1,0) -- (1.4,0) node[label=(\cmdKV@general@rotate):\cmdKV@gLabels@eLD] {};
    \draw[label distance=0,\cmdKV@gTypes@iA] (0,-0.4) -- node[label=(-135+\cmdKV@general@rotate):\cmdKV@gLabels@iLA] {} (-0.4,0);
    \draw[label distance=0,\cmdKV@gTypes@iB] (-0.4,0) -- node[label=(135+\cmdKV@general@rotate):\cmdKV@gLabels@iLB] {} (0,0.4);
    \draw[label distance=0,\cmdKV@gTypes@iC] (0,0.4) -- node[label=(45+\cmdKV@general@rotate):\cmdKV@gLabels@iLC] {} (0.4,0);
    \draw[label distance=0,\cmdKV@gTypes@iD] (0.4,0) -- node[label=(-45+\cmdKV@general@rotate):\cmdKV@gLabels@iLD] {} (0,-0.4);
    \draw[label distance=0,\cmdKV@gTypes@iE] (0.4,0) --  node[label=(90+\cmdKV@general@rotate):\cmdKV@gLabels@iLE] {} (0.7,0);
    \draw[label distance=0,\cmdKV@gTypes@iF] (0.7,0) .. controls (0.7,0.26) and (1.1,0.26) .. node[label=(90+\cmdKV@general@rotate):\cmdKV@gLabels@iLF] {} (1.1,0);
    \draw[label distance=0,\cmdKV@gTypes@iG] (1.1,0) .. controls (1.1,-0.26) and (0.7,-0.26) ..  node[label=(-90+\cmdKV@general@rotate):\cmdKV@gLabels@iLG] {} (0.7,0);
    #2
  \end{tikzpicture}
  \gSetStdTypes{line}
}

\def\gPentaTad{\@ifnextchar[\@gPentaTad{\@gPentaTad[]}}
\def\@gPentaTad[#1]#2{%
  \setkeys*{pre}{#1}%
  \setrmkeys{general,gTypes,gLabels}%
  \begin{tikzpicture}
    [line width=1pt,
    baseline={([yshift=-0.5ex+\cmdKV@general@yshift]current bounding box.center)},
    scale=\cmdKV@general@scale,
    rotate=\cmdKV@general@rotate,
    font=\cmdKV@general@font]
    \draw[label distance=-0.5,\cmdKV@gTypes@eA] (108:0.35) -- (108:0.75) node[label=(180+\cmdKV@general@rotate):\cmdKV@gLabels@eLA] {};
    \draw[\cmdKV@gTypes@eB] (36:0.35) -- (36:0.75) node[label=(36+\cmdKV@general@rotate):\cmdKV@gLabels@eLB] {};
    \draw[\cmdKV@gTypes@eC] (-36:0.35) -- (-36:0.75) node[label=(-36+\cmdKV@general@rotate):\cmdKV@gLabels@eLC] {};
    \draw[label distance=-0.5,\cmdKV@gTypes@eD] (-108:0.35) -- (-108:0.75) node[label=(-180+\cmdKV@general@rotate):\cmdKV@gLabels@eLD] {};
    \draw[label distance=0,\cmdKV@gTypes@iA] (108:0.35) -- node[label=(72+\cmdKV@general@rotate):\cmdKV@gLabels@iLA] {} (36:0.35);
    \draw[label distance=0,\cmdKV@gTypes@iB] (36:0.35) -- node[label=(\cmdKV@general@rotate):\cmdKV@gLabels@iLB] {} (-36:0.35);
    \draw[label distance=0,\cmdKV@gTypes@iC] (-36:0.35) -- node[label=(-72+\cmdKV@general@rotate):\cmdKV@gLabels@iLC] {} (-108:0.35);
    \draw[label distance=0,\cmdKV@gTypes@iD] (-108:0.35) -- node[label=(-144+\cmdKV@general@rotate):\cmdKV@gLabels@iLD] {} (-180:0.35);
    \draw[label distance=0,\cmdKV@gTypes@iE] (180:0.35) -- node[label=(144+\cmdKV@general@rotate):\cmdKV@gLabels@iLE] {} (108:0.35);
    \draw[label distance=0,\cmdKV@gTypes@iF] (180:0.35) -- node[label=(90+\cmdKV@general@rotate):\cmdKV@gLabels@iLF] {} (180:0.75);
    \draw[label distance=-0.5,\cmdKV@gTypes@iG] (180:0.75) .. controls ($(180:0.75)+(0,-0.18)$) and ($(180:1)+(0,-0.18)$) .. (180:1) node[label=(180+\cmdKV@general@rotate):\cmdKV@gLabels@iLG] {} .. controls ($(180:1)+(0,0.18)$) and ($(180:0.75)+(0,0.18)$) .. (180:0.75);
    #2
  \end{tikzpicture}
  \gSetStdTypes{line}
}

\def\gBoxTadA{\@ifnextchar[\@gBoxTadA{\@gBoxTadA[]}}
\def\@gBoxTadA[#1]#2{%
  \setkeys*{pre}{#1}%
  \setrmkeys{general,gTypes,gLabels}%
  \begin{tikzpicture}
    [line width=1pt,
    baseline={([yshift=-0.5ex+\cmdKV@general@yshift]current bounding box.center)},
    scale=\cmdKV@general@scale,
    rotate=\cmdKV@general@rotate,
    font=\cmdKV@general@font]
    \draw[label distance=-0.5,\cmdKV@gTypes@eA] (0,0.4) -- (0,0.7) node[label=(\cmdKV@general@rotate):\cmdKV@gLabels@eLA] {};
    \draw[\cmdKV@gTypes@eB] (0.4,0) -- (0.7,0) node[label=(\cmdKV@general@rotate):\cmdKV@gLabels@eLB] {};
    \draw[label distance=-0.5,\cmdKV@gTypes@eC] (0,-0.4) -- (0,-0.7) node[label=(\cmdKV@general@rotate):\cmdKV@gLabels@eLC] {};
    \draw[\cmdKV@gTypes@eD] (-0.7,0) -- (-0.7,-0.3) node[label=(-90+\cmdKV@general@rotate):\cmdKV@gLabels@eLD] {};
    \draw[label distance=0,\cmdKV@gTypes@iA] (0,0.4) -- node[label=(45+\cmdKV@general@rotate):\cmdKV@gLabels@iLA] {} (0.4,0);
    \draw[label distance=0,\cmdKV@gTypes@iB] (0.4,0) -- node[label=(-45+\cmdKV@general@rotate):\cmdKV@gLabels@iLB] {} (0,-0.4);
    \draw[label distance=0,\cmdKV@gTypes@iC] (0,-0.4) -- node[label=(-135+\cmdKV@general@rotate):\cmdKV@gLabels@iLC] {} (-0.4,0);
    \draw[label distance=0,\cmdKV@gTypes@iD] (-0.4,0) -- node[label=(135+\cmdKV@general@rotate):\cmdKV@gLabels@iLD] {} (0,0.4);
    \draw[label distance=0,\cmdKV@gTypes@iE] (-0.4,0) --  node[label=(90+\cmdKV@general@rotate):\cmdKV@gLabels@iLE] {} (-0.7,0);
    \draw[label distance=0,\cmdKV@gTypes@iF] (-0.7,0) -- node[label=(90+\cmdKV@general@rotate):\cmdKV@gLabels@iLF] {} (-1.1,0);
    \draw[label distance=-0.5,\cmdKV@gTypes@iG] (-1.1,0) .. controls (-1.1,-0.18) and (-1.35,-0.18) .. (-1.35,0) node[label=(180+\cmdKV@general@rotate):\cmdKV@gLabels@iLG] {} .. controls (-1.35,0.18) and (-1.1,0.18) .. (-1.1,0);
    #2
  \end{tikzpicture}
  \gSetStdTypes{line}
}

\def\gBoxBoxBoxA{\@ifnextchar[\@gBoxBoxBoxA{\@gBoxBoxBoxA[]}}
\def\@gBoxBoxBoxA[#1]#2{%
  \setkeys*{pre}{#1}%
  \setrmkeys{general,gTypes,gLabels}%
  \begin{tikzpicture}
    [line width=1pt,
    baseline={([yshift=-0.5ex+\cmdKV@general@yshift]current bounding box.center)},
    scale=\cmdKV@general@scale,
    rotate=\cmdKV@general@rotate,
    font=\cmdKV@general@font]
    \draw[\cmdKV@gTypes@eA] (-0.3,0.3) -- (-0.6,0) node[label=(-180+\cmdKV@general@rotate):\cmdKV@gLabels@eLA] {};
    \draw[\cmdKV@gTypes@eB] (-0.3,0.9) -- (-0.6,1.2) node[label=(180+\cmdKV@general@rotate):\cmdKV@gLabels@eLB] {};
    \draw[\cmdKV@gTypes@eC] (1.5,0.9) -- (1.8,1.2) node[label=(\cmdKV@general@rotate):\cmdKV@gLabels@eLC] {};
    \draw[\cmdKV@gTypes@eD] (1.5,0.3) -- (1.8,0) node[label=(\cmdKV@general@rotate):\cmdKV@gLabels@eLD] {};
    \draw[label distance=0,\cmdKV@gTypes@iA] (-0.3,0.3) -- node[label=(180+\cmdKV@general@rotate):\cmdKV@gLabels@iLA] {} (-0.3,0.9);
    \draw[label distance=0,\cmdKV@gTypes@iB] (-0.3,0.9) -- node[label=(90+\cmdKV@general@rotate):\cmdKV@gLabels@iLB] {} (0.3,0.9);
    \draw[label distance=0,\cmdKV@gTypes@iC] (0.3,0.9) -- node[label=(90+\cmdKV@general@rotate):\cmdKV@gLabels@iLC] {} (0.9,0.9);
    \draw[label distance=0,\cmdKV@gTypes@iD] (0.9,0.9) -- node[label=(90+\cmdKV@general@rotate):\cmdKV@gLabels@iLD] {} (1.5,0.9);
    \draw[label distance=0,\cmdKV@gTypes@iE] (1.5,0.9) -- node[label=(\cmdKV@general@rotate):\cmdKV@gLabels@iLE] {} (1.5,0.3);
    \draw[label distance=0,\cmdKV@gTypes@iF] (1.5,0.3) -- node[label=(-90+\cmdKV@general@rotate):\cmdKV@gLabels@iLF] {} (0.9,0.3);
    \draw[label distance=0,\cmdKV@gTypes@iG] (0.9,0.3) -- node[label=(-90+\cmdKV@general@rotate):\cmdKV@gLabels@iLG] {} (0.3,0.3);
    \draw[label distance=0,\cmdKV@gTypes@iH] (0.3,0.3) -- node[label=(-90+\cmdKV@general@rotate):\cmdKV@gLabels@iLH] {} (-0.3,0.3);
    \draw[label distance=0,\cmdKV@gTypes@iI] (0.3,0.9) -- node[label=(\cmdKV@general@rotate):\cmdKV@gLabels@iLI] {} (0.3,0.3);
    \draw[label distance=0,\cmdKV@gTypes@iJ] (0.9,0.9) -- node[label=(\cmdKV@general@rotate):\cmdKV@gLabels@iLJ] {} (0.9,0.3);
    #2
  \end{tikzpicture}
  \gSetStdTypes{line}
}

\def\gBoxBoxBoxB{\@ifnextchar[\@gBoxBoxBoxB{\@gBoxBoxBoxB[]}}
\def\@gBoxBoxBoxB[#1]#2{%
  \setkeys*{pre}{#1}%
  \setrmkeys{general,gTypes,gLabels}%
  \begin{tikzpicture}
    [line width=1pt,
    baseline={([yshift=-0.5ex+\cmdKV@general@yshift]current bounding box.center)},
    scale=\cmdKV@general@scale,
    rotate=\cmdKV@general@rotate,
    font=\cmdKV@general@font]
    \draw[\cmdKV@gTypes@eA] (0.3,0.3) -- (0,0) node[label=(-180+\cmdKV@general@rotate):\cmdKV@gLabels@eLA] {};
    \draw[\cmdKV@gTypes@eB] (0.3,1.5) -- (0,1.8) node[label=(180+\cmdKV@general@rotate):\cmdKV@gLabels@eLB] {};
    \draw[\cmdKV@gTypes@eC] (1.5,1.5) -- (1.8,1.8) node[label=(\cmdKV@general@rotate):\cmdKV@gLabels@eLC] {};
    \draw[\cmdKV@gTypes@eD] (1.5,0.3) -- (1.8,0) node[label=(\cmdKV@general@rotate):\cmdKV@gLabels@eLD] {};
    \draw[label distance=0,\cmdKV@gTypes@iA] (0.3,0.3) -- node[label=(180+\cmdKV@general@rotate):\cmdKV@gLabels@iLA] {} (0.3,0.9);
    \draw[label distance=0,\cmdKV@gTypes@iB] (0.3,0.9) -- node[label=(180+\cmdKV@general@rotate):\cmdKV@gLabels@iLB] {} (0.3,1.5);
    \draw[label distance=0,\cmdKV@gTypes@iC] (0.3,1.5) -- node[label=(90+\cmdKV@general@rotate):\cmdKV@gLabels@iLC] {} (0.9,1.5);
    \draw[label distance=0,\cmdKV@gTypes@iD] (0.9,1.5) -- node[label=(90+\cmdKV@general@rotate):\cmdKV@gLabels@iLD] {} (1.5,1.5);
    \draw[label distance=0,\cmdKV@gTypes@iE] (1.5,1.5) -- node[label=(\cmdKV@general@rotate):\cmdKV@gLabels@iLE] {} (1.5,0.3);
    \draw[label distance=0,\cmdKV@gTypes@iF] (1.5,0.3) -- node[label=(-90+\cmdKV@general@rotate):\cmdKV@gLabels@iLF] {} (0.9,0.3);
    \draw[label distance=0,\cmdKV@gTypes@iG] (0.9,0.3) -- node[label=(-90+\cmdKV@general@rotate):\cmdKV@gLabels@iLG] {} (0.3,0.3);
    \draw[label distance=0,\cmdKV@gTypes@iH] (0.9,1.5) -- node[label=(\cmdKV@general@rotate):\cmdKV@gLabels@iLH] {} (0.9,0.9);
    \draw[label distance=0,\cmdKV@gTypes@iI] (0.9,0.9) -- node[label=(\cmdKV@general@rotate):\cmdKV@gLabels@iLI] {} (0.9,0.3);
    \draw[label distance=0,\cmdKV@gTypes@iJ] (0.9,0.9) -- node[label=(90+\cmdKV@general@rotate):\cmdKV@gLabels@iLJ] {} (0.3,0.9);
    #2
  \end{tikzpicture}
  \gSetStdTypes{line}
}

\def\cTree{\@ifnextchar[\@cTree{\@cTree[]}}
\def\@cTree[#1]#2{%
  \setkeys*{pre}{#1}%
  \setrmkeys{general,gTypes,gLabels,blobs}%
  \begin{tikzpicture}
    [line width=1pt,
    baseline={([yshift=-0.5ex+\cmdKV@general@yshift]current bounding box.center)},
    scale=\cmdKV@general@scale,
    rotate=\cmdKV@general@rotate,
    font=\cmdKV@general@font]
    \fill[\cmdKV@blobs@blobA] (0,0) ellipse (0.25 and 0.25);
    \draw[\cmdKV@gTypes@eA] ($ (0,0) + (-135:0.25) $) -- ($ (0,0) + (-135:0.6) $) node[label=(-180+\cmdKV@general@rotate):\cmdKV@gLabels@eLA] {};
    \draw[\cmdKV@gTypes@eB] ($ (0,0) + (135:0.25) $) -- ($ (0,0) + (135:0.6) $) node[label=(180+\cmdKV@general@rotate):\cmdKV@gLabels@eLB] {};
    \draw[\cmdKV@gTypes@eC] ($ (0,0) + (45:0.25) $) -- ($ (0,0) + (45:0.6) $) node[label=(\cmdKV@general@rotate):\cmdKV@gLabels@eLC] {};
    \draw[\cmdKV@gTypes@eD] ($ (0,0) + (-45:0.25) $) -- ($ (0,0) + (-45:0.6) $) node[label=(\cmdKV@general@rotate):\cmdKV@gLabels@eLD] {};
    \node[label=(90+\cmdKV@general@rotate):\cmdKV@gLabels@iLA] at (0,0.3) {};
    \node[label=(\cmdKV@general@rotate):\cmdKV@gLabels@iLB] at (0.3,0) {};
    #2
  \end{tikzpicture}
  \gSetStdTypes{line}
}

\def\cBoxA{\@ifnextchar[\@cBoxA{\@cBoxA[]}}
\def\@cBoxA[#1]#2{%
  \setkeys*{pre}{#1}%
  \setrmkeys{general,gTypes,gLabels,dots,blobs}%
  \begin{tikzpicture}
    [line width=1pt,
    baseline={([yshift=-0.5ex+\cmdKV@general@yshift]current bounding box.center)},
    scale=\cmdKV@general@scale,
    rotate=\cmdKV@general@rotate,
    font=\cmdKV@general@font]
    \fill[\cmdKV@blobs@blobA] (-0.4,0) ellipse (0.2 and 0.4);
    \fill[\cmdKV@blobs@blobB] (0.4,0) ellipse (0.2 and 0.4);
    \draw[\cmdKV@gTypes@eA] ($ (-0.4,0) + (-135:0.26) $) -- ($ (-0.4,0) + (-135:0.7) $) node[label=(-180+\cmdKV@general@rotate):\cmdKV@gLabels@eLA] {};
    \draw[\cmdKV@gTypes@eB] ($ (-0.4,0) + (135:0.26) $) -- ($ (-0.4,0) + (135:0.7) $) node[label=(180+\cmdKV@general@rotate):\cmdKV@gLabels@eLB] {};
    \draw[\cmdKV@gTypes@eC] ($ (0.4,0) + (45:0.26) $) -- ($ (0.4,0) + (45:0.7) $) node[label=(\cmdKV@general@rotate):\cmdKV@gLabels@eLC] {};
    \draw[\cmdKV@gTypes@eD] ($ (0.4,0) + (-45:0.26) $) -- ($ (0.4,0) + (-45:0.7) $) node[label=(\cmdKV@general@rotate):\cmdKV@gLabels@eLD] {};
    \draw[label distance=1,\cmdKV@gTypes@iA] ($ (-0.4,0) + (45:0.26) $) -- node[label=(90+\cmdKV@general@rotate):\cmdKV@gLabels@iLA] {}($ (0.4,0) + (135:0.26) $);
    \draw[label distance=1,\cmdKV@gTypes@iB] ($ (0.4,0) + (-135:0.26) $) -- node[label=(-90+\cmdKV@general@rotate):\cmdKV@gLabels@iLB] {}($ (-0.4,0) + (-45:0.26) $);
    \node[label=(180+\cmdKV@general@rotate):\cmdKV@gLabels@iLC] at (-0.6,0) {};
    \node[label=(\cmdKV@general@rotate):\cmdKV@gLabels@iLD] at (0.6,0) {};
    \ifKV@dots@lDots
      \draw[dotted] ($(-0.4,0) + (-150:0.7)$) to[bend left] ($(-0.4,0) + (150:0.7)$);%
    \fi
    \ifKV@dots@rDots
      \draw[dotted] ($(0.4,0) + (30:0.7)$) to[bend left] ($(0.4,0) + (-30:0.7)$);%
    \fi
    #2
  \end{tikzpicture}
  \gSetStdTypes{line}
}

\def\cBoxAA{\@ifnextchar[\@cBoxAA{\@cBoxAA[]}}
\def\@cBoxAA[#1]#2{%
  \setkeys*{pre}{#1}%
  \setrmkeys{general,gTypes,gLabels,dots,blobs}%
  \begin{tikzpicture}
    [line width=1pt,
    baseline={([yshift=-0.5ex+\cmdKV@general@yshift]current bounding box.center)},
    scale=\cmdKV@general@scale,
    rotate=\cmdKV@general@rotate,
    font=\cmdKV@general@font]
    \fill[\cmdKV@blobs@blobA] (-0.4,0) ellipse (0.2 and 0.4);
    \fill[\cmdKV@blobs@blobB] (0.4,0) ellipse (0.2 and 0.4);
    \draw[\cmdKV@gTypes@eA] ($ (-0.4,0) + (-135:0.26) $) -- ($ (-0.4,0) + (-135:0.7) $) node[label=(-180+\cmdKV@general@rotate):\cmdKV@gLabels@eLA] {};
    \draw[\cmdKV@gTypes@eB] ($ (-0.4,0) + (135:0.26) $) -- ($ (-0.4,0) + (135:0.7) $) node[label=(180+\cmdKV@general@rotate):\cmdKV@gLabels@eLB] {};
    \draw[\cmdKV@gTypes@eC] ($ (0.4,0) + (45:0.26) $) -- ($ (0.4,0) + (45:0.7) $) node[label=(\cmdKV@general@rotate):\cmdKV@gLabels@eLC] {};
    \draw[\cmdKV@gTypes@eD] ($ (0.4,0) + (-45:0.26) $) -- ($ (0.4,0) + (-45:0.7) $) node[label=(\cmdKV@general@rotate):\cmdKV@gLabels@eLD] {};
    \draw[label distance=1,\cmdKV@gTypes@iA] ($ (-0.4,0) + (60:0.32) $) to[bend left] node[label=(90+\cmdKV@general@rotate):\cmdKV@gLabels@iLA] {}($ (0.4,0) + (120:0.32) $);
    \draw[label distance=1,\cmdKV@gTypes@iB] ($ (0.4,0) + (-120:0.32) $) to[bend left]  node[label=(-90+\cmdKV@general@rotate):\cmdKV@gLabels@iLB] {}($ (-0.4,0) + (-60:0.32) $);
    \draw[dotted] (0,0.25) -- (0,-0.25);
    \node[label=(180+\cmdKV@general@rotate):\cmdKV@gLabels@iLC] at (-0.6,0) {};
    \node[label=(\cmdKV@general@rotate):\cmdKV@gLabels@iLD] at (0.6,0) {};
    \ifKV@dots@lDots
      \draw[dotted] ($(-0.4,0) + (-150:0.7)$) to[bend left] ($(-0.4,0) + (150:0.7)$);%
    \fi
    \ifKV@dots@rDots
      \draw[dotted] ($(0.4,0) + (30:0.7)$) to[bend left] ($(0.4,0) + (-30:0.7)$);%
    \fi
    #2
  \end{tikzpicture}
  \gSetStdTypes{line}
}

\def\cBoxB{\@ifnextchar[\@cBoxB{\@cBoxB[]}}
\def\@cBoxB[#1]#2{%
  \setkeys*{pre}{#1}%
  \setrmkeys{general,gTypes,gLabels,blobs}%
  \begin{tikzpicture}
    [line width=1pt,
    baseline={([yshift=-0.5ex+\cmdKV@general@yshift]current bounding box.center)},
    scale=\cmdKV@general@scale,
    rotate=\cmdKV@general@rotate,
    font=\cmdKV@general@font]
    \fill[\cmdKV@blobs@blobA] (0,-0.3) ellipse (0.4 and 0.2);
    \fill[\cmdKV@blobs@blobB] (0,0.3) ellipse (0.4 and 0.2);
    \draw[\cmdKV@gTypes@eA] ($ (0,-0.3) + (180:0.4) $) -- ($ (0,-0.3) + (-170:0.7) $) node[label=(180+\cmdKV@general@rotate):\cmdKV@gLabels@eLA] {};
    \draw[\cmdKV@gTypes@eB] ($ (0,0.3) + (180:0.4) $) -- ($ (0,0.3) + (170:0.7) $) node[label=(180+\cmdKV@general@rotate):\cmdKV@gLabels@eLB] {};
    \draw[\cmdKV@gTypes@eC] ($ (0,0.3) + (0:0.4) $) -- ($ (0,0.3) + (10:0.7) $) node[label=(\cmdKV@general@rotate):\cmdKV@gLabels@eLC] {};
    \draw[\cmdKV@gTypes@eD] ($ (0,-0.3) + (0:0.4) $) -- ($ (0,-0.3) + (-10:0.7) $) node[label=(\cmdKV@general@rotate):\cmdKV@gLabels@eLD] {};
    \draw[label distance=1,\cmdKV@gTypes@iA] ($ (0,-0.3) + (135:0.26) $) -- node[label=(-180+\cmdKV@general@rotate):\cmdKV@gLabels@iLA] {}($ (0,0.3) + (-135:0.26) $);
    \draw[label distance=1,\cmdKV@gTypes@iB] ($ (0,0.3) + (-45:0.26) $) -- node[label=(\cmdKV@general@rotate):\cmdKV@gLabels@iLB] {}($ (0,-0.3) + (45:0.26) $);
    \node[label=(90+\cmdKV@general@rotate):\cmdKV@gLabels@iLC] at (0,0.5) {};
    \node[label=(-90+\cmdKV@general@rotate):\cmdKV@gLabels@iLD] at (0,-0.5) {};
    #2
  \end{tikzpicture}
  \gSetStdTypes{line}
}

\def\cBoxBoxA{\@ifnextchar[\@cBoxBoxA{\@cBoxBoxA[]}}
\def\@cBoxBoxA[#1]#2{%
  \setkeys*{pre}{#1}%
  \setrmkeys{general,gTypes,gLabels,blobs}%
  \begin{tikzpicture}
    [line width=1pt,
    baseline={([yshift=-0.5ex+\cmdKV@general@yshift]current bounding box.center)},
    scale=\cmdKV@general@scale,
    rotate=\cmdKV@general@rotate,
    font=\cmdKV@general@font]
    \fill[\cmdKV@blobs@blobA] (-0.4,0) ellipse (0.2 and 0.4);
    \fill[\cmdKV@blobs@blobB] (0.4,0) ellipse (0.2 and 0.4);
    \fill[\cmdKV@blobs@blobC] (1.2,0) ellipse (0.2 and 0.4);
    \draw[\cmdKV@gTypes@eA] ($ (-0.4,0) + (-135:0.26) $) -- ($ (-0.4,0) + (-135:0.7) $) node[label=(-180+\cmdKV@general@rotate):\cmdKV@gLabels@eLA] {};
    \draw[\cmdKV@gTypes@eB] ($ (-0.4,0) + (135:0.26) $) -- ($ (-0.4,0) + (135:0.7) $) node[label=(180+\cmdKV@general@rotate):\cmdKV@gLabels@eLB] {};
    \draw[\cmdKV@gTypes@eC] ($ (1.2,0) + (45:0.26) $) -- ($ (1.2,0) + (45:0.7) $) node[label=(\cmdKV@general@rotate):\cmdKV@gLabels@eLC] {};
    \draw[\cmdKV@gTypes@eD] ($ (1.2,0) + (-45:0.26) $) -- ($ (1.2,0) + (-45:0.7) $) node[label=(\cmdKV@general@rotate):\cmdKV@gLabels@eLD] {};
    \draw[label distance=1,\cmdKV@gTypes@iA] ($ (-0.4,0) + (45:0.26) $) -- node[label=(90+\cmdKV@general@rotate):\cmdKV@gLabels@iLA] {}($ (0.4,0) + (135:0.26) $);
    \draw[label distance=1,\cmdKV@gTypes@iB] ($ (0.4,0) + (45:0.26) $) -- node[label=(90+\cmdKV@general@rotate):\cmdKV@gLabels@iLB] {}($ (1.2,0) + (135:0.26) $);
    \draw[label distance=1,\cmdKV@gTypes@iC] ($ (1.2,0) + (-135:0.26) $) -- node[label=(-90+\cmdKV@general@rotate):\cmdKV@gLabels@iLC] {}($ (0.4,0) + (-45:0.26) $);
    \draw[label distance=1,\cmdKV@gTypes@iD] ($ (0.4,0) + (-135:0.26) $) -- node[label=(-90+\cmdKV@general@rotate):\cmdKV@gLabels@iLD] {}($ (-0.4,0) + (-45:0.26) $);
    \node[label=(180+\cmdKV@general@rotate):\cmdKV@gLabels@iLE] at (-0.6,0) {};
    \node[label=(\cmdKV@general@rotate):\cmdKV@gLabels@iLF] at (0.6,0) {};
    \node[label=(\cmdKV@general@rotate):\cmdKV@gLabels@iLG] at (1.4,0) {};
    #2
  \end{tikzpicture}
  \gSetStdTypes{line}
}

\def\cBoxBoxB{\@ifnextchar[\@cBoxBoxB{\@cBoxBoxB[]}}
\def\@cBoxBoxB[#1]#2{%
  \setkeys*{pre}{#1}%
  \setrmkeys{general,gTypes,gLabels}%
  \begin{tikzpicture}
    [line width=1pt,
    baseline={([yshift=-0.5ex+\cmdKV@general@yshift]current bounding box.center)},
    scale=\cmdKV@general@scale,
    rotate=\cmdKV@general@rotate,
    font=\cmdKV@general@font]
    \fill[\cmdKV@blobs@blobA] (0,0.3) ellipse (0.4 and 0.2);
    \fill[\cmdKV@blobs@blobB] (0,-0.3) ellipse (0.4 and 0.2);
    \fill[\cmdKV@blobs@blobC] (1,0) ellipse (0.2 and 0.4);
    \draw[\cmdKV@gTypes@eA] ($ (0,-0.3) + (180:0.4) $) -- ($ (0,-0.3) + (-170:0.7) $) node[label=(180+\cmdKV@general@rotate):\cmdKV@gLabels@eLA] {};
    \draw[\cmdKV@gTypes@eB] ($ (0,0.3) + (180:0.4) $) -- ($ (0,0.3) + (170:0.7) $) node[label=(180+\cmdKV@general@rotate):\cmdKV@gLabels@eLB] {};
    \draw[\cmdKV@gTypes@eC] ($ (1,0) + (45:0.26) $) -- ($ (1,0) + (45:0.7) $) node[label=(\cmdKV@general@rotate):\cmdKV@gLabels@eLC] {};
    \draw[\cmdKV@gTypes@eD] ($ (1,0) + (-45:0.26) $) -- ($ (1,0) + (-45:0.7) $) node[label=(\cmdKV@general@rotate):\cmdKV@gLabels@eLD] {};
    \draw[label distance=1,\cmdKV@gTypes@iA] ($ (0,-0.3) + (135:0.26) $) -- node[label=(-180+\cmdKV@general@rotate):\cmdKV@gLabels@iLA] {}($ (0,0.3) + (-135:0.26) $);
    \draw[label distance=1,\cmdKV@gTypes@iB] ($ (0,0.3) + (0:0.4) $) -- node[label=(90+\cmdKV@general@rotate):\cmdKV@gLabels@iLB] {} ($ (1,0) + (135:0.26) $);
    \draw[label distance=1,\cmdKV@gTypes@iC] ($ (1,0) + (-135:0.26) $) -- node[label=(-90+\cmdKV@general@rotate):\cmdKV@gLabels@iLC] {} ($ (0,-0.3) + (0:0.4) $);
    \draw[label distance=1,\cmdKV@gTypes@iD] ($ (0,0.3) + (-45:0.26) $) -- node[label=(\cmdKV@general@rotate):\cmdKV@gLabels@iLD] {}($ (0,-0.3) + (45:0.26) $);
    \node[label=(90+\cmdKV@general@rotate):\cmdKV@gLabels@iLE] at (0,0.5) {};
    \node[label=(\cmdKV@general@rotate):\cmdKV@gLabels@iLF] at (1.2,0) {};
    \node[label=(-90+\cmdKV@general@rotate):\cmdKV@gLabels@iLG] at (0,-0.5) {};
    #2
  \end{tikzpicture}
  \gSetStdTypes{line}
}

\def\cNPBox{\@ifnextchar[\@cNPBox{\@cNPBox[]}}
\def\@cNPBox[#1]#2{%
  \setkeys*{pre}{#1}%
  \setrmkeys{general,gTypes,gLabels,blob}%
  \begin{tikzpicture}
    [line width=1pt,
    baseline={([yshift=-0.5ex+\cmdKV@general@yshift]current bounding box.center)},
    scale=\cmdKV@general@scale,
    rotate=\cmdKV@general@rotate,
    font=\cmdKV@general@font]
    \fill[\cmdKV@blobs@blobA] (-0.6,0) ellipse (0.2 and 0.4);
    \fill[\cmdKV@blobs@blobB] (0.4,0) ellipse (0.2 and 0.4);
    \draw[\cmdKV@gTypes@eA] ($ (-0.6,0) + (-180:0.2) $) -- ($ (-0.6,0) + (-180:0.6) $) node[label=(-180+\cmdKV@general@rotate):\cmdKV@gLabels@eLA] {};
    \draw[\cmdKV@gTypes@eB] ($ (-0.6,0) + (0:0.2) $) -- (-0.18,0) node[label=(\cmdKV@general@rotate):\cmdKV@gLabels@eLB] {};
    \draw[\cmdKV@gTypes@eC] ($ (0.4,0) + (45:0.26) $) -- ($ (0.4,0) + (45:0.7) $) node[label=(\cmdKV@general@rotate):\cmdKV@gLabels@eLC] {};
    \draw[\cmdKV@gTypes@eD] ($ (0.4,0) + (-45:0.26) $) -- ($ (0.4,0) + (-45:0.7) $) node[label=(\cmdKV@general@rotate):\cmdKV@gLabels@eLD] {};
    \draw[label distance=1,\cmdKV@gTypes@iA] ($ (-0.6,0) + (60:0.3) $) -- node[label=(90+\cmdKV@general@rotate):\cmdKV@gLabels@iLA] {}($ (0.4,0) + (135:0.26) $);
    \draw[label distance=1,\cmdKV@gTypes@iB] ($ (0.4,0) + (-135:0.26) $) -- node[label=(-90+\cmdKV@general@rotate):\cmdKV@gLabels@iLB] {}($ (-0.6,0) + (-60:0.3) $);
    \node[label=(\cmdKV@general@rotate):\cmdKV@gLabels@iLD] at (0.6,0) {};
    #2
  \end{tikzpicture}
  \gSetStdTypes{line}
}

\def\cFourPtRec{\@ifnextchar[\@cFourPtRec{\@cFourPtRec[]}}
\def\@cFourPtRec[#1]#2{%
  \setkeys*{pre}{#1}%
  \setrmkeys{general,gTypes,gLabels,blobs}%
  \begin{tikzpicture}
    [line width=1pt,
    baseline={([yshift=-0.5ex+\cmdKV@general@yshift]current bounding box.center)},
    scale=\cmdKV@general@scale,
    rotate=\cmdKV@general@rotate,
    font=\cmdKV@general@font]
    \fill[\cmdKV@blobs@blobA] (-0.5,0) ellipse (0.2 and 0.3);
    \fill[\cmdKV@blobs@blobB] (0.5,0) ellipse (0.3 and 0.4);
    \filldraw[fill=white,draw=black] (0.55,0.15) circle (0.1);
    \filldraw[fill=white,draw=black] (0.45,-0.1) circle (0.1);
    \draw[\cmdKV@gTypes@eA] ($(-0.5,0) + (135:0.23)$) -- ($(-0.5,0) + (135:0.5)$) node[label=(180+\cmdKV@general@rotate):\cmdKV@gLabels@eLA] {};
    \draw[\cmdKV@gTypes@eB] ($(-0.5,0) + (-135:0.23)$) -- ($(-0.5,0) + (-135:0.5)$) node[label=(180+\cmdKV@general@rotate):\cmdKV@gLabels@eLB] {};
    \draw[\cmdKV@gTypes@eC] ($(0.5,0) + (45:0.35)$) -- ($(0.5,0) + (45:0.7)$) node[label=(\cmdKV@general@rotate):\cmdKV@gLabels@eLC] {};
    \draw[\cmdKV@gTypes@eD] ($(0.5,0) + (-45:0.35)$) -- ($(0.5,0) + (-45:0.7)$) node[label=(\cmdKV@general@rotate):\cmdKV@gLabels@eLD] {};
    \draw[label distance=1,\cmdKV@gTypes@iA] ($(-0.5,0) + (30:0.23)$) -- node[label=(90+\cmdKV@general@rotate):\cmdKV@gLabels@iLA] {} ($(0.5,0) + (157:0.3)$);
    \draw[label distance=1,\cmdKV@gTypes@iB] ($(0.5,0) + (-157:0.3)$) -- node[label=(-90+\cmdKV@general@rotate):\cmdKV@gLabels@iLB] {} ($(-0.5,0) + (-30:0.23)$);
    #2
 \end{tikzpicture}
  \gSetStdTypes{line}
}

\makeatother

\numberwithin{equation}{section}

\DeclareMathOperator{\Arcsinh}{arcsinh}
\DeclareMathOperator{\Arctan}{arctan}
\DeclareMathOperator{\arcosh}{arcosh}
\newcommand{\nn}{\nonumber}

\newcommand\beq{\begin{equation}}
\newcommand\eeq{\end{equation}}

\newcommand\bea{\begin{eqnarray}}
\newcommand\eea{\end{eqnarray}}

\newcommand\dd{{\mathrm d}}

\newcommand{\bb}{{\boldsymbol b}}
\newcommand{\bk}{{\boldsymbol k}}
\newcommand{\bl}{{\boldsymbol l}}

\newcommand{\bp}{{\boldsymbol p}}
\newcommand{\bq}{{\boldsymbol q}}
\newcommand{\br}{{\boldsymbol r}}
\newcommand{\bw}{{\boldsymbol w}}
\newcommand{\bx}{{\boldsymbol x}}

\newcommand{\bz}{{\boldsymbol z}}

\newcommand\cM{\mathcal{M}}
\newcommand\cO{\mathcal{O}}
\newcommand\cE{\mathcal{E}}
\newcommand\cS{\mathcal{S}}
\newcommand\cP{\mathcal{P}}

\makeatletter
\newcommand{\Biggg}{\bBigg@{3.5}}
\makeatother

\begin{document}
\preprint{DESY\,19-167 \hskip 0.2cm UUITP-40/19}

\title{\center From Boundary Data to Bound States}
\author[a]{\large Gregor K\"alin}
\author[b,c]{\large and Rafael A. Porto}
\affiliation[a]{Department of Physics and Astronomy, Uppsala University,\\ Box 516, 751 20 Uppsala, Sweden}
\affiliation[b]{Deutsches Elektronen-Synchrotron DESY,\\ Notkestrasse 85, 22607 Hamburg, Germany}
\affiliation[c]{The Abdus Salam International Center for Theoretical Physics,\\ Strada Costiera 11, Trieste 34151, Italy}
\emailAdd{gregor.kaelin@physics.uu.se,\,rafael.porto@desy.de}
\abstract{
  We introduce a --- somewhat {\it holographic} --- dictionary between gravitational observables for scattering processes (measured at the {\it boundary}) and adiabatic invariants for bound orbits (in the {\it bulk}), to all orders in the Post-Minkowskian (PM) expansion. Our~map relies on remarkable connections between the relative momentum of the two-body problem,~the classical limit of the scattering amplitude and the deflection angle in hyperbolic motion. These relationships allow us to compute observables for generic orbits (such as the periastron advance $\Delta\Phi$) through analytic continuation, via a radial~action depending only on boundary data. A~simplified (more geometrical) map can be obtained for~circular orbits, enabling us to extract the orbital frequency as a function of the (conserved) binding energy, $\Omega(E)$, directly from scattering information. As~an example, using the results in Bern~et~al. [\href{https://arxiv.org/abs/1901.04424}{\texttt{1901.04424}}, \href{https://arxiv.org/abs/1908.01493}{\texttt{1908.01493}}], we readily derive $\Omega(E)$ and $\Delta\Phi(J,E)$ to two-loop orders. We also provide closed-form expressions for the orbital frequency and periastron advance at tree-level and one-loop order, respectively, which capture a series of exact terms in the Post-Newtonian expansion. We~then perform a partial PM resummation, using  a \emph{no-recoil} approximation for the amplitude. This limit is behind the map between the scattering angle for a test-particle and the two-body dynamics to 2PM. We show that it also captures a subset of higher order terms beyond the test-particle limit.
  While a (rather lengthy) Hamiltonian may be derived as an intermediate step, our map applies directly between gauge invariant quantities. Our~findings provide a starting point for an alternative approach to the binary problem.
  We~conclude with future directions and some speculations on the classical~double~copy.}
\maketitle
\newpage

\section{Introduction} \label{sec:introduction}

The nascent field of gravitational wave (GW) science and multi-messenger astronomy~\cite{GBM:2017lvd,LIGOScientific:2018mvr,Venumadhav:2019lyq,Zackay:2019tzo} will be a truly interdisciplinary subject, enriching different branches of physics.
Yet, to fully exploit the discovery potential in GW observations, precise theoretical predictions for the binary problem in General Relativity will be mandatory.
The computational challenges in \emph{precision gravity}, however, are enormous~\cite{Porto:2017lrn,Porto:2016zng}.
Numerical methods cover mostly the late stages and merger regime of the binary's dynamics, where the gravitational interaction becomes strong.
However, numerical codes are incapable of solving for the entirety of the observed cycles, which may be of the order of $10^5$ with third generation detectors~\cite{Punturo:2010zz} and for several astrophysically motivated sources, such as neutron star binaries~\cite{ns}.
In those cases, the majority of the cycles in the detector's band will occur during the inspiral phase, which is instead described using analytic methods such as the Post-Newtonian (PN) expansion~\cite{blanchet,Schafer:2018kuf,review}.\vskip 4pt

The PN regime involves perturbative calculations where ideas from particle physics have already played a prominent role.
The Effective Field Theory (EFT) approach put forward in~\cite{Goldberger:2004jt}  has introduced a series of quantum field theory techniques into classical General Relativity, which have successfully reduced the two-body problem in gravity into a computation of Feynman diagrams~\cite{Goldberger:2004jt,iragrg,Foffa:2013qca,review}.
Using the powerful EFT machinery, as well as other methodologies~\cite{Bernard:2017bvn,Marchand:2017pir,Damour:2014jta,Jaranowski:2015lha}, the current level of accuracy in the PN framework has achieved the next-to-next-to-next-to-next-to-leading order (N$^4$LO) in the conservative sector, or four-loops~\cite{tail,lamb,apparent,Foffa:2019rdf,nrgr4}, and up to five-loops in the static limit \cite{5pn1,5pn2}.
The advantage of the PN formalism is the separation of scales in the non-relativistic regime, which in the EFT approach allows the use of the method of regions~\cite{Beneke:1997zp} to compute otherwise intractable integrals.
The disadvantage is the non-relativistic truncation (as opposite to a relativistic result), as well as the need of gauge choices and gauge dependent objects (e.g. lengthy potentials) on the way to derive measurable quantities, such as the binding energy for circular orbits.
The latter, on the other hand, displays a much simpler analytic form, which suggests that a formalism involving only gauge invariant quantities must exist. We develop such framework in this paper, by relating gravitational observables from scattering processes directly to adiabatic invariants for bound~states.\vskip 4pt

In a parallel development, the field of scattering amplitudes has taken a leading role at the frontier of theoretical physics~\cite{jj,cliff,bcj}.
It is then not surprising that, endowed with novel ideas such as the \emph{spinor helicity formalism} (see e.g.~\cite{Dixon:1996wi,Elvang:2013cua}), the \emph{double copy}~\cite{Bern:2008qj,Bern:2010ue}, \emph{generalized unitarity}~\cite{Bern:1994zx,Bern:1994cg}, and other tools from quantum field theory, scattering amplitudes have found their way into the (classical) two-body problem in gravity, e.g.~\cite{zvi1,zvi2,ira1,Galley:2013eba,Vaidya:2014kza,Monteiro:2014cda,Goldberger:2017vcg,donal,donalvines,Plefka:2018dpa,Li:2018qap,Cheung:2018wkq,Chung:2018kqs,Caron-Huot:2018ape,Plefka:2019hmz,cristof1,cristof2,Chung:2019duq,Bautista:2019tdr,Bautista:2019evw,KoemansCollado:2019ggb,Johansson:2019dnu}.
Some of the basic ideas are very simple, going back to the computation by Iwasaki of the 1PN correction using a scattering process~\cite{iwasaki}.
The scattering matrix carries the information about the \emph{kick} deflecting the particles, which may be extracted iteratively.
In its modern incarnation, a judicious classical limit of the quantum amplitudes --- taking the large impact parameter and heavy-mass limits --- enables the extraction of the gravitational potential.
The~latter is obtained through a matching computation to an effective (relativistic) theory with only local interactions~\cite{Cheung:2018wkq}.
In these approaches, the standard perturbative regime of field theory turns into what is known as the Post-Minkowskian (PM) expansion of the two-body problem.\vskip 4pt

While obtaining physical information out of the scattering amplitude by matching to an effective potential is a valid approach --- only specifying a gauge by Fourier transforming in the center of mass frame with respect to the transfer momentum --- it still relies on deriving a gauge dependent object as an intermediate step:
The Hamiltonian. This is of course useful, and it was used in~\cite{zvi1,zvi2} to derive the scattering angle to 3PM.
Once applied to bound states, it can also be used to compute the binding energy~\cite{Antonelli:2019ytb}.
However, as we argue, going through the Hamiltonian is unnecessarily complicated (and of increasing complexity, given the value of the coefficients of the PM expanded potential), in comparison with the much simpler form of the scattering amplitude.
Moreover, the Hamiltonian also hinders the power of the PM approach versus the PN formalism, mainly because a truncated velocity expansion may be needed to solve for, e.g. the binding energy, in a self-consistent fashion.\footnote{For instance, in the analysis of~\cite{Antonelli:2019ytb} this was circumvented by assuming the 3PM Hamiltonian provides the \emph{true answer}, and then resorting to a numerical solution.
As we show here, one can compute an analytic expression which naturally incorporates the gauge invariant relativistic information in the scattering data.}
Motivated by the on-shell philosophy of modern approaches to scattering amplitudes, we show that a map between gauge invariant quantities can be constructed instead, to all orders in velocity. This map neatly illustrates how the physical information is encoded in the amplitude, and how to translate it into observables for bound states.\vskip 4pt

The connection between scattering data and two-body dynamics has also been explored elsewhere, e.g. via an effective matching to the scattering angle~\cite{damour1,damour2,donal}.
An alternative venue was also pursued in~\cite{Vines:2018gqi}, where it was shown how the scattering angle to 2PM can be obtained from the test-particle limit, and later used in~\cite{justin2} to construct a (local-in-time) Hamiltonian for circular orbits. In all of these cases, a Hamiltonian (or equivalent) has played a central role; or a resummed version in the effective one-body formalism~\cite{eob}. As we show here, the gravitational potential can be derived from the scattering data without resorting to a matching computation to an effective description. However, motivated by the simplicity of the physical information encoded in both --- the scattering problem and bound orbits --- we have instead constructed a map that directly connects the two in a gauge invariant fashion.\vskip 4pt

Our approach relies on a remarkable connection between the momentum for the (conservative) two-body problem in the center of mass frame, and the (Fourier transform of the) scattering amplitude in the classical limit.
This relationship --- which we refer to as the \emph{impetus formula}  --- allows us, for instance, to relate the amplitude to the deflection angle to all orders in the PM expansion.
This is achieved by first inverting the well-known equation that determines the scattering angle from the radial momentum, a problem solved by Firsov many years ago~\cite{firsov}. The direct connection between the scattering matrix and deflection angle illustrates the gauge invariant information encoded in the former. Having the analytic form of the relative momentum, extracted from scattering data, we then move to the case of bound orbits.
By analytic continuation in the energy, we construct a radial action from which adiabatic invariants for elliptic motion can be computed. For instance, from the scattering amplitude we can readily obtain the periastron advance. As an example, bypassing the need of an effective Hamiltonian, we obtain the precession of the perihelion from the knowledge of the scattering amplitude to two-loops \cite{Cheung:2018wkq,zvi1,zvi2}, which recovers the known result in General Relativity to 2PN order. We give also a closed-form expression for the precession at one-loop order, which captures a series of exact terms in the PN expansion. \vskip 4pt

Our  framework is universal, and can be used to relate scattering information to observables for bound states in generic~configurations (so far ignoring spin effects). For circular orbits, however, a more geometrical program can be pursued. Using an interplay between Firsov's inversion formula and our impetus formula, we extract the orbital elements for hyperbolic motion from the scattering process, to all orders in the PM expansion. Hence, after performing an analytic continuation in the energy (or \emph{rapidity}) and impact parameter (as well as for the eccentric anomaly), we obtain the orbital elements for elliptic motion.
In particular, we determine the eccentricity, which must vanish for circular orbits.
This condition allows us to extract the (reduced) angular momentum of the orbit as a function of the binding energy, from which we can derive the orbital frequency using the first law for binary dynamics~\cite{letiec}.
As an example, using the results from~\cite{zvi1,zvi2}, we provide an expression for the orbital frequency as a function of the binding energy valid to 3PM order, and to all orders in velocity.
We also provide a closed-form expression for the contribution at tree-level, or 1PM order, which encapsulates an infinite series of exact terms in the PN expansion. \vskip 4pt  
Motivated by the relationship between the test-particle limit and two-body problem to 2PM discovered in~\cite{Vines:2018gqi}, we use the impetus formula to perform a resummation of PM effects. This is achieved by implementing the \emph{no-recoil} approximation for the scattering amplitude.
In~this limit, the impetus formula allows us to compute the amplitude by boosting to the center of mass frame from the rest frame of one of the particles.
In the rest frame, the relative momentum is then given by the expression in Schwarzschild, up to ${\cal O}(\nu)$ corrections. The~boost transforms the (reduced) energy of the test-particle into a factor of $\gamma = p_1\cdot p_2/(m_1m_2)$, while the additional overall $\Gamma^{-1} \equiv (2M)/(2E)$ discovered in~\cite{Vines:2018gqi} is due to the relativistic normalization ($(2E)^{-1}$) of the amplitude (needed to obtain scalar quantities).
As we demonstrate, the no-recoil approximation is exact to 2PM, while recovering the Schwarzschild expression for the scattering angle in the limit $\Gamma \to 1$.\footnote{The connection between test-particle limit and two-body scattering angle to 2PM can also be understood with the aid of Feynman diagrams, see~\S\ref{sec:disc}.}
We also show that it captures an infinite subset of terms at higher PM orders.
As a whole, however, the no-recoil approximation fails at 3PM. Yet, together with the impetus formula it provides a playground to explore other (potentially more accurate) resummation schemes, and a novel starting point for a gauge invariant approach to the binary inspiral problem. (See~\cite{damour3} for other recent developments.) \vskip 4pt

There are several venues to continue developing our framework. First of all, we have only considered the conservative sector and ignored radiation-reaction effects. These can be incorporated following the analysis in \cite{donal}, or using the EFT approach for radiation modes~\cite{review,Goldberger:2009qd,tail,natalia1,natalia2,lamb,nrgr4,apparent}.
We have also concentrated entirely on the non-spinning case. The inclusion of spin, both in the EFT approach~\cite{nrgrs,prl,nrgrso,nrgrs1,nrgrs2,Vines:2016qwa,levi,Vines:2017hyw} and in scattering amplitudes~\cite{Vaidya:2014kza,Chung:2018kqs,Chung:2019duq,Arkani-Hamed:2019ymq,Guevara:2018wpp,Guevara:2019fsj,Guevara:2017csg,donalvines}, introduces rich new structures. In particular, the map from hyperbolic into elliptic motion, once no longer restricted to a plane, deserves a more careful study.
We have focused also on scattering data from the amplitude and deflection angle, yet it would be useful to explore other possibilities to characterize invariant information. In principle, our map applies in the non-perturbative regime. It would be useful to refocus numerical efforts toward scattering processes in General Relativity, to use our dictionary for the --- numerically more challenging --- binary problem. We will conclude our paper with further discussions on some of these issues. Finally --- and in light of the impetus formula --- we will also indulge in some speculations about the classical double copy between gauge theory and gravity.\vskip 4pt

This paper is organized as follows.
In~\S\ref{sec:1} we review the determination of the scattering angle from the Hamiltonian.
In~\S\ref{sec:angy} we review Firsov's approach to infer the scattered momentum in the center of mass from the scattering angle, and vice versa.
We also describe how to obtain the gravitational potential without resorting to matching to an effective theory.
In~\S\ref{sec:scat} we demonstrate the impetus formula, which relates the two-body relative momentum to the Fourier transform of the classical scattering amplitude, in the conservative sector.
Furthermore, we show how radiation-reaction effects are encoded in terms quadratic in the amplitude.
Subsequently, we use the impetus formula to relate the amplitude and deflection angle, highlighting the gauge invariant information encoded in the scattering matrix.
As an example, given the amplitude to one-loop order, we provide exact expressions for the scattering angle.
In~\S\ref{sec:invariants} we develop a map to transform scattering data into adiabatic invariants for bound orbits.
We construct the radial action and illustrate the needed steps to compute observables.
Using the new results in Bern et. al. \cite{zvi1,zvi2} we then obtain the periastron advance to two-loops, which reproduces the result in General Relativity to 2PN.
We also give a closed-form expression at one-loop, which encodes a series of exact PN terms. 
Afterwards, we show how to obtain the orbital elements for elliptic motion from the hyperbolic case, via analytic continuation in impact parameter and energy.
Imposing the vanishing of the eccentricity, we obtain the orbital frequency as a function of the binding energy.
The results in~\cite{zvi1,zvi2} then allow us derive an expression valid to 3PM, and to all orders in velocity.
We also provide an exact result for the orbital frequency to 1PM, which also captures an infinite series of terms in the PN expansion.
The no-recoil approximation is described in~\S\ref{sec:resum}.
We show how it exactly recovers the 2PM dynamics, and a subset of higher order terms.
We conclude in~\S\ref{sec:disc} with a collection of main results, further discussions on open problems and some speculative explorations into the non-perturbative regime and the classical double copy.
Appendix~\ref{appA} contains a (more direct) proof of the impetus formula restricted to the conservative sector.
Throughout the paper we use $\hbar=c=1$ units, unless otherwise noted.

\section{From Dynamics to Scattering Angles}\label{sec:1}
We start by reviewing the standard procedure to compute the scattering angle given a Hamiltonian. 

\subsection{Hamiltonian Approach}\label{sec:Ham}
In classical physics we can compute the scattering angle in the center of mass frame from the radial momentum, \beq p_r^2(r,E,J) = \bp^2(r,E)-J^2/r^2\,,\eeq 
where $J$ is the conserved angular momentum. Using the Hamiltonian,
\beq
H(r,\bp^2) = \sqrt{\bp^2+m_1^2}+\sqrt{\bp^2+m_2^2} + V(r,\bp^2)\,,
\eeq
to solve for $\bp^2(r,H=E)$, the scattering angle is then given by (see e.g. \cite{zvi2})
\beq
\chi(J,E)  = -\pi + 2J \int_{r_{\rm min}}^\infty \frac{\dd r}{r^2\sqrt{\bp^2(r,E)-J^2/r^2}}\,.\label{eq:chi1}
\eeq
The closest distance between the particles $r_{\rm min}(E,J)$ is obtained by imposing $p_r(r_{\rm min})=0$.
Introducing $p^2_\infty = \bp^2(r \to \infty)$ and the impact parameter $b=J/p_\infty$, we can re-write the above expression as:
\beq
\chi(b,E)  = -\pi + 2b \int_{r_{\rm min}}^\infty \frac{\dd r}{r\sqrt{r^2\widebar\bp^2(r,E)-b^2}}\,,\label{fris1}
\eeq
with $\widebar\bp = \bp/p_\infty$. Notice that, assuming the interaction turns off at infinity $V(r,\bp^2) \xrightarrow{r\to\infty} 0$, we have
\beq
\begin{aligned}
  E &= E_1 + E_2 = \sqrt{p_\infty^2+m_1^2}+\sqrt{p_\infty^2+m_2^2} \, ,\\ p_\infty^2&= \frac{1}{4E^2}\left(E^2-(m_1-m_2)^2)(E^2-(m_1+m_2)^2\right)\,.\label{pinf1}
\end{aligned}
\eeq
The scattering angle can be computed following these simple steps. For example, in Newtonian mechanics, $H_N= \frac{\bp^2}{2\mu} -\frac{GM\mu}{r}$, we have (see e.g. \cite{damour1})
\beq
\tan \frac{\chi_{\rm N}}{2} = \frac{1}{\sqrt{2 \cE j^2}}\,,\label{eqN}
\eeq
where
\beq
j= \frac{J}{GM\mu}\,, \quad E= M(1+\nu {\cal E})\,,\label{eq:Ee}
\eeq
with $\mu = {m_1m_2 \over m_1+m_2}$, $M=m_1+m_2$, the reduced and total mass, and $\nu = {m_1m_2 \over (m_1+m_2)^2}$ is the symmetric mass ratio.

\subsection{Post-Minkowskian Expansion}

In General Relativity, on the other hand, the computation is much more challenging due to the non-linearities involved.  For large impact parameter, $b \gg GM$, the scattering angle can be computed as a series in $GM/b$, or $1/j$, what is known as the Post-Minkowskian (PM) expansion: 
\beq
\frac{1}{2} \chi(b,E) =  \sum_n \chi^{(n)}_b (E)  \left(\frac{GM}{b}\right)^n  =\sum_n \chi^{(n)}_{j} (E)\frac{1}{j^n}\,, \label{eq:chipm}
\eeq
with \beq 
\chi^{(n)}_j= \hat p_\infty^n \chi_{b}^{(n)}\,, \label{eq:bvsj}
\eeq and $\hat p_\infty = p_\infty/\mu$. While it is straightforward to read off from~\eqref{eqN} the $\chi^{(n)}_b$'s for the Newtonian case~\cite{damour1}, the scattering angle up to second order~\cite{Westpfahl:1979gu}, 
\beq
\begin{aligned}
\frac{\chi^{(1)}_b}{\Gamma} &=  \, \frac{2\gamma^2-1}{\gamma^2-1}  \,,\\ 
\frac{\chi^{(2)}_b}{\Gamma}&=\frac{3\pi}{8}\, \frac{5\gamma^2-1}{\gamma^2-1} \,,\ \label{2pmchi}
\end{aligned}
\eeq
was the state-of-the-art in General Relativity for quite some time. In the above expressions we introduced \bea
\gamma &\equiv& \frac{1}{2} \frac{E^2-m_1^2-m_2^2}{m_1m_2} = 1+ {\cal E}+\frac{1}{2}\nu{\cal E}^2 \,, \\ \Gamma &\equiv&  E/M = \sqrt{1+2\nu(\gamma-1)} = 1+\nu {\cal E} \,.
\eea
In terms of these variables, we have
\beq
\hat p_\infty^2  =  \frac{{\gamma^2-1}}{\Gamma^2}\,.
\eeq
The scattering angle to 3PM order was derived more recently, using novel tools from the theory of scattering amplitudes \cite{Bern:2008qj,Bern:2010ue,jj,cliff,bcj} via the 3PM Hamiltonian \cite{zvi1,zvi2}.
The result reads:
\bea
\chi^{(3)}_b &=& \frac{\Gamma^3}{(\gamma^2-1)^{3/2}}\left[\frac{64\gamma^6-120\gamma^4+60\gamma^2-5}{3(\gamma^2-1)^{3/2}}-\frac{4}{3}\frac{\nu}{\Gamma^2}\gamma\sqrt{\gamma^2-1}(14\gamma^2+25)
\right.\label{chi3zvi}\\
&& \left. \quad-8\frac{\nu}{\Gamma^2}(4\gamma^4-12\gamma^2-3)\Arcsinh\sqrt{\frac{\gamma-1}{2}}\,\right]\,.\nn
\eea
We will discuss the use of scattering amplitudes, and other approaches, which borrow from particle physics later on in \S\ref{sec:scat}, when we introduce a derivation of the scattering angle which does not require computing an explicit Hamiltonian.\footnote{It has been noticed that the result of the two-body scattering to 2PM order in impact parameter space resembles the test-body limit \cite{Vines:2018gqi}. We will return to this point in \S\ref{sec:resum}.}\vskip 4pt

It is somewhat convenient to introduce also the concept of \emph{rapidity}
\beq \label{rapidity}
\beta \equiv \arcosh \gamma  \,,
\eeq which will be useful later on when we study the map to bound states (for which $\gamma <1$). For example, 
\bea
\frac{\chi^{(1)}_b}{\Gamma} &=&  \, \frac{\cosh(2\beta)}{\sinh^2(\beta)}\,.
\eea
The expression of the scattering angle as a function of energy and impact parameter will be the starting point to obtain adiabatic invariants for a two-body system with bound orbits.   

\section{From Scattering Angles to Dynamics}\label{sec:angy}
In this section we proceed in reverse order, and obtain the Hamiltonian of the system in terms of the scattering angle. 

\subsection{Inversion Formula}

The inverse problem was solved some time ago by Firsov, who obtained the following formula \cite{firsov,Landau:1969}
\beq
\label{firsov1}
\widebar\bp^2(r,E) = \exp\left[ \frac{2}{\pi} \int_{r|\widebar\bp(r,E)|}^\infty \frac{\chi_b(\tilde b,E)\dd\tilde b}{\sqrt{\tilde b^2-r^2\widebar\bp^2(r,E)}}\right]\,,
\eeq
for the relative momentum.
Inserting the PM expansion of the scattering angle~\eqref{eq:chipm} into Firsov's formula and expanding the exponential in a Taylor series, we obtain
\beq
\bp^2(r,E) =  p_\infty^2(E) +\sum_i P_i(E) \left(\frac{G}{r}\right)^i  = p_\infty^2(E) \left(1+\sum_i f_i(E) \left(\frac{GM}{r}\right)^i \right)\,.\label{eq:exp}
\eeq 
The $f_n$'s are given in terms of the $\chi_b^{(n)}$'s as:
\beq
\label{eq:fi}
f_n = \sum_{\sigma\in\mathcal{P}(n)}g_\sigma^{(n)} \prod_{\ell} \left(\widehat{\chi}_b^{(\sigma_{\ell})}\right)^{\sigma^{\ell}}\,.
\eeq
For notational convenience we have introduced\footnote{This is simply motivated by the integral appearing in Firsov's formula:
\beq
\int_a^\infty \frac{\dd x}{x^n\sqrt{x^2-a^2}} = \frac{\sqrt{\pi}}{2a^n} \frac{\Gamma[n/2]}{\Gamma[(n+1)/2]}\label{integral}\,,
\eeq
with the overall factor of $2$ from the definition in \eqref{eq:chipm}.}
\beq
\widehat{\chi}_b^{(n)} = \frac{2}{\sqrt{\pi}}\frac{\Gamma(\frac{n}{2})}{\Gamma(\frac{n+1}{2})}\chi^{(n)}_b\,.
\eeq
In the expression in \eqref{eq:fi}, $\mathcal{P}(n)$ is the set of all integer partitions of $n$. Each partition is described by $n = \sigma_{\ell} \sigma^{\ell}$ (implicit summation) with mutually different $\sigma_{\ell}$'s.
In other words, the number $\sigma_{\ell}$ appears $\sigma^{\ell}$ times in the partition.
The coefficients in the above expansion are given by (with $\sigma_{\ell}\sigma^{\ell} = n$)\footnote{For the special case $n=2\cdot1$ we define $0^0=1$, such that $g_{2\cdot 1}\neq 0$, while $g_{1\cdot 2}=0^1=0$.}
\beq
g_\sigma^{(n)} = \frac{2(2-n)^{\Sigma^{\ell} - 1}}{\prod_{\ell} (2\sigma^{\ell})!!}\,.
\eeq
where $\Sigma^\ell \equiv \sum_{\ell} \sigma^{\ell}$. The relationship between the scattering angle and the coefficients of the PM expansion of the momentum suggests that both carry the relevant boundary data. As we show here, this is indeed the case, and moreover scattering information can be used to compute observables for binary systems in bound orbits.\vskip 4pt

Of course, the expressions in \eqref{eq:fi} can also be obtained following the derivation in \S\ref{sec:Ham}, order by order, by finding the value of the $\chi^{(n)}_b$'s as a function of the $f_n$'s. (For instance, \eqref{eq:fi} reproduces Eq.~(11.35) in~\cite{zvi2} to 4PM.) It is then possible, if so desired, to construct a Hamiltonian. As we show next, this can be accomplished without matching to a (local) effective theory. 

\subsection{Gravitational Potential}\label{sec:gravP}
Given the relationship between the relative momentum and the scattering angle, it is straightforward to construct the gravitational potential.
Using~\eqref{pinf1} and~\eqref{eq:exp}, we can find the Hamiltonian iteratively in powers of $G$, by solving the equation:
\bea
\sqrt{\bp^2-\sum_{i=1}^{\infty} P_i(E) \left(\frac{ G}{r}\right)^i+m_1^2}+\sqrt{\bp^2-\sum_{i=1}^\infty P_i (E) \left(\frac{ G}{r}\right)^i+m_2^2}\label{eq:E1}
= \sum_{i=0}^{\infty}  \frac{c_i(\bp^2)}{i!} \left(\frac{G}{r}\right)^i\,.
\eea 
(Notice we use a slightly different normalization for the $c_i$ coefficients than the one used in~\cite{Cheung:2018wkq,zvi1,zvi2}.) 
Clearly, at zeroth order we have:
\beq
c_0(\bp^2) = E(\bp^2) = E_1(\bp^2)+E_2(\bp^2) \equiv \sqrt{\bp^2+m_1^2} + \sqrt{\bp^2+m_2^2}\,.
\eeq
At higher orders, we find a recursion relation of the type:
\beq
c_i({\bp}^2) = \sum_{k=1}^{k=i} \frac{\sqrt{\pi}}{2\Gamma\left(\frac{3}{2}-k\right)}
\frac{E_1({\bp}^2)^{2k-1}+E_2({\bp}^2)^{2k-1}}{(E_1({\bp}^2)E_2({\bp}^2))^{2k-1}}B_{i,k}\left({\cal G}_1({\bp}^2),\dots, {\cal G}_{i-k+1}({\bp}^2)\right)\,.\label{eq:ci}
\eeq
The $B_{i,k}$'s are partial Bell polynomials, and ${\cal G}_m(\bp^2)$ is given by\footnote{Formally, one can show that the ${\cal G}_m$'s can be written as ${\cal G}_m (\bp^2) = \partial {\cal G} (\bp^2)/\partial G$ evaluated at $G=0$, with a generating function given by \beq
{\cal G} (\bp^2):=- \sum_n P_n\left(\sum_i c_i(\bp^2) \frac{G^i}{i!} \right) G^n\,.
\eeq
}
\beq
{\cal G}_m({\bp}^2) = -\sum_{s=0}^m\sum_{\ell=0}^s\frac{m!}{s!}P_{m-s}^{(\ell)}(c_0({\bp}^2))B_{s,\ell}\left(c_1({\bp}^2),\dots,c_{s-\ell+1}({\bp}^2)\right)\,.
\eeq
with $P^{(\ell)}_n = \frac{\partial^\ell}{\partial E^\ell}P_n(E)$. By the properties of the Bell polynomial, only factors of $c_k$ with $k<i$ appear on the RHS of~\eqref{eq:ci}. Notice that the pattern discovered in~\cite{zvi2}
\beq
 \frac{c_k}{k!}=  -\frac{1}{2} \left(\frac{1}{E_1}+ \frac{1}{E_2}\right) P_k(E)+\cdots\,,
\eeq
follows directly from the first term in the expansion of the square-root in~\eqref{eq:E1}.\vskip 4pt
The above formulae  reproduce the expressions given to 4PM order in Eqs.~(11.27)-(11.30) of~\cite{zvi2}, when inverted to solve for $c_i$'s as a function of $P_i$'s (and accounting for the different normalizations).
Hence, once the $P_i$'s (or $f_i$'s) are written in terms of the scattering angle through~\eqref{eq:fi}, the $c_i$'s agree with the explicit expressions as a function of $\bp^2$ given in Eq.~(10.10) of~ \cite{zvi2}.
For instance, as expected at leading order,
\beq
c_1 (\bp^2)= -\frac{1}{2} \frac{E(\bp)}{E_1(\bp)E_2(\bp)} P_1(E(\bp)) =-\frac{\bp^2}{\xi}\frac{\chi_b^{(1)}}{\Gamma}=
-\frac{\bp^2}{\xi}\frac{2\gamma^2-1}{\gamma^2-1} = -M\mu + \cO(\bp^2)\,,
\eeq
where $\xi \equiv E_1(\bp)E_2(\bp)/E^2(\bp)$. \vskip 4pt
Let us emphasize that the coefficients of the expanded Hamiltonian are obtained directly in terms of the coefficients of the PM series for the scattering angle, without performing a matching computation.
This suggests that we may bypass the use of a Hamiltonian to compute observable quantities, as we demonstrate here.
\section{From Amplitudes to Scattering Angles}\label{sec:scat}

In the context of the two-body problem in gravity, up to now the (classical limit of the) scattering amplitude has been computed as a series expansion in the PM framework and been used to derive the gravitational Hamiltonian.
We refer the reader to the comprehensive analysis in~\cite{zvi2} for further details.
In this section we explore the possibility to use the gauge invariant information encoded in the scattering amplitude to directly compute the scattering angle, without resorting to a gauge dependent Hamiltonian.
Along the way we will find a remarkable connection between the amplitude and the relative momentum of the two-body problem, from which we can readily obtain the scattering data.\vskip 4pt

Since intermediate IR divergences are expected to cancel out in observables quantities, we will restrict ourselves to IR-safe quantities.
Moreover, in the classical limit of the conservative sector we can simply retain the contributions which are non-analytic in $\bq$, and real.\footnote{There are of course also imaginary parts which account for internal gravitons going on-shell. In what follows we will concentrate on the conservative sector. See~\S\ref{sec:disc} for more on radiation-reaction.} For the sake of notation, in this and subsequent sections we will simply denote (unless otherwise noted)
\beq 
{\cal M}(\bq,\bp) \equiv  \Re \, {\cal M}^{\rm cl}_{\rm IR\text{-}fin}(\bq,\bp),
\eeq
the relativistically normalized IR finite piece of the classical amplitude in the center of mass frame. We will add, however, a few comments about isolating the IR finite piece of the amplitude in a systematic fashion.

\subsection{Amplitude \texorpdfstring{$\rightarrow$}{->} Impetus \dots}\label{sec:impetus}

As it was shown in~\cite{Cheung:2018wkq,zvi1,zvi2}, a non-relativistic EFT can be constructed to read off the gravitational potential through a matching condition: $ {\cal M}_n = 4E_1 E_2 {\cal M}^{\rm EFT}_n$,\footnote{The relativistic amplitude is related to the non-relativistic version by an overall factor of $(4E_1E_2)$.} where the ${\cal M}_n$ are obtained as a series expansion in powers of $G$:
\bea
{\cal M}(\bq,\bp) &=& \sum_n {\cal M}_n(\bq,\bp) G^n\,\\
&=& M_1 (\bp^2) \frac{G}{\bq^2} + M_2 (\bp^2) \frac{G^2}{|\bq|} + M_3 (\bp^2) G^3\log\bq^2 + \cdots\,. \nn
 \eea 
 In principle, it is not obvious how to obtain the scattering angle from the amplitude without going through a Hamiltonian.
 A key observation in this direction was made in~\cite{zvi2}, where it was speculated that the ${\cal M}^{\rm EFT}_n$'s are proportional to the $P_n$'s (or $f_n$'s) entering in the expansion of the momentum (see \S\ref{sec:angy}).\footnote{This  was explicitly checked to 5PM order. (We thank Mikhail Solon for sharing this with us.)} In fact, introducing
\beq
\label{miracle00}
{\widetilde{\cal M}}(r,E) \equiv \frac{1}{2E}\int \frac{\dd^3\bq}{(2\pi)^3}\, {\cal M}(\bq,\bp^2=p_\infty^2(E)) e^{-i\bq\cdot \br} \,,
\eeq
the Fourier transform of the amplitude (up to a relativistic normalization factor), we observe that the proportionality is  exactly what is needed to obtain the relation
\beq
\widetilde{\cal M}_n(E) = P_n(E)= p_\infty^2 M^n f_n(E) \label{eq:fM0}\,,
\eeq
where $\widetilde{\cal M}_n(E)$ are the coefficients in the PM expansion
\beq 
\widetilde{\cal M}(r,E) = \sum_{n=1}^\infty \widetilde{\cal M}_n(E) \left(\frac{G}{r}\right)^n \,.
\eeq
Provided Eq.~\eqref{eq:fM0} holds to all orders, it implies the following --- remarkably simple --- formula:
\beq 
\bp^2(r,E) = p_\infty^2(E) + \widetilde{\cal M}(r,E)\label{miracle}\,,
\eeq
relating the (Fourier transform of the) full amplitude and the momentum in the center of mass frame.
This expression then allows for the computation of the scattering angle directly from the amplitude.
We will refer to it as the \emph{impetus formula}.\vskip 4pt  Following the steps outlined in~\cite{zvi2} (see also~\cite{cristof1}), one could in principle demonstrate that~\eqref{eq:fM0} persists to all PM orders.
However, the validity of~\eqref{miracle} deserves a better understanding than just an accident of the computation in the EFT side.
We provide a derivation below which highlights the (physical) origin of~\eqref{miracle}, as well as how to correct it to incorporate radiation-reaction effects.
We give an alternative (more explicit) proof in Appendix~\ref{appA}, where we also discuss the subtleties involved in isolating the IR finite pieces.\vskip 4pt 

Let us start by mapping the computation of the scattering amplitude in the relativistic field theory with that of \emph{potential} scattering in standard quantum mechanics.
To that purpose, we re-write the evolution equation of the relativistic state,
\beq
H_{\rm PM} |\psi\rangle = \left(c_0(\bp^2) + \sum_{i=1}^\infty \frac{c_i(\bp^2)}{i!} \frac{G^i}{r^i} \right) | \psi\rangle = E |\psi \rangle\,,
\eeq
in terms of an effective Schr\"odinger-like equation, 
\beq
\left(\bp^2+ U_{\rm eff}(\bp^2,r)\right)|\psi \rangle = p_\infty^2(E) |\psi \rangle\,.\label{eq:Schr}
\eeq
The effective potential is constructed by solving
\beq
\bp^2=p_\infty^2\left(E-\sum_{i=1}^\infty \frac{c_i(\bp^2)}{i!} \frac{G^i}{r^i}\right)\,, 
\eeq
iteratively in powers of $G$, and replacing $F(\bp^2,r) E|\psi\rangle \to F(\bp^2,r) H_{\rm PM}(\bp^2,r) |\psi\rangle$ to the given PM order.
(Notice we can choose any ordering, since the commutators do not contribute in the $\hbar \to 0$ limit.)
As long as we keep all the PM corrections, we expect the solution to the original and effective Hamiltonians to match into each other in the classical limit.\footnote{ Let us stress that, while these manipulations require the existence of a local evolution equation, {\it i}) we do not need to know the explicit form of the Hamiltonian, and {\it ii}) following \cite{donal} we can rephrase the entire analysis in terms of the scattering matrix. Nevertheless, we find the connection with potential scattering extremely useful to prove the validity of \eqref{miracle}.}  \vskip 4pt 

We compute the following expectation value: 
\beq
 \langle \psi_\bp(p_\infty)| {\hat P}^2 - p_\infty^2|\psi_\bp(p_\infty)\rangle = \int \dd^3 \br\, \psi^\dagger _\bp(\br,p_\infty)\, (-\nabla^2 - p_\infty^2) \psi_\bp(\br,p_\infty)\,,\label{hatp}
\eeq
with $\hat P$ the momentum operator, and we have inserted the identity in the form $\int |\br\rangle \langle \br| \dd^3\br= \mathbb{1}$. 
The full solution with `energy' $p_\infty^2$ of the scattering problem, $\psi_\bp(p_\infty)$, is obtained in terms of the  amplitude for potential scattering, defined as
\beq
\frac{4\pi}{{\rm Vol}} f(\bp,\bp') = -  \langle \bp' | U_{\rm eff} |\psi_\bp(p_\infty)\rangle\,,\label{eq:fpp}
\eeq
where $\langle \br | \bp\rangle =  \phi _\bp = \frac{e^{i\bp\cdot \br}}{ \sqrt {\rm Vol}}$ are the free momentum states (with $`{\rm Vol}$' the usual volume factor).\footnote{In principle we should introduce wave-packets, as described in~\cite{donal}.
  However, the formula in~\eqref{miracle} can also be derived using plane waves.}
It is clear that the total energy/momentum is conserved.
However, we can use the above expression in \eqref{hatp} to define a \emph{classical, localized and instantaneous} (square of the) scattered momentum as follows:\footnote{This definition resembles the one used for the momentum in terms of the flux $\boldsymbol{J} = -\frac{i}{2}\left(\psi^\dagger \nabla\psi - \psi\nabla\psi^\dagger\right)$.
  Notice that, since the $\psi_\bp(\br,p_\infty)$ are solutions of \eqref{eq:Schr}, our expression is real by construction.} 
\beq
\label{eq:psc}
\frac{1}{{\rm Vol}}\, \bp_{\rm sc}^2(r,p_\infty^2) =\, \psi^\dagger _\bp(\br,p_\infty)\, (-\nabla^2 - p_\infty^2) \psi_\bp(\br,p_\infty)\,.
\eeq
We can justify this definition by considering the classical limit.
Re-installing the $\hbar$'s, the wave-function takes the form $\psi \simeq e^{iS_{\rm cl}/\hbar}/\sqrt{\rm Vol}$, with $S_{\rm cl}$ the classical action.
Including the non-interacting piece, \eqref{eq:psc} implies $\bp_{\rm cl}^2 = (\nabla S_{\rm cl})^2 + {\cal O}(\hbar)$, as expected. \vskip 4pt
Following \cite{donal}, we split \eqref{eq:psc} into two contributions:
\beq
 \bp_{\rm sc}^2(r,E) =  I_{(1)}(r,E) + I_{(2)}(r,E)\,,\label{eq:i1i2}
\eeq
the purely conservative part~$I_{(1)}$ (from the `potential region') which is linear in the amplitude, and radiation-reaction effects~$I_{(2)}$ (due to `radiation modes'), which depend quadratically instead. Notice that the latter involves not only dissipative but also conservative contributions, e.g. \cite{tail}. See \S\ref{sec:disc} for more on this point.
For the first term,
\beq
I_{(1)}(r,E) = {\rm Vol} \,  \Re \left[\phi^\dagger _\bp(\br,p_\infty)\, (-\nabla^2 - p_\infty^2) \psi_\bp(\br,p_\infty)\right] \,,\label{eq:I1}
\eeq
we keep only the real part of the RHS in \eqref{eq:psc} to linear order in $\psi_\bp(p_\infty)$, since the imaginary parts must cancel out by construction. The remaining term, 
\beq
I_{(2)}(r,E) = {\rm Vol} \,  \Re \left[ \left(\psi^\dagger_\bp(\br,p_\infty)-\phi^\dagger_\bp(\br,p_\infty)\right)\, (-\nabla^2 - p_\infty^2) \psi_\bp(\br,p_\infty)\right] \,,\label{eq:I2}
\eeq
 quadratic in $f(\bp,\bp')$, carries information about radiation-reaction effects (also in the conservative sector), which we do not incorporate at this moment. See \S\ref{sec:disc} for further details.\vskip 4pt 

Before we proceed, let us make a few important remarks about intermediate IR divergences. In general, IR poles exponentiate into an overall phase, which cancels out in observables quantities (such as the cross section). By the definition in \eqref{eq:psc}, an overall phase would also cancel out in $\bp^2_{\rm sc}$, making it an IR-safe observable.
On the other hand, both the $I_{(1)}$ and $I_{(2)}$ terms can in principle have (spurious) IR divergences. They may also contain imaginary parts from intermediate soft modes. Yet the IR poles must cancel out in the sum
\beq
I^{\rm IR}_{(1)}(r,E) + I^{\rm IR}_{(2)}(r,E) =0\,,
\eeq
similarly to what happens in the EFT approach \cite{tail,apparent,lamb}. While the explicit cancelation requires knowledge of the second term in \eqref{eq:I2}, we can still isolate the physical contribution from $I_{(1)}$ to $\bp_{\rm sc}^2$ by removing its IR divergent piece.
This however, must be done in a systematic fashion. In \cite{ira1,zvi1,zvi2}, the cancelation of divergences was manifest after matching to an EFT description with a local potential. As~we demonstrate in Appendix~\ref{appA}, a simple procedure can also be adapted to our case, directly obtaining (unambiguously) the relationship given by the impetus formula involving the finite part of the amplitude.\vskip 4pt

Notice that the expression in \eqref{eq:I1} involves an infinite series of terms, from iterations of the Lippmann-Schwinger equation.
However, as we demonstrate below, these terms are precisely the combination which enters in the scattering amplitude.
To show this we use the fact that $\psi _\bp(\br,p_\infty)$ obeys the Schr\"odinger equation in \eqref{eq:Schr}.
Hence,
\beq
\langle \phi_\bp| {\hat P}^2 - p_\infty^2|\psi_\bp(p_\infty)\rangle =  -\langle \bp| U_{\rm eff} |\psi_\bp(p_\infty)\rangle =  \frac{4\pi}{\rm Vol} \int \frac{\dd^3\bk}{(2\pi)^3} \int \dd^3\br \,e^{i(\bk-\bp)\cdot \br} \,f(\bk,\bp)\,,
\eeq
where we inserted the momentum identity $\int |\bk\rangle \langle \bk| \dd^3\bk= \mathbb{1}$ and used \eqref{eq:fpp}.
Gathering the pieces together, and identifying the momentum transfer $\bq=\bp-\bk$, we find: 
\beq
\label{psc2}
\bp_{\rm sc}^2(r,p_\infty^2) = 4\pi \Re  \int \frac{\dd^3\bq}{(2\pi)^3} e^{-i\bq\cdot \br}\, f_{\rm IR\text{-}fin}(p_\infty^2,\bq)\,.
\eeq
(As we explained above, IR divergences cancel out between the two contributions in \eqref{eq:i1i2}, and therefore we keep only the IR finite part of the scattering amplitude.)\vskip 4pt
To finish the proof, we must identify the effective Schr\"odinger-like scattering amplitude with the (relativistic) one from the original problem.
This can be done in a number of ways, for instance by matching the (gauge- and coordinate-invariant) cross section to all PM orders:
\beq
\frac{d\sigma}{d\Omega}= |f(p_\infty^2,\bq)|^2 = \frac{1}{(4\pi)^2(2E)^2}|{\cal M}(p_\infty^2,\bq)|^2\,.
\eeq
From here, taking the classical limit on both sides,\footnote{The perturbative expansion can also generate so-called \emph{super-classical} terms, which scale with inverse powers of $\hbar$, see e.g.~\cite{zvi2}. Similarly to the IR poles, super-classical contributions cancel out in~\eqref{eq:psc}. Therefore we can also discard them from the amplitude. See Appendix~\ref{appA} for more details.}
we conclude
\beq
4\pi \, \Re\, f^{\rm cl}_{\rm IR\text{-}fin}(p_\infty^2,\bq)=\frac{1}{2E}\Re\, {\cal M}^{\rm cl}_{\rm IR\text{-}fin}(p^2_\infty,\bq)\,,\label{eq:fM1}
\eeq
which together with~\eqref{psc2} leads to the impetus formula in~\eqref{miracle} (without radiation-reaction terms). See Appendix~\ref{appA} for a more explicit derivation.

\subsection{\dots~\texorpdfstring{$\rightarrow$}{->} Deflection Angle}  
Once the relationship between the scattered momenta and amplitude is established, we can then swiftly remove the scaffolding we use in the form of the Hamiltonian evolution.
Applying the impetus formula, it is now straightforward to compute the scattering angle by simply using~\eqref{fris1},\footnote{The $``1"$ guarantees that the unperturbed solution has $\chi=0$.}
\beq
\chi(b,E)  = -\pi + 2b \int_{r_{\rm min}}^\infty \frac{\dd r}{r\sqrt{r^2 (1+\widebar{\cal M}(r,E))-b^2}}\,,\label{eqchisim}
\eeq
where $\widebar{\cal M}(r,E)= \widetilde{\cal M}(r,E)/p_\infty^2$.
The point of closest approach can be obtained from the condition
\beq
b^2= r_{\rm min}^2\,{\widebar\bp}^2(r_{\rm min}) = r_{\rm min}^2(E,b) \left(1+ \widebar{\cal M}(r_{\rm min}(E,b),E)\right)\,.
\eeq
The above equations provide a non-perturbative map between $\chi(b,E)$ and ${\cal M}(\bp,\bq)$. This relationship can also be expressed in terms of Firsov's formula in Eq.~\eqref{firsov1}.
First we notice that $r_{\mathrm{min}}$ is given by
\beq
\label{firsov2}
r_{\rm min}^2 =  b^2 \exp\left[ -\frac{2}{\pi} \int_{b}^\infty \frac{\chi(\tilde b,E)\dd\tilde b}{\sqrt{\tilde b^2-b^2}}\right]\,,
\eeq 
where we used $\widebar\bp^2(r_{\mathrm{min}},E) = b^2/r_{\mathrm{min}}^2$.
This implies, returning to the scattering amplitude, that
\beq
 \frac{1}{2E\,p_\infty^2} \int \frac{\dd^3\bq}{(2\pi)^3} {\cal M}(\bq,\bp)e^{-i\bq\cdot\br_{\rm min}(b,E)} = \exp\left[ \frac{2}{\pi} \int_{b}^\infty \frac{\chi(\tilde b,E)\dd\tilde b}{\sqrt{\tilde b^2-b^2}}\right]-1\,.
\eeq
The reader will immediately notice that this expression closely resembles the eikonal approximation.
Indeed, taking $r_{\mathrm{min}} =b + \cdots$, we find
\beq
 \frac{1}{2E\,p_\infty^2} \int \frac{\dd^3\bq}{(2\pi)^3} {\cal M}(\bq,\bp)e^{-i\bq\cdot\bb} +\cdots = \exp\left[ \frac{2}{\pi} \int_{b}^\infty \frac{\chi (\tilde b,E)\dd\tilde b}{\sqrt{\tilde b^2-b^2}}\right]-1\,.
\eeq
Expanding the exponential in the PM approximation, we have 
\beq
\chi^{(1)}_b  = \frac{1}{2 M p_\infty^2} {\widetilde M}_1(E) = \frac{f_1}{2}\,,
\eeq
as expected.
Higher orders terms can be determined iteratively.\vskip 4pt

Alternatively, we can invert the relationship in \eqref{eq:fi}, obtaining for the coefficients of the scattering angle in impact parameter space:\footnote{\label{foot6} Notice that in some cases the $n$PM order coefficient, with  $n=\sigma_{\ell} \sigma^{\ell}$, contains a factor of $\Gamma^{-1}(p_n(\sigma)) \to 0$, whenever $p_n(\sigma)=1+\tfrac{n}{2}-\Sigma^{\ell}$ (recall $\Sigma^{\ell} = \sum_{\ell} \sigma^\ell$) is a non-positive integer.  This leads to a vanishing contribution at $n$-th order from that decomposition. For instance, the contribution from $n=1\cdot 2k$ (with $k$ a positive integer) has $p_{2k}=1+k-2k=1-k$, and therefore the  $f_1^{2k}$'s are missing from $\chi^{(2k)}_b$, while the odd powers do contribute to~$\chi^{(2k+1)}_b$.} 
  \beq
\chi_b^{(n)} = \frac{\sqrt{\pi}}{2} \Gamma\left(\frac{n+1}{2}\right)\sum_{\sigma\in\mathcal{P}(n)}\frac{1}{\Gamma\left(1+\frac{n}{2} -\Sigma^\ell\right)}\prod_{\ell} \frac{f_{\sigma_{\ell}}^{\sigma^{\ell}}}{\sigma^{\ell}!}\label{eq:fi2}\,,
\eeq
which can be written in terms of the scattering amplitude using
\beq
 f_n (E) = \frac{\widebar{\cal M}_n(E)}{M^n}\,.\label{eq:fnMt}
\eeq
For example, using eq.~\eqref{eq:fi2}, let us compute the scattering angle to 3PM order.
For $n=1$ there is only a single integer partition $1=1\cdot1$ and one obtains
\beq
\chi_b^{(1)}=\frac{\sqrt{\pi}}{2}\Gamma(1)\frac{1}{\Gamma(1/2)}\frac{f_1^1}{1!}=\frac{1}{2}f_1\,.
\eeq
At second order there are already two integer partitions, $2=1\cdot2=2\cdot1$, leading to
\beq
\chi_b^{(2)}=\frac{\sqrt{\pi}}{2}\Gamma\left(\frac{3}{2}\right)
\left(\frac{1}{\Gamma(0)}\frac{f_1^2}{2!}+\frac{1}{\Gamma(1)}\frac{f_2^1}{1!}\right)
=\frac{\pi}{4}f_2\,.
\eeq
(Notice, as we mentioned earlier, the $\tfrac{1}{\Gamma(0)}$ removes the factor of $f_1^2$ from the expansion.)  Finally, for $n=3$, we find three integer partitions of the form $3=3\cdot1=2\cdot1+1\cdot1=1\cdot3$ and we compute
\beq
\chi_b^{(3)}=\frac{\sqrt{\pi}}{2}\Gamma(2)\left(\frac{1}{\Gamma(3/2)}\frac{f_3^1}{1!}+\frac{1}{\Gamma(1/2)}\frac{f_2^1f_1^1}{1!1!}+\frac{1}{\Gamma(-1/2)}\frac{f_1^3}{3!}\right)
=f_3+\frac{1}{2}f_2f_1-\frac{1}{24}f_1^3\,.
\eeq
Assembling according to~\eqref{eq:chipm} yields
\beq
\label{eq:chi3pm}
\chi = f_1 \frac{GM}{b} + \frac{\pi}{2} f_2 \left(\frac{GM}{b}\right)^2 +\left( 2f_3 + f_2f_1 -\frac{f^3_1}{12}\right) \left(\frac{GM}{b}\right)^3 +\cdots \,,
\eeq
and so on and so forth. The relationships in \eqref{eq:fi} and \eqref{eq:fi2} ultimately illustrate the physical information encoded in the scattering amplitude, with the $f_n$'s representing the (gauge invariant) boundary data that is intimately linked to the scattering angle, and vice versa.\vskip 4pt
Notice that, in principle, the knowledge of the $f_i$'s allows us to read off an infinite series of PM terms for the deflection angle.
For example, for an `$f_1$-theory' --- obtained from ${\cal M}_1$  at 1PM --- we find from \eqref{eq:fi2}
\beq
\chi_b^{(n)}[f_1] = \frac{1}{n}(-1)^{\frac{n-1}{2}}\left(\frac{f_1}{2}\right)^n\,,\label{eq:chif1}
\eeq 
if $n$ is odd and zero otherwise (see footnote \ref{foot6}). This reproduces all the  $f_1$ terms in~\eqref{eq:chi3pm}, and beyond.
Needless to say, this sums into the Newtonian form (see~\eqref{eqN}),
\beq
\frac{\chi[f_1]}{2} = \sum_{n=1}^\infty \frac{1}{(2n+1)}(-1)^{n}\left(\frac{y}{2}\right)^{2n+1} = \Arctan(y/2)\,,
\eeq
with $y \equiv GMf_1/b$. We can also perform similar manipulations for the `$f_{1,2}$-theory', extracted from the 2PM scattering amplitude. We find from \eqref{eq:fi2},
\beq
\begin{aligned}
  \chi_b^{(2n)}[f_{1,2}] &= \frac{\sqrt{\pi}f_2^n\Gamma\left(n+\frac{1}{2}\right)}{2\Gamma(n+1)},\quad n=1,2,\dots\\
  \chi_b^{(2n+1)} [f_{1,2}] &= \frac{1}{2}f_1f_2^n \,{}_2F_1\left(\frac{1}{2},-n;\frac{3}{2};\frac{f_1^2}{4f_2}\right),\quad n=0,1,\dots \,,\label{chif12}
\end{aligned}
\eeq
and performing the sum we obtain
\beq
\frac{\chi[f_{1,2}]+\pi}{2} = \frac{1}{\sqrt{1-{\cal F}_2 y^2}}\left(\frac{\pi}{2} + \Arctan\left(\frac{y}{2\sqrt{1-{\cal F}_2 y^2}}\right)\right)\,,
\eeq
with ${\cal F}_2 \equiv f_2/f_1^2$ (see the next section for more details). It is also worth pointing out that the above expression can also be obtained directly from the integral in \eqref{eqchisim}.
\vskip 4pt

In the next section we will discuss how the boundary information from scattering processes can be used to construct adiabatic invariants for bound orbits.

\section{From Scattering Data to Adiabatic Invariants}\label{sec:invariants}
In this section we transform the information from the scattering process, with $E>M$, to the case of bound states, with $E<M$. In principle, this can be done by first inferring the Hamiltonian from the boundary data, as discussed in \S\ref{sec:gravP}, and afterwards searching for bound orbits.
As we demonstrate here, we can bypass the use of a Hamiltonian and proceed directly from gauge invariant quantities in scattering processes, such as the deflection angle, to adiabatic invariants for elliptic and circular orbits, such as the periastron advance and binding energy.
Throughout this section we use the \emph{non-relativistic} energy, 
\beq
{\cal E} = \frac{E-M}{\mu}\,,
\eeq
as well as the reduced angular momentum $j\equiv J/(GM\mu)$, introduced in \eqref{eq:Ee}. We will also use the definition $\epsilon = -2{\cal E}$, which is often standard in the PN literature.

\subsection{Radial Action}
We will follow the analysis in \cite{9912}, and introduce the radial action integral:
\beq
{\cal S}_r(J,{\cal E}) \equiv \frac{1}{\pi} \int_{r_-}^{r_+} p_r \dd r = \frac{1}{\pi} \int_{r_-}^{r_+} \sqrt{\bp^2(r,{\cal E})-J^2/r^2} \, \dd r\,,
\eeq
from which gravitational observables can be computed.
The points $r_\pm$ are the real positive roots of $p_r(r)=0$, with $0<r_- < r_+$. We will return to the issue of isolating the relevant roots in \S\ref{toellipses}. These solutions exist only for bound states, a condition which can be enforced once $\bp^2(r,E)$ is known.
As we discussed in the previous section, the functional form of $\bp^2(r,E)$ may be obtained from the knowledge of the scattering amplitude, through~\eqref{miracle}.
By analytic continuation to the region ${\cal E}<0$ (or in rapidity $\beta \to i\beta$), we can then compute gravitational observables for bound systems. Therefore, provided we consider classical processes (without anomalous thresholds), the radial action takes the form
\beq
{\cal S}_r (J,{\cal E}) = \frac{1}{\pi} \int_{r_-}^{r_+} \sqrt {p_\infty^2({\cal E}) + \widetilde{\cal M}(r,{\cal E})-J^2/r^2} \,\dd r\,,
\eeq
where the scattering amplitude is analytically continued to ${\cal E} < 0$.
This equation allows us to compute gravitational observables directly from the knowledge of the amplitude.
For instance, the periastron to periastron period,
\beq
\frac{T_p}{2\pi} \equiv \frac{1}{\mu} \frac{\partial {\cal S}_r(J,{\cal E})}{\partial {\cal E}}\,,\label{eq:Tp}
\eeq
as well as the periastron advance,
\beq
\frac{\Phi}{2\pi} = 1+ \frac{\Delta\Phi}{2\pi}= - \frac{\partial {\cal S}_r(J,{\cal E})}{\partial J}\,. \label{eq:Phi}
\eeq
At the end of the day, these expressions will be written in terms of analytic continuations of boundary data, i.e. the $f_n({\cal E})$'s or the scattering angle $\chi^{(n)}({\cal E})$.
As usual, the precise form of the analytic continuation requires a little work, as well as finding the $r_\pm$ endpoints for the radial motion.
We will give a more concrete procedure as we move along, in particular when we concentrate on circular orbits.
In general, since the $f_i$'s are functions of $\gamma$, and themselves functions of $\beta$ through~\eqref{rapidity}, we will adopt the prescription $\beta \to i\beta$, such that $\gamma \to \cos\beta$, with $0<\beta<\pi$. For example, at 1PM we have
\beq
f_1(\beta) = 2 \chi^{(1)}_b (\beta) = 2\Gamma \frac{\cosh(2\beta)}{\sinh^2\beta} \to - 2 (1+\nu{\cal E}) \frac{\cos(2\beta)}{\sin^2\beta}\,, \label{eq:f13}
\eeq
which is negative, allowing for bound orbits.\vskip 4pt

Before we conclude the general case, let us give a few useful formula to compute the radial action in the PM expansion.
As we see below, for some applications the precise knowledge of the boundary points, $r_\pm$, is not needed.

\subsubsection*{Post-Minkowskian Expansion}

Expanding the radial action in the PM framework using \eqref{eq:exp}, which are obtained through the scattering amplitude via \eqref{eq:fM0}, we encounter expressions of the type:
\beq
{\cal S}_r(J,{\cal E}) = \frac{1}{\pi} \int_{r_-}^{r_+} \dd r \sqrt{ Q(J,{\cal E},r) + \lambda \sum_{\ell=1}^\infty \frac{D_\ell({\cal E})}{r^{\ell+2}}}\,,\label{eq:radaction}
\eeq
where we introduced the split:
\bea
Q(J,{\cal E},r) &\equiv& A ({\cal E}) + \frac{2B({\cal E})}{r} + \frac{C(J,{\cal E})}{r^2}  \,,\\
A({\cal E}) &\equiv&\, p_\infty^2({\cal E})\,,\\
2B({\cal E}) &\equiv& \widetilde{M}_1({\cal E}) G \,\\
C(J,{\cal E}) &\equiv&  \widetilde{M}_2({\cal E}) G^2 -J^2\,,\\
D_n({\cal E}) &\equiv&  \widetilde{M}_{n+2} ({\cal E}) G^{n+2}\,,
\eea
with $\lambda$ a formal small parameter associated with the PM expansion.\vskip 4pt

In order to obtain the radial action, we follow~\cite{9912}, where it was demonstrated that~\eqref{eq:radaction} can be computed in terms of a contour integral with residues at $0$ and $\infty$, without the need of the values for the $r_\pm$, the turning points. First of all, we  expand the action in powers of $\lambda$,
\beq
{\cal S}_r(J,{\cal E}) =\sum_{n=0} \lambda^n {\cal S}^{(n)}_r(J,{\cal E}) \,.
\eeq
One can then show that the ${\cal S}^{(n)}_r(J,{\cal E})$'s are given as polynomials of the $D_i$'s, times a series of master integrals,
\beq
{\cal S}_{\{m,q\}} = \frac{1}{2\pi} \oint_C \frac{\dd r}{r^m}Q^{\tfrac{1}{2}-q}\,.
\eeq 
These master integrals can then be evaluated using residues:\footnote{The terms with the $\delta_{m,0}$ come from the residue at $\infty$, whereas the hypergeometric functions stem from the residue at $0$. See~\cite{9912} for more details.} 
\begin{align}
  \mathcal{S}_{\{2m,q\}} &= \begin{aligned}[t]
    &-i\,\delta_{m,0} (2q-1)B({\cal E}) A({\cal E})^{-q-\frac{1}{2}}\\
  	&+i \frac{(-1)^{m+q}A({\cal E})^{m-q}\Gamma\left(m-\frac{1}{2}\right)}
    {C(J,{\cal E})^{m-\frac{1}{2}}\Gamma(m-q+1)\Gamma\left(q-\frac{1}{2}\right)}
    {}_2F_1\left(m-\frac{1}{2},q-m;\frac{1}{2};\frac{B^2({\cal E})}{A({\cal E})C(J,{\cal E})}\right)\,,
  \end{aligned}\\
  \mathcal{S}_{\{2m+1,q\}} &= \begin{aligned}[t]
    &i\,\delta_{m,0}A({\cal E})^{\frac{1}{2}-q}\\
    &-2i\frac{(-1)^{m+q}A({\cal E})^{m-q}B({\cal E})\Gamma\left(m+\frac{1}{2}\right)}
    {C(J,{\cal E})^{m+\frac{1}{2}}\Gamma(m-q+1)\Gamma\left(q-\frac{1}{2}\right)}
    {}_2F_1\left(m+\frac{1}{2},q-m,\frac{3}{2};\frac{B^2({\cal E})}{A({\cal E})C(J,{\cal E})}\right)\,,
  \end{aligned}
\end{align}
where ${}_2F_1$ are hypergeometric functions.
Notice that the reality condition for the radial action is directly connected with the existence of bound states, for which $p_\infty^2({\cal E})<0$.\vskip 4pt

At the end of the day, the radial action takes the form:
\beq
\begin{aligned}
  \cS_r(J,\cE)
  &= -\sum_{k=0}^\infty \frac{(-1)^k\Gamma\left(k-\frac{1}{2}\right)}{2\sqrt{\pi}\Gamma(k+1)}\frac{1}{\pi}\int_{r_-}^{r_+}\dd r \,Q(J,\cE,r)^{\frac{1}{2}-k}\left(\sum_{\ell=1}^\infty \frac{D_\ell(\cE)}{r^{\ell+2}}\right)^k\\
 &=-\sum_{k=0}^\infty \frac{(-1)^k\Gamma\left(k-\frac{1}{2}\right)}{2\sqrt{\pi}\Gamma(k+1)}\frac{1}{\pi}\int_{r_-}^{r_+}\!\dd r \,Q(J,\cE,r)^{\frac{1}{2}-k}
  \!\!\!\sum_{k_1+\dots+k_\infty=k}\!\! \binom{k}{k_1,\dots,k_\infty}\prod_{\ell=1}^\infty \left(\frac{D_\ell(\cE)}{r^{\ell+2}}\right)^{k_\ell}\\
  &= -\sum_{n=0}^\infty \sum_{\sigma\in\cP(n)}
  \frac{(-1)^{\Sigma^\ell}\Gamma\left(\Sigma^\ell- \frac{1}{2}\right)}{2\sqrt{\pi}}
  \frac{1}{\pi}\int_{r_-}^{r_+}\dd r \,\frac{Q(J,\cE,r)^{\frac{1}{2}-\Sigma^\ell}}{r^{n +2\Sigma^\ell}}\prod_\ell \frac{D_{\sigma_\ell}^{\sigma^\ell}(\cE)}{\sigma^\ell!}\\
  &=-\sum_{n=0}^\infty \sum_{\sigma\in\cP(n)}
  \frac{(-1)^{ \Sigma^{\ell}}\Gamma\left(\Sigma^{\ell} - \frac{1}{2}\right)}{2\sqrt{\pi}}\cS_{\left\{n+2\Sigma^{\ell},\Sigma^{\ell}\right\}}(J,\cE)\prod_{\ell} \frac{D_{\sigma_{\ell}}^{\sigma^{\ell}}(\cE)}{\sigma^{\ell}!}
  \label{eq:srej}
\end{aligned}
\eeq
in terms of partitions of $n = \sigma_{\ell} \sigma^{\ell}$ (recall $\Sigma^{\ell} = \sum_{\ell} \sigma^{\ell}$). The gravitational observables are computed via \eqref{eq:Tp} and~\eqref{eq:Phi}. 

\subsection{Periastron Advance to Two-Loops}\label{sec:peri}

For example, we can use the radial action to compute the precession of the perihelion. From the scattering amplitude at one-loop order \cite{Cheung:2018wkq}, we have 
\beq
{\widetilde \cM}_2 = \frac{3}{2}M^2\mu^2\left(\frac{5\gamma^2-1}{\Gamma}\right)\,.
\eeq
In this case the radial action can be computed exactly, and \eqref{eq:Phi} automatically yields
\beq
\left(\frac{\Delta\Phi}{2\pi}\right)_{1\text{-}\rm loop}  =-1+ \frac{1}{\sqrt{1-\frac{{\widetilde \cM}_2G^2}{J^2}} } = \frac{{\widetilde \cM}_2G^2}{2J^2} + \cdots
=  \frac{3}{4j^2}\left(\frac{5\gamma^2-1}{\Gamma}\right) + \cdots\,,
\label{2pmperi} 
\eeq
for the periastron advance to 2PM order, and to all orders in velocity. As expected, using that $\gamma = 1+{\cal O}(v^2)$ this result reproduces the leading order value in General Relativity, $\Delta\Phi = {6\pi/j^2} + \cdots\,,$ at 1PN order.\vskip 4pt
The computation at higher orders is more involved. For instance, at two-loops we have  $D_1 \propto \widetilde{M}_3$, 
with \cite{zvi1,zvi2}
\bea
\widetilde{M}_3 (\gamma) &=&  -\frac{M^3\mu^2}{6\,\Gamma}\Biggg( 3-54\gamma^2  + \nu \left(-6+206 \gamma + 108 \gamma^2+4\gamma^3-\frac{18\Gamma(1-2\gamma^2)(1-5\gamma^2)}{(1+\Gamma)(1+\gamma)} \right) \nn \\
&& - 48 \nu(3+12\gamma^2-4\gamma^4)\frac{\Arcsinh\sqrt{\frac{\gamma-1}{2}}}{\sqrt{\gamma^2-1}}\Biggg)\,,\label{eq:tm3}
\eea
where we used the IR finite piece of the ${\cal M}_3$ scattering amplitude in Eq.~(9.3) of~\cite{zvi2} (after proper relativistic normalization). The radial action involves a series expansion, as shown in \eqref{eq:srej}, which can be written as
\beq
\begin{aligned}
  \mathcal{S}_r  =&\frac{iB}{\sqrt{A}}-\frac{i\sqrt{C}}{2\sqrt{\pi}}\sum_{n=0}^\infty\frac{(-1)^n\Gamma\left(3n-\frac{1}{2}\right)}{\Gamma(n+1)\Gamma(2n+1)}\left(\frac{AD_1^2}{C^3}\right)^n{}_2F_1\left(-n,3n-\frac{1}{2};\frac{1}{2};\frac{B^2}{AC}\right)\\
  &-\frac{i B D_1}{C^{\frac{3}{2}}\sqrt{\pi}}\sum_{n=0}^\infty\frac{(-1)^n\Gamma\left(3n+\frac{3}{2}\right)}{\Gamma(n+1)\Gamma(2n+2)}\left(\frac{AD_1^2}{C^3}\right)^n{}_2F_1\left(-n,3n+\frac{3}{2};\frac{3}{2};\frac{B^2}{AC}\right)\,.\label{eq:srd}
\end{aligned}
\eeq
Before we proceed, it is worthwhile noticing how the analytic continuation in energy works with the new term in \eqref{eq:tm3}. Implementing $\beta \to i\beta$ goes smoothly, except for the $\Arcsinh$, which leads to a complex number. However, there is also a complex denominator, such that \beq \frac{\Arcsinh\sqrt{\frac{\cosh\beta-1}{2}}}{\sqrt{\cosh^2\beta-1}} \to \frac{\Arcsinh \left[i\sqrt{\frac{1-\cos\beta}{2}}\right]}{i \sin\beta}= \frac{\arcsin \sqrt{\frac{1-\cos\beta}{2}}}{\sin\beta} \,. \eeq  
It is straightforward to show that  $G^3$ terms (and generically of the form $G^{2n+1}$) are not present. Therefore, we need to calculate the $G^4$ contribution, which includes also the amplitude at three-loops, through $\widetilde{M}_4$. Expanding the radial action to 4PM, \eqref{eq:Phi} yields
\beq
\frac{\Delta\Phi}{2\pi} = \frac{\widetilde\cM_2 G^2}{2J^2}+\frac{3(\widetilde\cM_2^2+2\widetilde\cM_1\widetilde\cM_3+2p_\infty^2\widetilde\cM_4)G^4}{8J^4} + \mathcal{O}(G^6)\,,\label{4pmperi}
\eeq
which in principle includes an infinite series of velocity corrections. By restricting to the contribution to two-loops, and performing a PN expansion, we have
\beq
\begin{aligned}
  \left(\frac{\Delta\Phi}{2\pi}\right)_{2\text{-}\rm loop} &= \frac{3}{j^2}+ \frac{3(35-10\nu)}{4j^4} + \frac{3}{4j^2}\left(10-4\nu+\frac{194-184\nu+23\nu^2}{j^2}\right)\mathcal{E}\\
  &+\frac{3}{4j^2}\left(5-5\nu+4\nu^2+\frac{3535-6911\nu+3060\nu^2-375\nu^3}{10j^2}\right)\mathcal{E}^2\\
  &+\frac{3}{4j^2}\left((5-4\nu)\nu^2+\frac{35910-126347\nu+125559\nu^2-59920\nu^3+7385\nu^4}{140j^2}\right)\mathcal{E}^3\\
  &+ \frac{3}{4j^2}\left(\left(5-20\nu+16\nu^2\right)\frac{\nu^2}{4}\right){\cal E}^4 + \cdots\,, \label{phi2loop}
\end{aligned}
\eeq
which reproduces the known result to 2PN order in General Relativity, see e.g. Eq.~(5.8) in~\cite{Bernard:2016wrg}, but it includes also a (partial) series of PN corrections.\footnote{It appears there is a typo inside the last term of Eq.~(4.16) in \cite{9912}. It should be $i_3(\nu)E^2$, without the factor of~$3$ (which is already outside the curly bracket).} In particular, the contribution from the one-loop amplitude in \eqref{2pmperi} yields the exact ${\cal O}(1/j^2)$ correction, to all orders in the binding energy. Therefore, the ${\cal O}({\cal E}^2/j^2)$ and ${\cal O}({\cal E}^3/j^2)$ terms at 3PN and 4PN, respectively, are already included in the above result, and we can already predict the 5PN contribution (shown in the last line of \eqref{phi2loop}). 
\subsection{From Hyperbolas to Ellipses \dots} \label{toellipses}
The above procedure is very generic. However, there is also a more geometrical approach to construct adiabatic invariants, which will be useful when we study the circular case. The~main observation is that the point of closest approach, $r_{\mathrm{min}}$, is a root of $r^2\widebar\bp^2=b^2$, namely  
\beq
r^2 \left(1+ \sum_i f_i({\cal E})\left(\frac{GM}{r}\right)^i\right) = b^2\,.\label{eq:roots}
\eeq
Our task is to take the boundary data from the scattering problem, encoded in the $f_i$'s, and find the two real (positive) solutions $r_\pm({\cal E},J)$ for the bound state.
Of course, we can find these solutions from~\eqref{eq:roots} after analytic continuation, with $b=J/p_\infty$, as we would do also with the Hamiltonian.
However, as we shall see, this can be done as well through an analytic continuation of the impact parameter.

\begin{figure}
  \centering
  \begin{tikzpicture}
      [scale=0.7]
    \draw[dashed] (-6,1.5) -- (6,1.5);
    \draw[dashed] (-6,-1.5) -- (6,-1.5);
    \draw[dashed] (-3.8,3) -- (0.2,-3);
	\draw[dashed] (-0.2,3) -- (3.8,-3);
    \draw[dashed] (0.8,1.5) -- (-0.8,-1.5);
    \draw[->,line width=1] (-6,1.5) .. controls (-0.6,1.5) and (0.8,1.5) .. (3.8,-3);
    \draw[->,line width=1] (6,-1.5) .. controls (0.6,-1.5) and (-0.8,-1.5) .. (-3.8,3);
    \draw[<->,line width=1,black!50!blue] (-0.363,-0.68) -- node[very near start, right=0.1] {$\tilde{r}_-$} (0.363,0.68);
    \draw[<->] (-6,1.35) -- node[right] {$b$} (-6,-1.45);
    \draw (1.4,1.5) arc (0:-57:0.6);
    \node at (1.5,1.2) {$\chi$};
    \draw (-1.4,-1.5) arc (180:123:0.6);
    \node at (-1.5,-1.2) {$\chi$};
    \filldraw (-6,1.5) circle (0.1);
    \filldraw (6,-1.5) circle (0.1);
    \filldraw[gray] (0,0) circle (0.1);
  \end{tikzpicture}
  \caption{The geometry of the scattering problem in the center of mass frame (gray dot). The motion of the bodies traces two hyperbolas, which are separated by $\tilde{r}_-$ at the point of closest approach.}
  \label{fig:scatteringGeom}
\end{figure}
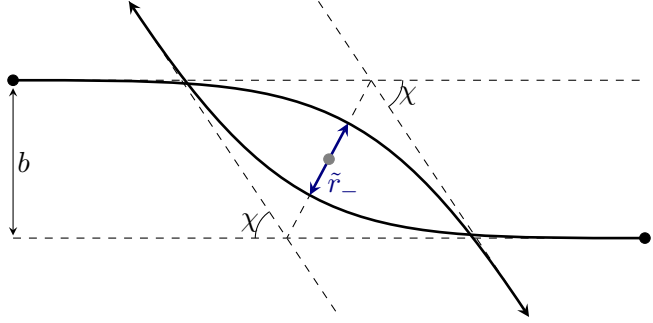

\begin{figure}
  \centering
  \begin{tikzpicture}
    [scale=0.65]
    \draw[line width=1] (0,0) ellipse (4 and 3);
    \draw[line width=1] (4.15,0) ellipse (2.5 and 2);
    \draw[] (-2.51,2.34) -- (5.86,-1.46);
    
    \draw[<->,black!30!blue] (-4,-0.15) -- node[below,near start] {$r_+$} (6.65,-0.15);
    \draw[<->,black!30!blue] (1.65,0.15) -- node[above] {$r_-$} (4,0.15);
    \draw[dashed,black!60!green] (1.71,-1.46) ellipse (6.5 and 5);
    \draw[dotted,black!60!green] (1.71,-1.46) circle (6.5);
    \draw[dashed,black!60!green] (-4.79,-1.46) -- (8.21,-1.46);
    \draw[<->,black!30!blue] (-4.79,-1.31) -- node[above] {$r_+$} (5.86,-1.31);
    \draw[<->,black!30!blue] (5.86,-1.31) -- node[above] {$r_-$} (8.21,-1.31);
    \draw[dashed,black!60!green] (-2.51,3.48) -- (-2.51,-1.46);
    \draw[dashed,black!60!green] (-2.51,3.48) -- (1.71,-1.46);

    \draw[<->,black!60!green] (1.71,-1.61) -- node[below,near end] {$a$} (8.21,-1.61);
    \draw[<->,black!60!red] (1.71,-1.85) -- node[below=0.15] {$e a$} (5.86,-1.85);

    \draw[black!60!green] (2.51,-1.46) arc (0:131:0.8);
    \node[black!60!green] at (2.25,-0.65) {$u$};

    \filldraw (-2.51,2.34) circle (0.1);
    \filldraw (5.86,-1.46) circle (0.1);
    \filldraw[gray] (2.65,0) circle (0.1);
    \filldraw[black!60!green] (1.71,-1.46) circle (0.05);
  \end{tikzpicture}
  \caption{Bound elliptic motion in the center of mass frame (gray dot). The black ellipses mark the paths of each individual body. The heavier one lies on the focus of the green dashed ellipse, which describes the worldline of the lighter body in the companion's reference frame. The dotted circle of radius $a$ defines the eccentric anomaly, $u$, which can be used to paramaterize the orbit.  See the text for a description of the orbital elements.}
  \label{fig:orbitGeom}
\end{figure}
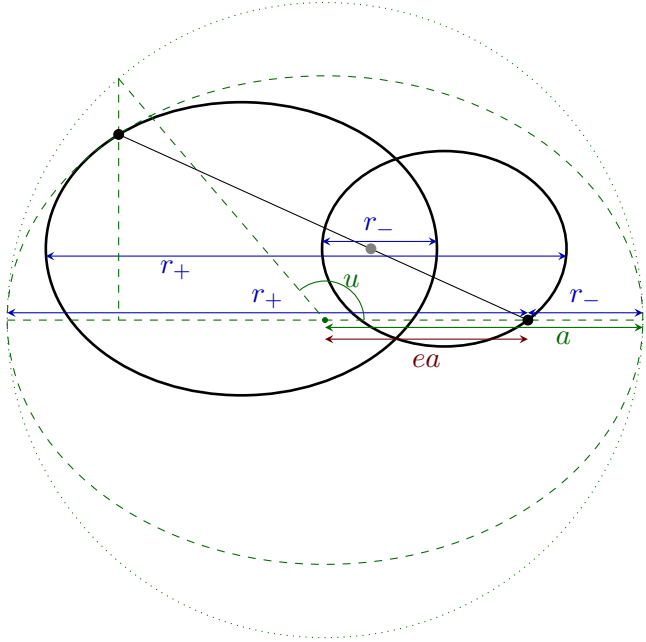

\vskip 4pt First of all we start with hyperbolic motion (see Fig.~\ref{fig:scatteringGeom}), described by the equation
\beq
r = \tilde a(\tilde e\cosh u-1)\quad \quad ({\rm Hyperbola})
\eeq
where $u$ is the {\it eccentric anomaly} and $\tilde a$ and $\tilde e$ the orbital elements.
These can be written as
\beq
-\tilde a = \frac{\tilde r_+ + \tilde r_-}{2}\,, \quad \tilde e = \frac{\tilde r_+ - \tilde r_-}{\tilde r_++\tilde r_-}\,,
\eeq
in terms of the two real solutions of~\eqref{eq:roots}, one of which is negative.
As we discuss momentarily, by performing a series of analytic continuations we can find the two roots, $r_\pm$, for bound orbits.
After these roots are found, we perform one last analytic continuation in the eccentric anomaly $u \to iu$~\cite{deruelle}, which transforms the hyperbolic into elliptic motion (see Fig.~\ref{fig:orbitGeom}): 
\beq
r= a(1-e\cos u), \label{ecc} \quad \quad ({\rm Ellipse})
\eeq
with 
\beq
a = \frac{r_+ + r_-}{2}\,, \quad e = \frac{r_+ - r_-}{r_++r_-}\,.
\eeq
The elliptic orbit can then be obtained as a result of analytic continuations from the two roots of the unbound problem.
The form in~\eqref{ecc} will be helpful later on to compute the binding energy for circular orbits, for which the eccentricity vanishes ($e=0$).

\subsubsection*{Analytic Continuation}

As a warm up, let us first consider the scattering problem to 1PM order.
The boundary information in this case is encoded in $f_1$, obtained from ${\cal M}_1$. In this ``$f_1$-theory", the condition for $r_{\min}$ follows from the second order equation
\beq
r^2 \left(1+f_1\left(\frac{G M}{r}\right)^1\right)=b^2\,.
\eeq
 The roots are given by
\beq\label{eq:f1E}
  \tilde r_\mp = -\frac{G M f_1}{2} \pm \sqrt{\frac{f_1^2G^2M^2}{4}+b^2}\,.
\eeq
with the reversed $\mp$ chosen for later convenience.  For the scattering problem we have  $f_1>0$ and $b^2>0$, such that we only have one real and positive solution: 
\beq
 r_{\rm min} = \tilde r_-= -\frac{G M f_1}{2} + \sqrt{\frac{f_1^2G^2M^2}{4}+b^2}\,. \label{eq:rmin0}
\eeq
On the other hand, for a bound orbit we have $f_1<0$. Moreover, $b^2=J^2/p_\infty^2$ takes on negative values.
Under these new conditions, it is clear that \eqref{eq:f1E} has two positive roots, obeying: $0< r_- < r_+$.
As we show below, the two real roots can also be obtained from the hyperbolic solution via an analytic continuation.
Using Firsov's formula, the procedure can be constructed entirely in terms of the scattering angle.\vskip 4pt We start with $r_-$, which can be readily obtained from $r_{\rm min}$, through the following analytic continuation
\beq
r_- (J,E)= r_{\rm min}(ib,i\beta) \,.
\eeq
Here we have: 
\beq
b =J/|p_\infty| = \frac{J  \Gamma}{\sin\beta}>0\,,
\eeq
and $\beta$ defined in \eqref{rapidity}.\vskip 4pt

To find the other solution, $r_+$, we proceed as follows.
(What we describe below can be equally applied to read off $\tilde r_+$ from the knowledge of $\tilde r_-$)
We first re-write $r_-$ in the form
\beq
r_-(b>0) = -\frac{G M f_1}{2} + b\, \sqrt{\frac{f_1^2G^2M^2}{4b^2}-1}\,. 
\eeq
To obtain the $r_+$ solution we perform the map $b \to - b$, which sends
\beq
r_+(b>0) = r_-(-b) = -\frac{G M f_1}{2} - b\, \sqrt{\frac{f_1^2G^2M^2}{4b^2}-1}\,. 
\eeq
\vskip 4pt
These manipulations can easily be extended to 2PM order.
For the ``$f_{1,2}$-theory" we have the same type of quadratic equation for the scattering problem:
\beq
  \left(1+f_1\left(\frac{G M}{r}\right)^1 + f_2\left(\frac{G M}{r}\right)^2\right) - \frac{b^2}{r^2} = 0\,.
\eeq
There are, of course, also two solutions
\beq\label{eq:f12Exact}
 \tilde r_\mp = -\frac{G M f_1}{2} \pm \sqrt{\frac{f_1^2G^2M^2}{4}-G^2M^2f_2+b^2}\,,
\eeq
and the same steps as above allow us to construct $r_\mp$ for the bound system:
\beq
r_\mp =  -\frac{G M f_1}{2} \pm \sqrt{\frac{f_1^2G^2M^2}{4}-G^2M^2f_2-b^2}\,. \label{rmf12}
\eeq
At this point the reader may wonder how the procedure will extend to higher orders ---
in particular, once the equation in~\eqref{eq:roots} becomes next-to-impossible to solve in closed analytic form.
As we discuss next,  Firsov's formula gives us a prescription which can be applied to find both the required $r_\pm$ solutions, to all PM orders. 

\subsubsection*{Via the Scattering Angle}
The main observation to construct a generalized map is the representation of ${\tilde r}_-$ in terms of the scattering angle in \eqref{firsov2}, which we reproduce here for the reader's convenience:
\beq
\label{firsov2new}
\tilde r_- =  b \exp\left[ -\frac{1}{\pi} \int_{b}^\infty \frac{\chi(\tilde b,E)\dd\tilde b}{\sqrt{\tilde b^2-b^2}}\right]\,.
\eeq
In the PM expansion of the scattering angle, it takes the form:
\beq
\tilde r_- =   b \prod_{n=1}^\infty e^{-\frac{(GM)^n\chi_b^{(n)}(\beta)\Gamma\left(\frac{n}{2}\right)}{b^n\sqrt{\pi}\Gamma\left(\frac{n+1}{2}\right)}}\,.\label{eq:r-b}
\eeq
As an example, let us show how this representation reproduces \eqref{eq:rmin0} for the $f_1$-theory.
In order to see this we need the $f_1$ contribution to the scattering angle at all orders, which is given in \eqref{eq:chif1}.
Plugging it into \eqref{eq:r-b} and performing the summation, we find 
\beq
\tilde r_-= b\, e^{\Arcsinh\left(-\frac{G M f_1}{2b}\right)} = b\left(-\frac{G M f_1}{2b}+ \sqrt{\left(\frac{G M f_1}{2b}\right)^2+1}\right)\,,
\eeq
after using the identity $\Arcsinh(x)=\log\left(x + \sqrt{x^2+1}\right)$.
According to our prescription, we obtain  $r_- (b,\beta)$ by analytic continuation,
\beq
 \tilde r_-(ib,i\beta) =  -\frac{G M f_1}{2} + {\rm sign}\, b\, \sqrt{\frac{(G M f_1)^2}{4}-b^2} = r_-(b>0,\beta)\,.
\eeq
It is also clear that $r_+(b>0) = r_-(-b,\beta)$.\vskip 4pt
The same manipulations can be applied to the $f_{1,2}$-theory.
Using~\eqref{chif12}, the representation in~\eqref{eq:r-b} yields
\beq
\begin{aligned}
  \tilde r_- 
  &=b\, \exp\left(
  \underbrace{-\sum_{n=1}^\infty \frac{(GM)^{2n}\chi_b^{(2n)}\Gamma(n)}{b^{2n}\sqrt{\pi}\Gamma\left(n+\frac{1}{2}\right)}}_{(1)}
  \underbrace{-\sum_{n=0}^\infty \frac{(GM)^{2n+1}\chi_b^{(2n+1)}\Gamma\left(n+\frac{1}{2}\right)}{b^{2n+1}\sqrt{\pi}\Gamma(n+1)}}_{(2)}
  \right)
\end{aligned}
\eeq
Let us compute one sum at a time.
The first one is straightforward:
\beq
\begin{aligned}
  (1) &= -\sum_{n=1}^\infty \frac{(GM)^{2n}\sqrt{\pi}f_2^n\Gamma\left(n+\frac{1}{2}\right)\Gamma(n)}{b^{2n}\sqrt{\pi}\Gamma\left(n+\frac{1}{2}\right)2\Gamma(n+1)}\\
  &= - \sum_{n=1}^\infty \frac{(GM)^{2n}f_2^n}{2n\,b^{2n}} = -\frac{1}{2}\sum_{n=1}^\infty \frac{1}{n}\left(\frac{G^2M^2}{b^2}f_2\right)^n =\frac{1}{2}\log\left(1-\frac{G^2M^2}{b^2}f_2\right)\,.
\end{aligned}
\eeq
The second sum is slightly more involved, \beq
\begin{aligned}
  (2)
 &= -\frac{G M f_1}{4b\sqrt\pi}\sum_{m,n=0}^\infty
  \frac{(-1)^m\Gamma\left(n+m+\frac{1}{2}\right)}{\Gamma(m+1)\Gamma(n+1)\left(m+\frac{1}{2}\right)}
  \left(\frac{G^2M^2f_2}{b^2}\right)^{n}\left(\frac{G^2M^2f_1^2}{4b^2}\right)^m\\
  &= -\frac{G M f_1}{4b\sqrt\pi}\sum_{m=0}^\infty
  \frac{(-1)^m}{\Gamma(m+1)\left(m+\frac{1}{2}\right)}
  \left(\frac{G^2M^2f_1^2}{4b^2}\right)^m
  \sum_{n=0}^\infty
  \frac{\Gamma\left(n+m+\frac{1}{2}\right)}{\Gamma(n+1)}
  \left(\frac{G^2M^2f_2}{b^2}\right)^{n}\,.
\end{aligned}
\eeq
Using the property
\beq
\sum_{n=0}^\infty \frac{\Gamma\left(n+m+\frac{1}{2}\right)}{\Gamma(n+1)}x^n = \frac{\Gamma\left(m+\frac{1}{2}\right)}{(1-x)^{m+\frac{1}{2}}}\,,
\eeq
we find
\beq
\begin{aligned}
  (2)
  &= -\frac{\sqrt{b^2}}{2b\sqrt{\pi}}\sum_{m=0}^\infty
  \frac{(-1)^m\Gamma\left(m+\frac{1}{2}\right)}{\Gamma(m+1)\left(m+\frac{1}{2}\right)}
  \left(\frac{G^2M^2f_1^2}{4(b^2-G^2M^2f_2)}\right)^{m+\frac{1}{2}}\,,
\end{aligned}
\eeq
which is nothing but the expansion of $\Arcsinh$:
\beq
\frac{1}{2\sqrt{\pi}}\sum_{m=0}^\infty \frac{(-1)^m\Gamma\left(m+\frac{1}{2}\right)}{\Gamma(m+1)\left(m+\frac{1}{2}\right)}x^{m+\frac{1}{2}} = \Arcsinh(\sqrt{x})\,,
\eeq
and we finally get
\beq
\begin{aligned}
  (2) 
  &=-\log\left(\sqrt{\frac{G^2M^2f_1^2}{4(b^2-G^2M^2f_2)}} + \sqrt{\frac{G^2M^2f_1^2}{4(b^2-G^2M^2f_2)}+1}\right)\,.
\end{aligned}
\eeq
Putting the two sums together, 
\beq
{\tilde r}_- = \frac{2b(b^2-G^2M^2f_2)}{\sqrt{b^2G^2M^2f_1^2}+\sqrt{4b^4+b^2G^2M^2(f_1^2-4f_2)}}\,,
\eeq
which agrees with \eqref{eq:f12Exact}, and can be shown to reproduce the correct $r_\pm$ via analytic continuation,
with
\beq
r_-(b,\beta) =  ib \prod_{n=1}^\infty e^{-\frac{(GM)^n\chi_b^{(n)}(i\beta)\Gamma\left(\frac{n}{2}\right)}{(ib)^n\sqrt{\pi}\Gamma\left(\frac{n+1}{2}\right)}}\,,
\eeq
and 
\beq
r_+(b,\beta) =   -ib \prod_{n=1}^\infty e^{-\frac{(GM)^n\chi_b^{(n)}(i\beta)\Gamma\left(\frac{n}{2}\right)}{(-ib)^n\sqrt{\pi}\Gamma\left(\frac{n+1}{2}\right)}} = r_-(-b,\beta)\,.
\eeq
\vskip 4pt

Of course, obtaining the solution to a quadratic equation is significantly simpler than the type of resummations we just performed.
However, finding the roots of higher order polynomials is a much more difficult problem, while Firsov's formula provides a compact representation which readily identifies the two roots we need to characterize the elliptic problem (something which is far less transparent in the generic form of the solution to~\eqref{eq:roots}).\footnote{
  We have not been able to find in the literature the representation given in~\eqref{firsov2new} for one of the roots of the polynomial in~\eqref{eq:roots}.
  We believe its remarkable simplicity deserves further study.}
This becomes more relevant for the case of circular orbits, which we study next.

\subsection{\dots to Circular Orbits}
In the limit where we collapse the ellipse into a circle, the eccentricity vanishes.
The condition $e=0$ implies that the roots are degenerate: $r_+=r_-$.
Given the representation we have for both roots, and promoting the impact parameter to a complex number~$z$ the circularity condition turns into
\beq
\begin{aligned}
  r_+ &= r_-\\
 & \Leftrightarrow \prod_{n=1}^\infty \exp\left(\frac{1}{\sqrt{\pi}}\left(\frac{GM}{z}\right)^n\frac{\Gamma\left(\frac{n}{2}\right)}{\Gamma\left(\frac{n+1}{2}\right)}\chi_b^{(n)}(-1+(-1)^n)\right) = -1\\
  & \Leftrightarrow \prod_{n=0}^\infty \exp\left(-\frac{2}{\sqrt{\pi}}\left(\frac{GM}{z}\right)^{2n+1}\frac{\Gamma\left(\frac{2n+1}{2}\right)}{\Gamma(n+1)}\chi_b^{(2n+1)}\right) = -1\\
&  \Leftrightarrow -2 \sum_{n=0}^\infty \left(\frac{1}{\sqrt{\pi}}\left(\frac{GM}{z}\right)^{2n+1}\frac{\Gamma\left(\frac{2n+1}{2}\right)}{\Gamma(n+1)}\chi_b^{(2n+1)}\right) = i \pi + 2\pi i \mathbb{N}\,. \label{eq:rprm}
\end{aligned}
\eeq
Notice that the LHS is complex for $z=ib$,  such that $z^2 < 0$ (for $b>0$), which is required to find solutions for bound orbits.
Once the above condition is met, we can solve for $b$, and subsequently for  the reduced angular momentum $j({\cal E})= |p_\infty| b/(GM\mu)$, as a function of the $f_n$'s.
Notice that, for circular orbits, the above steps bypass the need to compute the radial action.
In this special configuration, we can then read off the orbital frequency as a function of the binding energy from the first law of binary dynamics \cite{letiec},
\beq
GM\Omega_{\rm circ} = \left(\frac{d\, j({\cal E})}{d\, {\cal E}}\right)^{-1}\,.\label{eq:omegac}
\eeq
Since the PM computations depend on $\gamma$, it is natural to introduce a new object
\beq 
GM \Omega_\gamma \equiv \left(\frac{d\, j(\gamma)}{d\, \gamma} \right)^{-1} \,.
\eeq
Using that $d\gamma/d{\cal E} = \Gamma$, we find
\beq \Omega_{\rm circ}= \Omega_\gamma/\Gamma = \frac{1}{GE} \left(\frac{d\, j(\gamma)}{d\, \gamma} \right)^{-1} \,,\label{eq:orbital}\eeq
which characterizes the gauge invariant information for the bound state in a circular orbit.
As we shall see, it can be obtained systematically at any PM order, and to all orders in velocity, from the scattering data.
 
\subsubsection*{The $f_{1,2}$-Theory}

Let us demonstrate the necessary steps for the derivation to 2PM order.
We have already computed this sum for the $f_{1,2}$-theory, which now runs over the odd values, $\chi^{(2n+1)}[f_{1,2}]$ in~\eqref{chif12}.
The result, as expected, is the $\Arcsinh$ function.
The circular orbit condition becomes
\beq
\Arcsinh \left[ \sqrt{\frac{G^2M^2f_1^2}{4(z^2-G^2M^2f_2)}}\right] =  i \frac{\pi}{2} + i \pi  \mathbb{N}\,,
\eeq
which agrees with the condition of degenerate roots from~\eqref{rmf12}. In terms of $j$ we have to 2PM,\footnote{\label{foot18} Notice that, in principle, the same expression can be obtained directly from the vanishing of the radial action for circular orbits to 2PM, which becomes $(-A)(-C)=B^2$, see \S\ref{sec:invariants}. In terms of the amplitude, this can be written as 
\beq
|p_\infty|^2(J^2-G^2\widetilde{M_2}) = \left(\frac{G\widetilde{M_1}}{2}\right)^2\,,\nn
\eeq
which reproduces \eqref{eq:j2}, after using $f_n=\widetilde{M}_n/(p_\infty^2 M^n)$ and noticing $p_\infty^2<0$ for a bound state. While imposing ${\cal S}_r=0$ for circular orbits is straightforward at one-loop, it becomes much more cumbersome at higher orders (see the next subsection).}
\beq
j^2_{\rm 2PM} = |\hat p_\infty|^2 \left(\left(\frac{f_1}{2}\right)^2-f_2\right)+\cdots\,,\label{eq:j2}
\eeq
with 
\bea
f_1 &=& 2 \chi_b^{(1)} = 2\Gamma \frac{2\gamma^2-1}{\gamma^2-1}\,,\label{f12} \quad
f_2 = \frac{4}{\pi} \chi_b^{(2)} = \frac{3}{2} \Gamma \frac{5\gamma^2-1}{\gamma^2-1}\,,
\eea
obtained directly either from the scattering amplitude or deflection angle (via~\eqref{eq:fi}).
The ellipsis includes contributions from higher $f_n$'s, which we will discuss momentarily.
To find a solution to this equation we must also evaluate the $f_{1}$ and $f_2$ functions with $\gamma<1$, through the analytic continuation $\beta \to i\beta$. \vskip 4pt

Using the values for $f_{1,2}$ in~\eqref{f12}, together with~\eqref{eq:j2}, we readily obtain:
\beq
\tilde j_{\rm circ} \equiv \epsilon j^2 = \epsilon(1-\gamma^2)\left( \left(\frac{2\gamma^2-1}{\gamma^2-1}\right)^2 - \frac{3}{2\Gamma}  \frac{5\gamma^2-1}{\gamma^2-1}\right)\,,\label{tildejota}
\eeq
where we followed the convention in~\cite{blanchet} with  $\epsilon = -2 {\cal E}$.
The result in~\eqref{tildejota} is valid at 2PM, to all orders in velocity.
However, it does not include all effects needed in the PN framework for the bound system.
Nevertheless, it does capture relevant information, while also including a partial resummation of higher order terms.
To see this, we perform a PN expansion in powers of $\epsilon \sim v^2$, resulting in
\beq
\tilde j_{\rm circ}= 1 + \frac{9+\nu}{4} \epsilon + \frac{1}{16}(-55 + 48\nu + \nu^2)\epsilon^2 +\cdots\,. \label{tjota}
\eeq
We immediately notice that~\eqref{tjota} reproduces the 1PN term (see p.~140 of~\cite{blanchet}).
From here we can readily compute the orbital frequency, using the first law of binary dynamics \cite{letiec}.
The full 2PM result is a little messy, however, we can introduce the standard PN parameter,
\beq
x \equiv (GM \Omega_{\rm circ})^{2/3} =  \left(\frac{1}{\Gamma} \left(\frac{d\, j(\gamma)}{d\, \gamma} \right)^{-1}\right)^{2/3} = \epsilon + \frac{\epsilon^2}{12}(9+\nu)+{\cal O}(\epsilon^3)\,,
\label{eq:xpn}
\eeq
which is written here in powers of $\epsilon$, the binding energy. The above relationship can be inverted,
\beq
\epsilon = x - x^2\left(\frac{3}{4} +\frac{\nu}{12}\right) + {\cal O}(x^3)\,, 
\eeq
 precisely reproducing the value of the binding energy as a function of frequency to 1PN order (see e.g. Eq.~(232) in~\cite{blanchet}).
\subsection{Orbital Frequency to 3PM}\label{orbital}

In order to incorporate the newly obtained 3PM effects to the orbital frequency, we need to include the $f_3(\gamma)$ contribution.
Using~\eqref{eq:fM0} (or via the deflection angle through~\eqref{eq:fi}), we have
\bea
f_3 (\gamma) &=& \frac{r^3}{2E p_\infty^2 M^3} \int \frac{\dd^3\bq}{(2\pi)^3}\, {\cal M}_3(\bq,\bp^2=p_\infty^2(E)) e^{-i\bq\cdot \br}\, \label{eq:f3}\\
 &=& -\frac{\Gamma}{6(\gamma^2-1)}\Biggg( 3-54\gamma^2  + \nu \left(-6+206 \gamma + 108 \gamma^2+4\gamma^3-\frac{18\Gamma(1-2\gamma^2)(1-5\gamma^2)}{(1+\Gamma)(1+\gamma)} \right)\nn\\
&& - 48 \nu(3+12\gamma^2-4\gamma^4)\frac{\Arcsinh\sqrt{\frac{\gamma-1}{2}}}{\sqrt{\gamma^2-1}}\Biggg)\,,\nn
\eea
In principle, the manipulations in~\eqref{eq:rprm} can be extended to add the $f_3$ part. However, while numerically straightforward, the resulting equations are somewhat analytically cumbersome.
When restricted to circular orbits, it is useful to develop a hybrid approach, where we input the fact that the roots in Frisov's form are also the zeros of the original equation in~\eqref{eq:roots}.
In this case, the existence of a circular orbit requires that all the roots are equal, and therefore the discriminant of the cubic equation must vanish.
Furthermore, we will assume that the solution matches the 2PM result for small $f_3$, which uniquely fixes the relevant roots.
(We discuss the power counting below.)
Under these conditions we find (for $z=ib$, with $b>0$)
\beq
z^2 = \frac{1}{24}\left(-2\hat f_1^2 + 24 \hat f_2 + \frac{2 e^{\frac{2i\pi}{3}}\hat f_1(\hat f_1^3+216 \hat f_3)}{w^{\frac{1}{3}}}+2e^{\frac{2i\pi}{3}}w^{\frac{1}{3}}\right)\,,\label{b2f3}
\eeq
where  $\hat f_i = (GM/z)^i f_i$ (no summation over repeated indices).
The factor of $w$ is given by:
\beq
\begin{aligned}
  w &= -\hat f_1^6+540\hat f_1^3 \hat f_3+5832 \hat f_3^2+24i\sqrt{3\hat f_3(\hat f_1^3-27\hat f_3)^3}\\
  &=  (\hat f_1(\hat f_1^3+216\hat f_3))^{\frac{3}{2}}e^{i\arg(w)}\,.
\end{aligned}
\eeq
In general, there are three solutions for $z^2$ in~\eqref{b2f3}.
As we mentioned, we chose the one that reduces to the known solution for the 2PM theory in the limit $f_3 \to 0$, yielding
\beq
\begin{aligned}
  z^2  = \frac{1}{12}\left[-\hat f_1^2+12\hat f_2 -2 \sqrt{\hat f_1(\hat f_1^3+216\hat f_3)}e^{\frac{2i\pi}{3}}\cos\left(\frac{1}{3}\arctan\left(\frac{\Im(w)}{\Re(w)}\right)\right)\right]\,.
\end{aligned}
\eeq
This expression can be rewritten and solved for $j$, obtaining:\footnote{As for the 2PM case (see footnote \ref{foot18}), one can show that the expression in \eqref{j3pm} implies the vanishing of the radial action, including the $D_1$ term. However, notice that solving for $j({\cal E})$ (or ${\cal E}(j)$) to all orders using the radial action, see e.g. \eqref{eq:srd}, becomes much more difficult than the procedure we outlined for the specific case of circular orbits.} 
\beq
\begin{aligned}
 j^2_{\rm 3PM}    &=j^2_{\rm 2PM}+|\hat p_\infty^2| \frac{f_1^2}{6}\left({}_2F_1\left(-\frac{2}{3},-\frac{1}{3};\frac{1}{2};27{\cal F}_3\right)-1\right)\\
 &=j_{\rm 2PM}^2+3|\hat p_\infty^2| f_1^2 \sum_{m=0}^\infty \frac{4^{m+1}\Gamma(3m)}{(2(m+1))!\Gamma(m)}{\cal F}_3^{m+1}\,\\
 &=  j_{\rm 2PM}^2 + |\hat p_\infty^2| \frac{f_3}{f_1}\big( 2 +4 {\cal F}_3+32{\cal F}_3^2+384{\cal F}_3^3+\cdots\big)\,,\label{j3pm}
\end{aligned}
\eeq
as a series expansion in ${\cal F}_3 \equiv f_3/f^3_1$; with $j_{\rm 2PM}^2$ the solution in~\eqref{eq:j2}.
The above result encapsulates the full 3PM information for the bound state system to all orders in velocity (through $\gamma$), extracted from the scattering data.\vskip 4pt

By power counting we observe that  ${\cal F}_3 \simeq {\cal O}(\epsilon^2)$.
Therefore, in search of 2PN accuracy, we would only need to keep the first extra term in~\eqref{j3pm}, such that
\beq
\frac{j^2_{\rm 3PM}-j^2_{\rm 2PM}}{j^2_{\rm 1PM}} \sim {\cal O}(\epsilon^2)\,,
\eeq
with $j^2_{\rm 1PM} = |\hat p_{\infty}^2| f_1^2/4$. It is straightforward to compute $\tilde j_{\rm circ}$, keeping the leading term in~\eqref{eq:f3}, and we find
\beq
\tilde j_{\rm circ}= 1 + \frac{9+\nu}{4} \epsilon + \frac{1}{16}(81 - 32\nu + \nu^2)\epsilon^2 + \frac{\epsilon^3}{64}\left(433+847\nu-192\nu^2+\nu^3\right)+{\cal O}(\epsilon^4) \,, \label{tjota2}
\eeq
which reproduces the 2PN result in~\cite{blanchet}.
Notice that the $\nu^2\epsilon^2$ term in~\eqref{tjota} remains the same, such that the 2PM result already carried part of the information for the 2PN dynamics.
That is the case because $f_3/f_1 \sim \alpha +\beta \nu^1$ at leading order in $\epsilon$.
(In fact, the $(\nu\epsilon)^n$ terms are entirely driven by the $f_1$-theory, see below.)\vskip 4pt

The orbital frequency is obtained via~\eqref{eq:orbital}. Taking a derivative of \eqref{j3pm} with respect to $\gamma$, we have
\beq
\begin{aligned}
\left(\frac{j}{\Omega}\right)_{\rm 3PM} = \frac{GE}{2}\frac{\partial}{\partial\gamma}j^2_{\rm 3PM}  &=  \left(\frac{j}{\Omega}\right)_{\rm 2PM} + \frac{2}{3} \left(\frac{j}{\Omega}\right)_{\rm 1PM}  \left({}_2F_1\left(-\frac{2}{3},-\frac{1}{3};\frac{1}{2};27{\cal F}_3\right)-1\right) \\
& \,\,\,\,\,\,\, + j^2_{\rm 1PM} \frac{4 GE}{\sqrt{3{\cal F}_3}}\sin\left[\frac{1}{3} \arcsin\left(3\sqrt{3{\cal F}_3}\right)\right]\frac{\partial {\cal F}_3}{\partial \gamma}\,.
\end{aligned}
\eeq
\vskip 4pt \noindent Using the standard PN parameter in \eqref{eq:xpn}, we find
\beq
 \begin{aligned}
\label{eq:x3pm}
 \frac{x}{\epsilon} &= 1 + \frac{\epsilon}{12}(9+\nu) + \frac{\epsilon^2}{ 2}\left(9-\frac{17\nu}{4} + \frac{\nu^2}{9}\right) + \frac{5\epsilon^3}{48}\left(115+214\nu-\frac{191}{4}\nu^2+\frac{7}{27}\nu^3\right) \\  
&+ \frac{\epsilon^4}{12} \left(1109 - \frac{11893\nu}{30}  +\frac{10927\nu^2}{24}  -\frac{10663\nu^3}{144} + \frac{25\nu^4}{162}\right)+{\cal O}(\epsilon^5)\,.
\end{aligned}
\eeq
From here we can invert to obtain the binding energy, 
\beq
 \begin{aligned}
\epsilon =&x \left[1 - \frac{x}{12}(9+\nu) - \frac{x^2}{8}\left(27 -19\nu + \frac{\nu^2}{3}\right) +\frac{x^3}{32}\left(\frac{535}{6}-\frac{5585\nu}{6}+135\nu^2-\frac{35\nu^3}{162}\right)\right. \\
&+ \left.\frac{x^4}{384}\left(-10171+\frac{559993}{15}\nu-\frac{34027\nu^2}{3}+\frac{11354\nu^3}{9}+\frac{77\nu^4}{81}\right) + {\cal O}(x^5)\right]\,.
 \label{eq:e3pm}
\end{aligned}
\eeq
These results agree with the known value to 2PN (see Eqs.~(232) and~(349) in~\cite{blanchet}), while they include also an infinite series of velocity terms.
However, still missing are the PM corrections necessary to complete higher levels of PN accuracy, even to recover the correct test-particle limit at 3PN. (Of course, this is expected since the ${\cal O}(G^4)$ term is needed.)\vskip 4pt

\subsubsection*{The Exact $f_1$-Theory}

The 2PN order is as far as we can go with the 3PM amplitude/angle. Yet, notice that the expression in~\eqref{eq:e3pm} (or~\eqref{eq:x3pm}) still captures a subset of higher PN corrections. For example, the same pattern we found before reappears. Namely, the terms with the highest powers of $\nu$ at a given PN order in~\eqref{eq:e3pm} reproduce the correct answer.
For instance the $\frac{77}{31104}\nu^4x^4$ at 4PN order (see e.g.~\cite{Schafer:2018kuf}).
However, this is not that surprising, and it is entirely driven by the 1PM theory (the $f_1^2$ term in~\eqref{eq:j2}), as it was pointed out in~\cite{Foffa:2013gja}. In the PM framework, this is due to the scaling of the $f_n$ terms in powers of $\epsilon$ and $\nu$, relative to $f_1$. This observation allows us to incorporate all of the ${\cal O}\left(\nu^n x^n\right)$ terms in the binding energy, which are controlled by the 1PM contribution to the orbital frequency, given in closed-form by the expression:
\beq
x_{\rm 1PM}= -\frac{\epsilon \tilde\nu (16+\epsilon\tilde\nu)}
{\big(-(2+\tilde\nu)(8+\epsilon\tilde\nu)(32-\epsilon\tilde\nu(16+\epsilon\tilde\nu))\big)^{2/3}}\,,
\eeq
where $\tilde\nu=\epsilon\nu-4$. This translates into the following series of terms in the binding energy,
  \beq
  \epsilon = x \sum_{n=0}^\infty \cos\left(\frac{(n+1)\pi}{3}\right)\Gamma\left(\frac{5+2n}{6}\right)\Gamma\left(\frac{1+4n}{6}\right)\frac{(x\nu)^{n}}{\pi (n+1)!} +\cdots\,,
\eeq
which captures the exact ${\cal O}\left(\nu^n x^n\right)$ at each PN order, and is not modified by higher~PM~effects.
 
\subsubsection*{Power-Counting}
At higher PM orders we expect the above power-counting to be generic, namely the $n$-th order contribution will scale as powers of ${\cal F}_n \equiv f_n/f_1^n$ ($n>1$) relative to the leading order.
The reason is that, for a bound state we have $GM/r \sim p_\infty^2 \sim v^2$, and therefore the PM corrections are naturally down by powers of $\epsilon$, such that the scaling of the (reduced) angular momentum obeys:
\beq
j^2 \sim \epsilon f_1^2 (\underbrace{\underbrace{\underbrace{\underbrace{ 1}_{\rm 1PM}+{\cal O} \left({\cal F}_2\right)}_{\rm 2PM} +{\cal O} \left({\cal F}_3\right)+ {\cal O} \left({\cal F}^2_3\right)+\cdots}_{\rm 3PM} + \cdots + {\cal O} \left({\cal F}_n\right) + {\cal O} \left({\cal F}^2_n\right)+\cdots}_{n\rm PM} + \cdots)\,.
\eeq
Each PM order includes a series of terms, each of which can be PN expanded.
At the end of the day the result takes the form:
\beq
\tilde j = \epsilon j^2 \sim \left( 1+{\cal O} \left(\epsilon\right) +{\cal O} \left( \epsilon^2 \right) + \cdots + {\cal O} \left(\epsilon^n\right) + \cdots\right)\,,
\eeq
to any desired PN order.

\section{No-Recoil Resummation}\label{sec:resum}

In this section we perform a partial resummation of PM contributions to the impetus formula, by applying a no-recoil approximation. 
\subsection{From Test Particle Limit \dots}

The impetus formula provides a unique opportunity to explore the resummation of PM effects.
The natural candidate is the no-recoil (or test particle) limit, at leading order in $\nu$ (the symmetric mass ratio).
Since the amplitude in momentum space is a (dimensionless) relativistic scalar, we can compute it in any frame.
For instance, in the rest frame of one of the particle, in which we assume $m_1 \gg m_2$, and therefore we can ignore \emph{self-force} effects.
In this limit we have $m_2 \to \mu$ and $m_1 \to M$, as well as $E= M (1+ {\cal O}(\nu))$.
The (Fourier transform of the) amplitude can then be read off directly from the impetus formula, multiplying~\eqref{miracle00} by $2E=2M$, 
\beq
2M {\widetilde{\cal M}}_{\rm no\text{-}rec}(r,{\cal E}_0) \to  2M\left(\bp^2_{\rm Sch}(r,{\cal E}_0)-\mu^2({\cal E}_0^2-1)\right)\,,\label{eq:Mnr}
\eeq
and after identifying $p_\infty^2 \to \mu^2({\cal E}^2_0-1)$.
The value of the momentum in a Schwarzschild background is given by
\beq
{\hat\bp}_{\rm Sch}^2 = \frac{\left(1+\frac{GM}{2r}\right)^6}{\left(1-\frac{GM}{2r}\right)^2}{\cal E}^2_0- \left(1+\frac{GM}{2r}\right)^{4} \label{motsch} \,,
\eeq 
where we introduced $\hat\bp=\bp/\mu$, ${\cal E}_0=E_0/\mu$, and $E_0$ is the energy of the small body in the rest frame of the heavier particle (of mass~$M$).
By expanding in powers of $G$ we obtain the ``Schwarzschild amplitude", which reproduces ${\cal M}^{\rm Sch}_{1,2,3}$ in Eq.~(11.12) of~\cite{zvi2} to 3PM (after inserting the non-relativistic normalization factor). Notice that the above  formula, in combination with Firsov's~\eqref{firsov1}, also neatly reproduces the scattering angle in Schwarzschild.
For instance, to 2PM order: 
\bea
\chi_{b,\,{\rm Sch}}^{(1)} &=& \frac{2{\cal E}_0^2-1}{{\cal E}_0^2-1}\,, \\
\chi_{b,\, {\rm Sch}}^{(2)} &=& \frac{3\pi}{8} \frac{5{\cal E}_0^2-1}{{\cal E}_0^2-1}\,.
\eea

\subsection{\dots to Two-Body Dynamics}
We now move to the center of mass frame of the two-body problem. While the amplitude is invariant, it is useful to write it in a covariant fashion, which we can then evaluate in any frame.
This is straightforward, by simply transforming the expression in~\eqref{eq:Mnr} into a relativistic invariant, noticing 
\beq
{\cal E}_0 = \frac{M p_2^0}{\mu M} \to \frac{p_1\cdot p_2}{m_1m_2} = \gamma\,.\label{boost}
\eeq
Hence, in the center of mass frame we have
\beq
 \widetilde {\cal M}_{\rm no\text{-}rec}(r,E) = \frac{1}{2E}\left(2M \widetilde{\cal M}_{\rm no\text{-}rec}(r,{\cal E}_0 \to \gamma)\right)\,,
\eeq
such that
\beq
 \bp^2_{\rm no\text{-}rec} =  p_\infty^2 + \widetilde {\cal M}_{\rm no\text{-}rec}(r,E) = p_\infty^2 + \frac{1}{\Gamma}\Delta \bp^2_{\rm Sch}(r,{\cal E}_0\to \gamma)\,,\label{eq:norec}
\eeq
in the no-recoil approximation, where 
\beq
\Delta \bp^2_{\rm Sch}(r,{\cal E}_0 \to \gamma) = \bp^2_{\rm Sch}(r,{\cal E}_0 \to \gamma) - \mu^2(\gamma^2-1)\,.
\eeq
Notice the factor of $1/\Gamma$ follows from the relativistic normalization.\vskip 4pt 

Clearly, the no-recoil approximation does not incorporate all of the PM effects.
However, it is straightforward to show that it reproduces the 2PM dynamics.
This can be seen directly from~\eqref{eq:norec}, by computing the $f^{\rm no\text{-}rec}_n(E)$'s.
From the definition in~\eqref{eq:exp}, we find 
\beq
\begin{aligned}
\frac{1}{\Gamma} f^{\rm no\text{-}rec}_1(E) &= 2  \frac{2\gamma^2-1}{\gamma^2-1} \,,\\
\frac{1}{\Gamma} f^{\rm no\text{-}rec}_2(E) &= \frac{3}{2} \frac{5\gamma^2-1}{\gamma^2-1} \,,\\
\frac{1}{\Gamma}  f^{\rm no\text{-}rec}_3(E) &= \frac{1}{2} \frac{18\gamma^2-1}{\gamma^2-1}\,,\label{f3ng}
\end{aligned}
\eeq
and so on and so forth. Using~\eqref{eq:fi}, the knowledge of the $f_n$'s leads directly to the scattering angle. Given that $\chi_b^{(1,2)} \propto f_{1,2}$, it is straightforward to show the value of $\chi$ for the two-body problem is recovered to 2PM, see \eqref{2pmchi}. Since the scattering angle (and/or the $f_n$'s) encode all the required information, we conclude that the no-recoil approximation reproduces the two-body dynamics to 2PM order. The impetus formula thus provide the backbone for the relationship between test-particle and two-body problem discovered in~\cite{Vines:2018gqi}.\vskip 4pt

At the next order, however, the no-recoil approximation fails due to self-force effects.
Nevertheless, the expression in~\eqref{f3ng} captures one of the contributions from the full~$f_3$. The no-recoil approximation amounts to the $\nu$-independent factor inside the parenthesis of~\eqref{eq:f3}. The series in principle continues ad infinitum. It is easy to see that in the limit $\Gamma=1$ we recover the exact Schwarzschild prediction for the scattering angle. For example, see~\eqref{eq:chi3pm}, we have 
\beq
\chi^{(3)}_{\rm no\text{-}rec} =  \frac{1}{2} \lim_{\Gamma\to 1} \left( 2f_3 + f_2f_1 -\frac{f^3_1}{12}\right)_{\rm no\text{-}rec}= \chi^{(3)}_{\rm Sch}({\cal E}_0 \to \gamma)\,,\label{eq:sch3}
\eeq
in impact parameter space, by construction.
However, when $\Gamma \neq 1$, the approximation does not provide a reliable proxy for the ${\cal O}(\nu)$ terms, as it does at lower PM orders.
This is related to the peculiar factors of $\Gamma$ that appear in the $f_n$'s at higher orders.
It is though possible to envision a resummation which also includes the factors of $\Gamma$, by finding the correct variable.
We will return to this point in future work.\vskip 4pt

Notice that the connection between two-body dynamics and the test-particle limit to 2PM order not only holds for the scattering angle, it also translates to physical observables for bound orbits.
This is expected, since there is a one-to-one map between the scattering angle and the binary dynamics.
The connection is in fact remarkably simple, also at the level of the orbital frequency for circular orbits.
Notice, that the overall factor of $\Gamma$ cancels out in~\eqref{eq:j2} to 1PM order (but remains at 2PM).
This means $j^2_{\rm Sch}({\cal E}_0 \to \gamma)$ is exact to 1PM, after taking $M=m_1+m_2$ for the mass.
This is yet another indication that General Relativity at 1PM order is recovered by the test-particle limit. 

\section{Discussion and Outlook}\label{sec:disc}

\subsubsection*{Summary \& Conclusions}
We have introduced a dictionary relating gravitational scattering data to observables for bound states in generic configurations. Our map can be described (schematically) as follows: \vskip 4pt
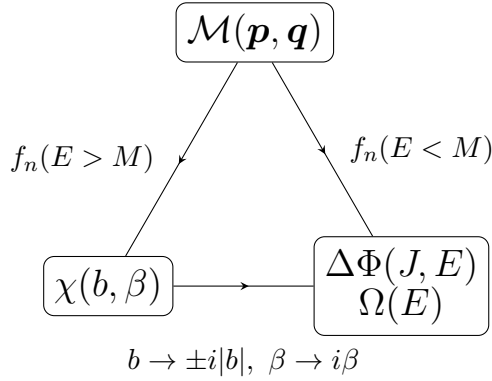
\begin{center}
  \begin{tikzpicture}
    [scale=1.5]
    \node[inner sep=4, rectangle, draw, rounded corners] (A) at (90:1.5) {\Large$\cM(\bp,\bq)$};
    \node[inner sep=4, rectangle, draw, rounded corners] (B) at (210:1.5) {\Large$\chi(b,\beta)$};
    \node[inner sep=4, rectangle, draw, rounded corners, align=center] (C) at (330:1.5) {\Large$\Delta\Phi(J,E)$\\\Large$\Omega(E)$};
    \draw[quark] (A) -- node[left=0.3] {$f_n(E>M)$} (B);
    \draw[quark] (B) -- node[below=0.75] {$b\rightarrow \pm i\lvert b \rvert,~\beta\rightarrow i \beta$} (C);
    \draw[quark] (A) -- node[right=0.3] {$f_n(E<M)$} (C);
  \end{tikzpicture}
\end{center}
Our dictionary relies on the remarkable connection we have shown exists between the relativistic scattering amplitude in the classical limit and the relative momentum of the two-body~system:
\beq 
\bp^2(r,E) = p_\infty^2(E) + \frac{1}{2E} \int \dd^3 \br\, {\cal M}(\bp,\bq)e^{i\bq\cdot\br} + {\rm R.R.}\,\label{mir1}
\eeq
The R.R. stands for radiation-reaction terms, quadratic in the amplitude (see~\eqref{eq:i1i2}), which may also contribute to the conservative sector (see below)~\cite{tail,apparent,lamb}.\footnote{Let us emphasize that, by construction, the RHS of the impetus formula is IR finite, such that intermedia IR divergences cancel out between the two terms, see \S\ref{sec:impetus} and Appendix \ref{appA}.} Together with Firsov's equation~\cite{firsov}, relating the momentum to the scattering angle, the above \emph{impetus formula} allowed us to relate the amplitude directly to the scattering angle: 
\beq
 \frac{1}{2E\,p_\infty^2} \int \frac{\dd^3\bq}{(2\pi)^3} {\cal M}(\bq,\bp)e^{-i\bq\cdot\br_{\rm min}(b,E)} = \exp\left[ \frac{2}{\pi} \int_{b}^\infty \frac{\chi(\tilde b,E)\dd\tilde b}{\sqrt{\tilde b^2-b^2}}\right]-1\,,
\eeq
with
\beq
r_{\rm min}^2(b,E) =  b^2 \exp\left[ -\frac{2}{\pi} \int_{b}^\infty \frac{\chi(\tilde b,E)\dd\tilde b}{\sqrt{\tilde b^2-b^2}}\right]\,.\label{rmindis}
\eeq
In the PM framework, we were then able to relate the coefficients in the expansion of the amplitude,
\beq
f_n = \frac{1}{2E p_\infty^2} \left(\frac{r}{M}\right)^n \int \frac{\dd^3\bq}{(2\pi)^3}\, {\cal M}_n(\bq,\bp^2=p_\infty^2(E)) e^{-i\bq\cdot \br}\,,
\eeq
to the coefficients for the deflection angle in impact parameter space, and vice versa. The general expression can be written as (see~\eqref{eq:fi} and \eqref{eq:fi2})\footnote{Expressions for the scattering angle (in angular momentum space) as a function of the $P_n$ coefficients (in our notation in \eqref{eq:exp}) were given more recently also in \cite{Bjerrum-Bohr:2019kec}, see e.g. their Table I (to 12PM order). One can check that \eqref{eq:fi2n} is (independently) confirmed by the results in \cite{Bjerrum-Bohr:2019kec}. Moreover, one can also show that the condition given in Eq. 4.39 of \cite{Bjerrum-Bohr:2019kec} for vanishing contributions is equivalent to the condition $p_n(\sigma)=1+\frac{n}{2}-\sum_{\ell} \sigma^{\ell}=0$ in \eqref{eq:fi2n}, see footnote \ref{foot6}. The additional cancelations we find, which occur in \eqref{eq:fi2n} whenever $p_n(\sigma)$ is a negative integer, are not included in their Eq. 4.39 but can be obtained following similar steps as described in \cite{Bjerrum-Bohr:2019kec}.} 
\beq
\chi_b^{(n)} = \frac{\sqrt{\pi}}{2} \Gamma\left(\frac{n+1}{2}\right)\sum_{\sigma\in\mathcal{P}(n)}\frac{1}{\Gamma\left(1+\frac{n}{2}-\sum_{\ell}\sigma^{\ell}\right)}\prod_{\ell} \frac{f_{\sigma_{\ell}}^{\sigma^{\ell}}}{\sigma^{\ell}!}\label{eq:fi2n}\,,
\eeq
For example, introducing ${\cal F}_n \equiv f_n/f_1^n$, for $n>1$, and
\beq
y\equiv \frac{GMf_1}{b} = \frac{2GE}{b}\frac{(2\gamma^2-1)}{(\gamma^2-1)} = \frac{2(2\gamma^2-1)}{\sqrt{(\gamma^2-1)j^2}}\,,
\eeq
we found to 3PM order:\footnote{Notice that the PM framework naturally introduces an expansion in $GMf_1/b$. Moreover, in the limit ${\cal F}_{2,3}=0$, \eqref{eq2:chi3pm} generalizes the Newtonian result in~\eqref{eqN}, away from $\nu = 0$ and to all orders in velocity.}
\beq
\label{eq2:chi3pm}
\frac{\chi+\pi}{2}=\frac{1}{\sqrt{1-{\cal F}_2 y^2}}\left(\frac{\pi}{2}+ {\Arctan}\left(\frac{y}{2\sqrt{1-{\cal F}_2 y^2}}\right)\right)+{\cal F}_3 y^3+\cdots\,,
\eeq
which includes a resummation of 1PM and 2PM terms. 
The relationship between $\chi$~and~$\cal M$, through the ${\cal F}_n$'s, suggests that the scattering amplitude by itself carries gauge invariant information in the form of \emph{asymptotic charges}. It would be useful to properly classify the scattering data, perhaps recasting~\eqref{eq:fi2n} (or equivalently \eqref{eq:fi}) and the above expressions in a geometrical (or algebraic) fashion. It would be also interesting to understand the connection with the eikonal~phase~\cite{KoemansCollado:2019ggb}.\vskip 4pt

By constructing a radial action using the impetus formula, together with a series of analytic continuations in the rapidity ($\gamma =  \cosh\beta$) and impact parameter, of the form $\beta \to i\beta,\, b \to ib $,\footnote{This transformation is intimately linked to the one relating hyperbolic to elliptic motion, via analytic continuation in the eccentric anomaly, $u \to iu$~\cite{deruelle}.} we are able to relate the scattering information to the computation of adiabatic invariants for bound orbits. As an example, we readily obtained the periastron advance to 4PM order,
\beq
\frac{\Delta\Phi}{2\pi} = \frac{\widetilde\cM_2 G^2}{2J^2}+\frac{3(\widetilde\cM_2^2+2\widetilde\cM_1\widetilde\cM_3+2p_\infty^2\widetilde\cM_4)G^4}{8J^4} + \mathcal{O}(G^6)\,,\label{4pmperin0}
\eeq
without the need of a Hamiltonian. The scattering amplitude is encoded in the coefficients of the PM expansion ($\widetilde{\cal M}_n = M^n p_\infty^2 f_n$) on the RHS of the impetus formula \eqref{mir1}.  The above expression, up to two-loops \cite{zvi1,zvi2}, matches the precession of the perihelion in General Relativity to 2PN order, while including also a series of velocity corrections, see~\S\ref{sec:peri}.\footnote{Notice that the agreement with the known value for the periastron advance provides --- yet another --- confirmation of the validity of the result presented in \cite{zvi1,zvi2} to two-loops. Moreover, from Eq. (5.8) of \cite{Bernard:2016wrg} to 4PN, the expression in \eqref{4pmperin0} can be used as a cross-check for the (instantaneous) value of the scattering amplitude to ${\cal O}(G^4v^2)$. In principle, at 4PM order radiation-reaction terms will also contribute (see below).} We~also found that the contribution from the one-loop amplitude is exact, and takes the form\footnote{The expression in \eqref{4pmperin} agrees with the result found in \cite{Caron-Huot:2018ape} to leading order, after taking into account the proper normalization factors and transforming the amplitude from momentum into real space. Our approach, however, is more generic and it applies to all loop orders via the radial action, without the introduction of a Hamiltonian, see \S\ref{sec:invariants}. Since only the triangle integral contributes to the IR finite part of the scattering amplitude at one-loop in the classical limit \cite{Cheung:2018wkq}, it is also evident that its vanishing would lead to no precession at leading order, as it was found in \cite{Caron-Huot:2018ape} for the case of ${\cal N}=8$ supergravity.}  
\beq
\left(\frac{\Delta\Phi}{2\pi}\right)_{1\text{-}{\rm loop}}  =-1+ \frac{1}{\sqrt{1-\frac{{\widetilde \cM}_2G^2}{J^2}} }  =  \frac{3}{4j^2}\left(\frac{5\gamma^2-1}{\Gamma}\right) + \cdots\,.\label{4pmperin}
\eeq
Given the structure of the PM expansion, it is easy to see the leading PM term in \eqref{4pmperin} captures the ${\cal O}(1/j^2)$ contributions to the periastron advance to all orders in the velocity expansion, see \S\ref{sec:peri}.\vskip 4pt 

The derivation of adiabatic invariants simplifies for the case of circular orbits, allowing us to solve for the (reduced) angular momentum of the bound state in terms of scattering data, from the condition of a vanishing eccentricity. We found
\beq
j_{\rm circ}  =j_{\rm 1PM} \left(1- 4 {\cal F}_2(i\beta)+\frac{2}{3} \left({}_2F_1\left[-\frac{2}{3},-\frac{1}{3};\frac{1}{2};{27{\cal F}_3(i\beta)}\right]-1\right)\right)^{1/2}\,,
\eeq
to 3PM order, where
\beq
 j_{\rm1PM}=\frac{(2\cos^2\beta-1)}{\sin\beta}\,,\label{j1pmdis}
 \eeq
 is the 1PM result.\footnote{Notice \eqref{j1pmdis} agrees with the value in Schwarzschild after replacing ${\cal E}_0\to\gamma$.} The orbital frequency can be derived from the first law of binary dynamics:
\beq \, \Omega_{\rm circ}=  \frac{1}{GE} \left(\frac{d\, j_{\rm circ}}{d\, \gamma} \right)^{-1} =  \Omega_\gamma/\Gamma \,,\eeq
which agrees with the known result to 2PN order. It also includes a resummation of all velocity terms scaling as ${\cal O}(G^3v^n)$, see \S\ref{orbital}. Our dictionary thus allows us to obtain the orbital frequency as a function of the binding energy, directly using the gauge invariant information from the scattering amplitude, without resorting to the (lengthier and gauge dependent) Hamiltonian. Following the steps outlined in this paper, we can translate the scattering data --- either from the amplitude or deflection angle --- directly to observables, at any order in the PM expansion and to all orders in velocity.\vskip 4pt

We have also performed a partial resummation of PM terms, using a \emph{no-recoil} approximation for the scattering amplitude. Using the impetus formula, we have
\beq
\Delta \bp^2_{\rm no\text{-}rec} = \frac{1}{\Gamma}\Delta \bp^2_{\rm Sch}(r,{\cal E}_0\to \gamma)\,,
\eeq
for the scattered momentum of the two-body problem in terms of the (boosted) scattered momentum in Schwarzschild.
The factor of $1/\Gamma$ arises from the relativistic normalization in~\eqref{mir1}.
This (PM-resummed) approximation exactly reproduces the 2PM dynamics, and also provides partial information for the higher order terms.
There is, of course, a lot of room for improvement.
Even though the no-recoil approximation carries some of the higher order terms (see~\eqref{f3ng}), while encoding at the same time the exact result in Schwarzschild at leading order in $\nu$ (see~\eqref{eq:sch3}), it fails to provide a good proxy for the exact result at 3PM.
For instance, while the first term in~\eqref{chi3zvi} mimics the Schwarzschild result, the overall factor of $\Gamma^3$ does not come out of the above approximation, which instead mixes different powers of $\Gamma$ through the different~$f_n$'s.
Given the nature of the computation, it appears that the structure $\Gamma^k \chi_j^{(n)}({\cal E}_0\to \gamma)$ for one of the terms at $n$PM order is quite generic~\cite{damour1,damour2}.
This suggests that we should be able incorporate the factors of $\Gamma$ inside the map between ${\cal E}_0$~and~$\gamma$, and to resum also all of these terms.
(For instance,  using $\Gamma  = 1/\Gamma + {\cal O}(\nu)$, it is easy to manipulate the expression in~\eqref{f3ng} to recover the first term in~\eqref{chi3zvi}.) We explore this possibility in future~work.\vskip 4pt

The 3PM state-of-the-art for the scattering amplitude/angle leads to the 2PN orbital-frequency/binding-energy. At this stage, even though a resummation of velocity corrections is included, this is not yet a very accurate result if compared with the present state-of-the-art in PN computations \cite{nrgr4}. However, a series of improvements can be easily incorporated. In principle we can fix the full ${\cal O}(\nu^0)$ contribution using the no-recoil approximation, which includes the Schwarzschild limit. We can also include all of the ${\cal O}\left(\nu^n x^n\right)$ terms, which are controlled by the $f_1$-theory. For the latter we found the exact result to 1PM (with  $\tilde\nu=\epsilon\nu-4$),
\beq
x_{\rm 1PM}= -\frac{\epsilon \tilde\nu (16+\epsilon\tilde\nu)}
{\big(-(2+\tilde\nu)(8+\tilde\nu)(32-\epsilon\tilde\nu(16+\epsilon\tilde\nu))\big)^{2/3}}\,.
\eeq
This closed-form expression can be inverted, to capture the following series of corrections to the binding energy,
  \beq
  \epsilon= x \sum_{n=0}^\infty \cos\left(\frac{(n+1)\pi}{3}\right)\Gamma\left(\frac{5+2n}{6}\right)\Gamma\left(\frac{1+4n}{6}\right)\frac{(x\nu)^{n}}{\pi (n+1)!} +\cdots\,,
  \eeq
in the PN expansion, which are solely determined by the 1PM theory. (In some sense, these terms are `tree-level exact' and do not get {\it renormalized} by higher PM orders.) These additional contributions will help in the construction of more accurate waveform models.\vskip 4pt

In conclusion, the dictionary we introduced in this paper provides a natural way to translate the gauge invariant information in scattering processes into observables for bound states in gravity, without the need of gauge dependent objects. The novel tools in the study of scattering amplitudes can potentially provide the necessary high-precision scattering data, bypassing the combinatorial hurdles of the Feynman technology, and in a manifestly relativistic framework. In principle, relativistic integration with internal massive particles --- which is needed to benefit from the powerful \emph{on-shell} techniques --- makes the problem of computing the relativistic amplitude significantly more laborious than the PN counter-part. That is the reason the 3PM order, at two-loops, has been tackled only very recently in \cite{zvi1,zvi2}, while the PN expansion is progressing towards the 5PN (five-loops) level of accuracy \cite{nrgr4,5pn1,5pn2}. Nevertheless, we expect further improvements will streamline the derivation of the required scattering amplitude (and/or independent derivations of the deflection angle) such that --- in conjunction with our novel framework and various other techniques --- the necessary high level of precision for the future of GW astronomy can be achieved.\vskip 4pt
 
In what follows we provide a brief discussion over a few directions in which our formalism can be explored further. We restrict ourselves, as for the entire present work, to the case of non-rotating bodies. We will explore the spinning scenario elsewhere.

\subsubsection*{Radiation-Reaction}
The impetus formula in~\eqref{eq:i1i2},
\beq
 \bp_{\rm sc}^2(r,E) =  I_{(1)}(r,E) + I_{(2)}(r,E)\,,\eeq
 also receives corrections due to radiation effects, encoded in the second term, $I_{(2)}$, which is quadratic in the amplitude.
 In principle, following the analysis in~\cite{donal}, we can systematize the necessary steps to add radiation-reaction. The latter includes so-called \emph{tail effects}~\cite{tail}, which in the  scattering amplitude can be incorporated through radiation modes, starting at 4PM order \cite{zvi2}. The advantage in the PM formalism is the lack of intermediate (spurious) IR divergences which pollute the PN computations due to the split into regions --- which would not be present in the PM framework.
 Since back-reaction terms involve long-distance modes, it is also possible to compute them using the EFT approach~\cite{review}, in the form of a radiation-reaction force~\cite{tail,natalia1,natalia2}.
 The latter also includes a conservative piece, which shifts the value of the binding energy~\cite{tail,apparent,lamb,review,nrgr4}.
 We can then re-do our analysis in the conservative sector by including the contribution to the binding energy due to the tail, e.g. $E \to E+ E_{\rm tail}$, which may be obtained independently (and resummed using the renormalization group evolution discussed in~\cite{tail}).
 We will study radiation-reaction in more detail, as well as absorption effects~\cite{dis1,dis2}, in future work.

\subsubsection*{Scattering Angle From the EFT Approach}

\begin{figure}
  \centering
  \begin{subfigure}[b]{0.1\textwidth}
    \centering
    \begin{tikzpicture}
      [scale=1]
      \draw[boson] (0,0) -- (0,-1.5);
      \filldraw (0,0) circle (0.05);
      \filldraw (0,-1.5) circle (0.05);
    \end{tikzpicture}
    \caption{}
  \end{subfigure}
  \begin{subfigure}[b]{0.15\textwidth}
    \centering
    \begin{tikzpicture}
      [scale=1]
      \draw[boson] (-0.7,0) -- (0,-1.5);
      \draw[boson] (0.7,0) -- (0,-1.5);
      \filldraw (-0.7,0) circle (0.05);
      \filldraw (0.7,0) circle (0.05);
      \filldraw (0,-1.5) circle (0.05);
    \end{tikzpicture}
    \caption{}
  \end{subfigure}
  \begin{subfigure}[b]{0.15\textwidth}
    \centering
    \begin{tikzpicture}
      [scale=1]
      \draw[boson] (-0.7,0) -- (0,-0.9);
      \draw[boson] (0.7,0) -- (0,-0.9);
      \draw[boson] (0,-0.9) -- (0,-1.5);
      \filldraw (-0.7,0) circle (0.05);
      \filldraw (0.7,0) circle (0.05);
      \filldraw (0,-1.5) circle (0.05);
    \end{tikzpicture}
    \caption{}
  \end{subfigure}
  \begin{subfigure}[b]{0.15\textwidth}
    \centering
    \begin{tikzpicture}
      [scale=1]
      \draw[boson] (-0.7,0) -- (-0.35,-0.5);
      \draw[boson] (0,0) -- (-0.35,-0.5);
      \draw[boson] (0.7,0) -- (0,-0.9);
      \draw[boson] (-0.35,-0.5) -- (0,-0.9);
      \draw[boson] (0,-0.9) -- (0,-1.5);
      \filldraw (-0.7,0) circle (0.05);
      \filldraw (0,0) circle (0.05);
      \filldraw (0.7,0) circle (0.05);
      \filldraw (0,-1.5) circle (0.05);
    \end{tikzpicture}
    \caption{}
  \end{subfigure}
  \begin{subfigure}[b]{0.15\textwidth}
    \centering
    \begin{tikzpicture}
      [scale=1]
      \draw[boson] (-0.7,0) -- (0,-0.9);
      \draw[boson] (0,0) -- (0,-0.9);
      \draw[boson] (0.7,0) -- (0,-0.9);
      \draw[boson] (0,-0.9) -- (0,-1.5);
      \filldraw (-0.7,0) circle (0.05);
      \filldraw (0,0) circle (0.05);
      \filldraw (0.7,0) circle (0.05);
      \filldraw (0,-1.5) circle (0.05);
    \end{tikzpicture}
    \caption{}
  \end{subfigure}
  \begin{subfigure}[b]{0.13\textwidth}
    \centering
    \begin{tikzpicture}
      [scale=1]
      \draw[boson] (-0.5,0) -- (-0.5,-0.75);
      \draw[boson] (-0.5,-0.75) -- (-0.5,-1.5);
      \draw[boson] (0.5,0) -- (0.5,-0.75);
      \draw[boson] (0.5,-0.75) -- (0.5,-1.5);
      \draw[boson] (-0.5,-0.75) -- (0.5,-0.75);
      \filldraw (-0.5,0) circle (0.05);
      \filldraw (0.5,0) circle (0.05);
      \filldraw (-0.5,-1.5) circle (0.05);
      \filldraw (0.5,-1.5) circle (0.05);
    \end{tikzpicture}
    \caption{}
    \label{eftFig1H}
  \end{subfigure}
  \caption{Sample of Feynman diagrams needed for the computation of the scattering angle in the EFT approach to 3PM order. Only the diagrams to one-loop order in (a)-(c) are needed to 2PM.}
  \label{eftfig1}
\end{figure}
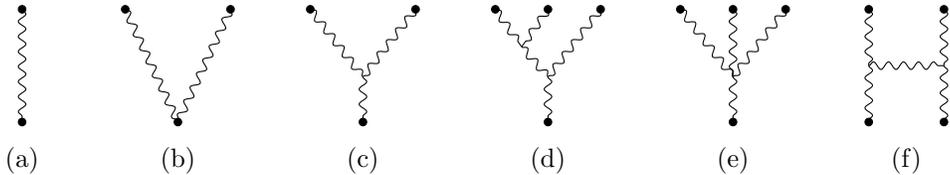

Using the EFT approach, in principle we can also derive the scattering angle in the PM framework (see also~\cite{damour1,damour2}).
The idea is simple. From the effective Lagrangian we can read off the total change in momentum,
\beq
\Delta \bp_a = \int_{-\infty}^{+\infty} \dd\sigma \frac{\partial L(x^\alpha,\dot x^\alpha,\cdots)}{\partial \dot \bx_a^\mu(\sigma)}\,,
\eeq
from which we obtain the scattering angle in the center of mass:
\beq
2 \sin \frac{\chi}{2} = \frac{|\Delta\bp_1|}{ p_\infty}\,,\label{sin2}
\eeq
with $|\Delta\bp_1|$ the total momentum change for particle 1, and the same for the companion.
The Lagrangian $L$ can be computed in the EFT approach to any order in the PM expansion.
By construction, the computation involves classical (point-like) sources, which simplifies the quantum problem from the onset.
However, we still rely on iterated Green's functions in the form of Feynman diagrams (see Fig.~\ref{eftfig1}) and, for the PM calculation, a series of relativistic integrals. Yet, once the deflection angle is known, we can plug it into our machinery to derive the $f_n$'s, thus reconstructing the scattering amplitude.
We can also perform the manipulations we describe in this paper to derive gauge invariant observables for bound orbits.
We will present the explicit derivation of the scattering angle in the EFT approach for relativistic sources in future work.
We will also further explore the connection between the EFT derivation and the one from the scattering amplitude elsewhere.\vskip 4pt

Let us add a comment on the map between test-particle and two-body dynamics. At 2PM, the computation of the deflection angle involves only diagrams to one-loop order, shown in (a)-(c) of Fig.~\ref{eftfig1}.
In the language of EFT, this is equivalent to computing the one-point function produced by e.g. particle 1, and evaluating it on the worldline effective action of particle 2, plus mirror image. For instance, the diagram in (b) comes from the expansion of the square-root in the point-particle action.
This one-point function can also be computed in the rest frame of particle 1, and then boosted to the center of mass.
(Alternatively, one can start with the boosted Schwarzschild solution.) The Lorentz transformation will naturally map ${\cal E}_0 \to \gamma$, as in~\eqref{boost}, while the mirror image will take care of the symmetrization.
It is clear then that the momentum change will be the same, obtained directly from the action, and the only difference in the scattering angle is the factor of $1/p_\infty$ in~\eqref{sin2}. Therefore, 
\beq
2 p_\infty(\gamma) \sin \frac{\chi_{\rm 1pt} (\gamma)}{2} =  2 p^{\rm test}_\infty({\cal E}_0 \to \gamma)\sin \frac{\chi_{\rm 
test}({\cal E}_0 \to \gamma)}{2} \,,
\eeq
where $\chi_{\rm 1pt}$ is the scattering angle obtained to one loop order from the one-point function, which is exact to 2PM (see Figs. \ref{eftfig1}(a)-(c)\,), and we used \beq  2p^{\rm test}_\infty({\cal E}_0 \to \gamma)/(2p_\infty(\gamma))=\Gamma\,.\eeq
(Notice this resembles the factors entering in the normalization of the amplitude in the no-recoil approximation.)
This implies,
\beq
\chi_{\rm 1pt} (\gamma) +{\cal O}(\chi_{\rm 1pt}^3) = \Gamma\, \chi_{\rm test}({\cal E}_0 \to \gamma) +{\cal O}(\chi_{\rm test}^3)\,,
\eeq
which leads to the map  found in~\cite{Vines:2018gqi}. The relationship clearly fails at 3PM, when terms such as the H-diagram (shown in Fig.~\ref{eftfig1}(f)) start to contribute. The no-recoil approximation, however, retains all of the higher order terms which involve the computation of the one-point function in the EFT approach, shown in diagrams (d) and (e) of Fig.~\ref{eftfig1} at two-loops. 

\subsubsection*{Non-Perturbative Dictionary}
The dictionary described in this paper is valid to all orders in the PM expansion. Therefore, knowledge of higher order terms in the scattering problem can be readily used to derive high-precision invariants for bound orbits.
While the steps we described were implemented in the context of the PM framework, in principle Firsov's formula in~\eqref{firsov1}, which can be parameterized as 
\beq
\begin{aligned}
r(\lambda,E) &= \lambda \, e^{-A(\lambda,E)}\,,\\
\widebar\bp^2(r(\lambda),E)&= e^{2A(\lambda,E)}\,,
\end{aligned}
\eeq
with
\beq
A(\lambda) \equiv \frac{1}{\pi} \int_\lambda^\infty \frac{\chi(\tilde b,E)\dd\tilde b}{\sqrt{\tilde b^2-\lambda^2}}\,,
\eeq
also holds in the non-perturbative regime. This means that, having a solution for the scattering angle as a function of the impact parameter and energy, for instance from numerical simulations, would translate into a solution for the dynamics of the two-body system in elliptic motion. Provided an ansatz for the dependence on the energy and impact parameter can be derived, one could perform the integral and analytically continue in the energy to construct the radial action, from which we can derive adiabatic invariants. As we mentioned earlier, the computation of the action does not require the precise knowledge of the $r_\pm$ endpoints, and therefore it would only be a matter of finding a suitable representation.\vskip 4pt

We could also use an ansatz for the non-perturbative scattering data to compute the minimum distance, see~\eqref{rmindis}, and analytically continue in impact parameter (as well as in the energy) to derive the two roots needed to characterize elliptic motion. Then, the condition for a circular orbit becomes
\beq
\left[ \int_{z}^\infty \frac{\chi(\tilde b,E)\dd\tilde b}{\sqrt{\tilde b^2-z^2}}-\int_{-z}^\infty \frac{\chi(\tilde b,E)\dd\tilde b}{\sqrt{\tilde b^2-z^2}}\right] = -i\frac{\pi^2}{2}\,.
\eeq
This expression must be understood as an analytic continuation in impact parameter space of the integral, evaluated at $b \to \pm z$, as described in~\S\ref{sec:invariants}. The (complex) solutions of the form $z=ib$ (with $b>0$) allow us to derive the reduced angular momentum, and subsequently the orbital frequency. Given that numerical simulations for the binary problem are time consuming, while scattering data appears significantly easier to collect, we think our formalism naturally opens up a new venue to explore the non-perturbative regime of the two-body problem in gravity, which deserves further exploration.

\subsubsection*{Classical Double Copy} 

It was discovered in \cite{Monteiro:2014cda} that classical spacetimes, such as Schwarzschild or Kerr black holes, can be shown to be double copies of gauge theory configurations. This means that test-particle (geodesic) motion in these background geometries can, in principle, also be mapped into each other.
In light of these developments, the impetus formula invites itself to speculations on its connection to the non-perturbative form of the double copy.
We will briefly comment on a few directions in what follows and return to this fascinating subject elsewhere.\vskip 4pt

In the no-recoil approximation, in the rest-frame of the heavy object, the impetus formula relates the scattering amplitude to motion in Schwarzschild's spacetime (see~\eqref{eq:Mnr} and~\eqref{motsch}), 
\beq
 {\bp}_{\rm Sch}^2(r,E) =  \mu^2({\cal E}_0^2-1) +  \widetilde{\cal M}_{\rm no\text{-}rec}(r,E)\,,
\eeq
 In turn, this can be written as geodesic motion,
\beq
g^{\mu\nu}_{\rm Sch} \hat p_\mu \hat p_\nu = 1\,,\label{eq:geo}
\eeq
with $g^{\mu\nu}_{\rm Sch}$ the Schwarzschild metric in isotropic coordinates, and $\hat p_0={\cal E}_0$. Therefore,
\beq
 \widetilde{\cal M}_{\rm no\text{-}rec}(r,E)=\left(g^{\mu\nu}_{\rm Sch}-\eta^{\mu\nu}\right)  p_\mu  p_\nu\,.
\eeq 
Since the equation in \eqref{eq:geo} is manifestly covariant, we can transform now to a different coordinate system.
We can then choose the Kerr-Schild coordinates that made the double copy manifest in~\cite{Monteiro:2014cda}.
In terms of the scattering amplitude, this would correspond to a Fourier transform with respect to a shifted momentum.\footnote{It would be interesting, using the fact that the amplitude is a Lorentz scalar, to explicitly construct the mapping between choices of coordinates and associated Fourier transforms of the amplitude.}
In these coordinates we have
\beq
g^{\mu\nu}_{\rm KS}-\eta^{\mu\nu} = \phi(r) k^\mu k^\nu\,,
\eeq 
with $\phi = 2GM/r$ and $k^\mu = (1,x^i/r)$, such that
\beq
\widetilde {\cal M}^{\rm KS}_{\rm test} = \frac{2GM}{r} (k\cdot p)^2\,.
\eeq
Hence, introducing
\beq
\widetilde{\cal A}^{\rm KS}_{\rm test} \equiv \frac{g c_a T^a}{r}  (k\cdot p)\,,
\eeq
with $c_a$ the color charge, the map found in \cite{Monteiro:2014cda} translates into a double copy relationship between scattering amplitudes in gravity and Yang-Mills, in the more traditional sense~\cite{bcj}. In fact, we can also look at this relationship in the other direction. In other words, taking the gauge theory amplitude and postulating the existence of a double copy map to classical gravity would imply the existence of the (linear in $G$!) Kerr-Schild solution for Schwarzschild. Moreover, following the approach in \cite{Harte:2016vwo}, one could also use the double copy to find other solutions of Einstein's equations. \vskip 4pt

Notice that the impetus formula also applies to gauge theory amplitudes.
Namely, for the relative momentum we have:
\beq
\bp^2(r,E) = p_\infty^2(E) + \frac{1}{2E} \int \dd^3 \br\, {\cal A}(\bp,\bq)e^{i\bq\cdot\br} + {\rm R.R.}\,,
\eeq
where ${\cal A}(\bp,\bq)$ is the classical Yang-Mills amplitude.
In principle, one can use the traditional double copy relating ${\cal M}$ to ${\cal A}$ to find the connection between the classical motion in both theories.
By mapping to the Yang-Mills case, this can potentially simplify the derivation of adiabatic invariants for the two body problem in gravity, including strong coupling.\vskip 4pt

Let us finish with yet another speculative idea.
In the derivation of the impetus formula in~\S\ref{sec:impetus} (see also the appendix~\ref{appA}) we map the problem into a Lippmann-Schwinger evolution equation for the case of potential scattering, resembling the Schr\"odinger problem.
As such, after analytic continuation to negative binding energies, the levels of the effective Hamiltonian in the Schr\"odinger-like equation correspond to the energies for elliptic orbits, through the identification of the adiabatic invariants.
(A similar idea was the spirit of the original effective one body map in~\cite{eob}.)
One can then imagine, following Dirac's steps, taking the square-root of the effective quantum problem.
Provided the double copy relation between gravity and gauge theory amplitude holds (schematically) $\sqrt{\cal M} \sim {\cal A}$, one could imagine then mapping the binding energies of the Dirac problem for Yang Mills to the binding energy for the two-body problem in gravity.
The precise form of this dictionary depends on the exact implementation of the double copy at the level of the classical amplitudes. We leave this as a spare time exercise for the reader.

\begin{acknowledgments}

  We thank the Munich Institute for Astro- and Particle Physics (MIAPP), supported by the DFG cluster of excellence ``Origin and Structure of the Universe'', and all the participants of the program ``Precision Gravity: From the LHC to LISA" for several fruitful discussions. In~particular, Zvi Bern, Poul Damgaard, Guillaume Faye, Donal O'Connell, Radu Roiban, Chia-Hsien Shen, Mikhail Solon, Jan Steinhoff, and Justin Vines. R.A.P. acknowledges financial support from the ERC Consolidator Grant ``Precision Gravity: From the LHC to LISA"  provided by the European Research Council (ERC) under the European Union's H2020 research and innovation programme (grant agreement No. 817791), as well as from the Deutsche Forschungsgemeinschaft (DFG, German Research Foundation) under Germany's Excellence Strategy (EXC 2121) `Quantum Universe' (390833306).
  The research of G.K. is supported by the Swedish Research Council under grant 621-2014-5722, the Knut and Alice Wallenberg Foundation under grant KAW 2013.0235, and the Ragnar S\"{o}derberg Foundation (Swedish Foundations' Starting Grant).

\end{acknowledgments}


\appendix

\section{The Impetus Formula}\label{appA}
There is another, more direct, way to prove the impetus formula in~\eqref{miracle}, when restricted to the `potential' region. Let us follow the same steps as in~\S\ref{sec:scat}, but take instead for the effective potential the expansion of $\bp^2$ as a function of the energy:
\beq
H_{\rm eff}|\psi_\bp(p_\infty)\rangle = \left(\bp^2 + V_{\rm eff}\right) |\psi_\bp(p_\infty)\rangle= p_\infty^2(E) |\psi_\bp(p_\infty)\rangle\,,\label{eq:eff2}
\eeq
with 
\beq
V_{\rm eff} =   -\sum_i P_i(E) \frac{G^i}{r^i}\,.\label{eq:veff2}
\eeq
Once again, we expect the solution to the full and effective problem to match to all orders in the PM expansion. Following the same steps as before, we have for the scattering amplitude of this (equivalent) Schr\"odinger problem 
\bea
\frac{4\pi}{\rm Vol} \tilde f(\bp^2,\bq)  =-  \langle \bp+\bq| V_{\rm eff} |\bp\rangle + \cdots = \frac{1}{\rm Vol} \sum_i \int \dd^3\br \left( P_i(E)\frac{G^i}{r^i}\right) e^{i\bq\cdot \br}+\cdots\,, \label{eq:Born}
\eea
written as a series of iterations of the Lippmann-Schwinger equation. From the relationship in~\eqref{eq:fM1}, adapted to~\eqref{eq:eff2}, we observe that the leading term, or Born approximation, already contains the information encoded in~\eqref{eq:fM0}.
Therefore, the impetus formula would hold, provided the additional iterations are composed of (super-classical) IR divergent terms. The latter do not contribute to the classical limit (since they cancel out between the two contributions described in \S\ref{sec:impetus}). We show that is the case below.\vskip 4pt

Before we proceed, let us add a few important comments regarding IR divergences and matching. Notice that in the effective theory of \eqref{eq:eff2}, the leading approximation for the scattering amplitude encodes all the physical information in the classical limit (thus far ignoring radiation-reaction effects). This is in contrast to the matching procedure in \cite{ira1,zvi1,zvi2}, where the iterations include terms which are crucial to read off the correct effective potential. Moreover, since the two problems are equivalent, the IR divergences must also cancel out in the matching of the scattering amplitudes between the full (relativistic) theory and effective description in \eqref{eq:eff2}. In other words, the divergences we uncover below for the series of iterations, must also appear in the relativistic scattering amplitude, modulo proper normalization. Indeed, by inspecting the form of the divergences obtained in \cite{zvi2}, we identify the same type of divergent integrals that appear in the effective theory. Once the two cancel against each other, the remaining finite terms --- which give us $V_{\rm eff}(E)$ in \eqref{eq:veff2} and ultimately the $P_n(E)$'s directly from the amplitude --- are therefore unambiguously obtained from the matching procedure. Following the manipulations in \cite{ira1,zvi1,zvi2}, it is also possible that a different choice of basis of master integrals could affect the resulting finite terms. Of course this is inconsequential. While some individual pieces may change, overall they must yield the same physical predictions.
This can be associated to the left-over freedom in the isotropic gauge, implicitly chosen by the Fourier transform of the amplitude. \vskip 4pt

We now move onto the proof that all iterations in the effective theory of \eqref{eq:eff2} do not contribute to the finite part in the classical limit, which goes as follows. 
(For the sake of notation, we omit below the volume factor and the overall $4\pi$ in the amplitude.) 
Since the $P_i(E)$'s are functions of the energy only, namely independent of the momentum,  the potential in the effective theory of~\eqref{eq:eff2} can be written as (in $d=3+\epsilon$ dimensions):
\beq
V_{\rm eff}(\bk,\bk',E) = -\sum_{n=1}^\infty P_n(E) \frac{G^n}{|\bk'-\bk|^{d-n}} \frac{(4\pi)^{d/2}\Gamma[d-n/2]}{2^n \Gamma[n/2]}\label{eq:veffn}\,.
\eeq
Moreover, the Green's function of the Schr\"odinger-like problem is simply given by 
\beq
G_0(\bk,E)= \frac{1}{H_0-E+i\epsilon} = \frac{1}{\bk^2-E+i\epsilon}\,. 
\eeq
Following the analysis in~\cite{zvi2}, we find that the $\ell$-iteration of the Lippmann-Schwinger equation, encoded in the ellipses in~\eqref{eq:Born}, can be written as
\beq
\sum_{\ell=1}^{n-1} \left[ \prod\limits_{i=1}^{\ell} \int \frac{\dd^d\bk_i}{(2\pi)^{d}}\right] \left[ \prod\limits_{i=1}^{\ell}\frac{1}{\bp^2-\bk_i^2+i\epsilon}\right]\left[ \prod\limits_{i=0}^{\ell}\frac{1}{|\bk_{i+1}-\bk_i|^2}\right]{\cal N}^{(\ell)}_{\rm eff}\,,\label{a6}
\eeq 
where $\bk_0=\bp$ and $\bk_{\ell+1}=\bp'$.
The numerator, ${\cal N}^{(\ell)}_{\rm eff}$, comes from expanding the potential in the classical limit as described in~\cite{zvi2}.
For instance, for the first iteration we have a constant ${\cal N}^{(1)}_{\rm eff}$, and it is straightforward to show the resulting integral is IR divergent (and purely imaginary):
\beq
\int \frac{\dd^d\bk}{(2\pi)^3} \frac{1}{(\bk^2-\bp^2+i\epsilon)|\bk-\bp|^2|\bk-\bp'|^2} = \frac{i}{8\pi}\frac{1}{|\bp|}\left(\frac{(\log\bq^2+2/(d-3)}{\bq^2}\right) \,, 
\eeq
with $\bq=\bp'-\bp$.
Notice it is also a super-classical contribution, since it scales with an extra power of $|\bq|^{-1}$ with respect to the classical term at this order. \vskip 4pt

We can now proceed by induction in the power counting.
We will show the rather intuitive fact that these super-classical terms cannot generate classical contributions through mixing.
Let us return to the Lippmann-Schwinger equation in~\eqref{eq:Born}. For the $n$-th order scattering amplitude in the effective Schr\"odinger problem, we have the recursion formula:
\beq
\tilde f^{(n)}(\bp,\bp') = -V^{(n)}_{\rm eff} (\bp,\bp') + \sum_\ell  \int \dd^d\bk \frac{V^{(\ell)}_{\rm eff} (\bp,\bk) \tilde f^{(n-\ell)}(\bp',\bk)}{\bp^2-\bk^2+i\epsilon}\,,
\eeq
which we can rewrite as
\bea
\tilde f^{(n)}(\bp,\bp') + V^{(n)}_{\rm eff} (\bp,\bp') &=&  -\underbrace{\sum_\ell  \int \dd^d\bk \frac{V^{(\ell)}_{\rm eff} (\bp,\bk) V_{\rm eff}^{(n-\ell)}(\bp',\bk)}{\bp'^2-\bk^2+i\epsilon}}_{(1)} \label{eq8}\\ 
&+& \underbrace{\sum_\ell  \int \dd^d\bk \frac{V^{(\ell)}_{\rm eff} (\bp,\bk) \left(\tilde f^{(n-\ell)}(\bp',\bk) + V^{(n-\ell)}_{\rm eff}(\bp',\bk)\right)}{\bp'^2-\bk^2+i\epsilon}}_{(2)}\nn\,.
\eea
Let us assume now that the lower amplitude is also super-classical, and it takes the form 
\beq 
\tilde f^{(m)}(\bk',\bk)+V_{\rm eff}^{(m)}(\bk',\bk) = \sum_{j>0} \frac{b_j(|\bk+\bk'|^2)}{|\bk-\bk'|^{d-m+j}}\, \label{eq9}\,.
\eeq
We have shown this is the case for the lowest iteration, scaling as $1/|\bq|^2$ rather than $1/|\bq|^{1}$, and we are allowing for more generic super-classical terms. 
Let us concentrate first on the second term, which scales as (up to numerical factors) 
\beq
 (2) = \sum_{\ell=2}^{n-1} \sum_{m=1}^{\ell} \sum_{j>0} b_j(\bp^2) \int \dd^d\bk \, \frac{1}{|\bp-\bk|^{\rho}} \frac{1}{|\bk-\bp'|^{\sigma}}\frac{1}{(\bk^2-\bp^2)^\gamma}  
\eeq
with $\rho=d-m, \sigma= d-n+\ell+j$ and $\gamma=1$, after using the form of the potential in~\eqref{eq:veffn} together with~\eqref{eq9}.
We now expand this integral in the classical limit, keep the scaling $\bk^2-\bp^2\sim \bq^2$, to transform it into the form in~\eqref{a6}.
Through a change of variables, $\bl=\bp-\bk$, we can map these integral into sums over integrals like Eq.~(7.8) of~\cite{zvi2} (with $\bw=\bp'-\bp=\bq$ and $\bz=\bp$ in their notation), 
\beq
\int \dd^d\bl \, \frac{f^{(\alpha\beta\gamma)}(\bl,\bp,\bq)}{|\bl|^{\alpha}|\bl+\bq|^{\beta}(2\bl\cdot\bp+\bl^2)^\gamma}\,,  
\eeq 
with  $\alpha>0,\beta>0,\gamma=1$, and $f^{(\alpha\beta\gamma)}(\bl,\bp,\bq)$ a polynomial.
As it was shown in~\cite{zvi2}, the condition $\gamma=1$ leads to super-classical contributions (which moreover are also IR divergent).
The remaining contribution from $(1)$ in~\eqref{eq9} also takes on the same form, with $j=0$.
Similarly to the first iteration of the Coulomb potential, and by the same token, the first term in~\eqref{eq8} also produces IR divergent super-classical contributions.
Since these terms cancel out in the classical limit, and also in the matching between our effective theory in \eqref{eq:eff2} and the full relativistic computation, this completes the proof of the impetus formula.
Notice, as a byproduct of the above result, we have shown that there are no correction beyond the Born approximation for the classical scattering in $1/r^n$ potentials, generalizing the Coulomb~case.\footnote{
  This resolves the naive conflict between the impetus formula and the case of $\bp$-independent coefficients ($dc_i/d\bp=0$) in a non-relativistic theory, for which $H=\bp^2+\sum_i c_iG^i/r^i$ implies $P_i=-c_i$, and the Born approximation already leads to~\eqref{miracle}.}
\newpage
\bibliographystyle{JHEP}
\bibliography{references}

\end{document}